\PassOptionsToPackage{table}{xcolor} 
\documentclass[preprint,12pt]{elsarticle}

\usepackage{url}
\usepackage[hidelinks]{hyperref}
\usepackage{amssymb}
\usepackage{amsmath}
\usepackage{lineno}
\usepackage{multirow}
\usepackage{longtable}
\usepackage{booktabs}
\usepackage{array}     
\usepackage{subcaption}  
\usepackage{makecell} 
\usepackage[table]{xcolor}
\usepackage{float}
\setcitestyle{open={},close={}}

\journal{Journal of Pathology Informatics}

\begin{document}

\begin{frontmatter}

\title{ADPv2: A Hierarchical Histological Tissue
Type-Annotated Dataset for Potential Biomarker Discovery of Colorectal Disease}

\author[inst1]{Zhiyuan Yang}
\author[inst2]{Kai Li}
\author[inst3]{Sophia Ghamoshi Ramandi}
\author[inst4]{Patricia Brassard}
\author[inst10]{Hakim Khellaf}
\author[inst5,inst6]{Vincent Quoc-Huy Trinh}
\author[inst2]{Jennifer Zhang}
\author[inst7,inst8]{Lina Chen}
\author[inst8]{Corwyn Rowsell}
\author[inst9]{Sonal Varma}
\author[inst2]{Kostas Plataniotis}
\author[inst1]{Mahdi S. Hosseini\fnref{cor1}}
\fntext[cor1]{Corresponding author.}
\ead{mahdi.hosseini@concordia.ca}

\address[inst1]{Department of Computer Science \& Software Engineering, Concordia University, 2155 Guy St, Montreal, QC H3H 2L9, Canada}
\address[inst2]{Department of Electrical \& Computer Engineering, University of Toronto, 10 King's College Rd, Toronto, ON M5S 3G8, Canada}
\address[inst3]{Department of Chemistry \& Biology, Toronto Metropolitan University, 350 Victoria St. Toronto, ON M5B 2K3, Canada}
\address[inst4]{Department of Medicine, Université de Montréal, Pavillon Roger-Gaudry, 2900 Edouard Montpetit Blvd, Montreal, QC H3T 1J4, Canada}
\address[inst5]{Axe Cancer, Centre de recherche du CHUM, 900 Saint-Denis St, Montréal, QC H2X 0A9, Canada}
\address[inst6]{Institut de recherche en immunologie et cancérologie, Université de Montréal, Marcelle-Coutu Pavilion, 2950 Chem. de Polytechnique, Montréal, QC H3T 1J4, Canada}
\address[inst7]{Anatomic Pathology, Sunnybrook Health Sciences Centre, 2075 Bayview Ave, Toronto, ON M4N 3M5, Canada}
\address[inst8]{Department of Laboratory Medicine \& Pathobiology, University of Toronto, Simcoe Hall, 1 King's College Circle, Toronto, ON M5S 3K3, Canada
}
\address[inst9]{Department of Pathology \& Molecular Medicine, Queen’s University, 88 Stuart Street
Queen's University
Kingston, ON K7L 3N6
Canada}
\address[inst10]{Department of Pathology \& Molecular Medicine, Université de Montréal, 2900 Édouard-Montpetit Blvd, Montréal, QC H3T 1J4, Canada}

\begin{abstract}
Computational pathology (CoPath) leverages histopathology images to enhance diagnostic precision and reproducibility in clinical pathology. However, publicly available datasets for CoPath that are annotated with extensive histological tissue type (HTT) taxonomies at a granular level remain scarce due to the significant expertise and high annotation costs required. Existing datasets, such as the Atlas of Digital Pathology (ADP), address this by offering diverse HTT annotations generalized to multiple organs, but limit the capability for in-depth studies on specific organ diseases. Building upon this foundation, we introduce ADPv2, a novel dataset focused on gastrointestinal histopathology. Our dataset comprises 20,004 image patches derived from healthy colon biopsy slides, annotated according to a hierarchical taxonomy of 32 distinct HTTs of 3 levels. Furthermore, we train a multilabel representation learning model following a two-stage training procedure on our ADPv2 dataset. We leverage the VMamba architecture and achieving a mean average precision (mAP) of 0.88 in multilabel classification of colon HTTs. Finally, we show that our dataset is capable of an organ-specific in-depth study for potential biomarker discovery by analyzing the model's prediction behavior on tissues affected by different colon diseases, which reveals statistical patterns that confirm the two pathological pathways of colon cancer development. Our dataset is publicly available here: \href{https://zenodo.org/records/15307021}{Part 1}, \href{https://zenodo.org/records/15312384}{Part 2} and \href{https://zenodo.org/records/15312792}{Part 3}. 
\end{abstract}

\begin{keyword}
Deep Learning \sep Multilabel Representation \sep Computational Pathology \sep ADPv2 Dataset \sep Biomarker Discovery
\end{keyword}

\end{frontmatter}


\section{Introduction}

In recent years, colorectal cancer has become the third most common cancer worldwide and a leading cause of cancer death in individuals under the age of 50. \cite{morgan2023global} \cite{bray2024global} Colorectal carcinomas typically originate from local precursor polyps or lesions through well-described pathways, including the classical adenoma-carcinoma sequence and the alternate serrated pathway. \cite{arends2013pathways} Lower gastrointestinal endoscopy is generally employed for screening in the general population and allows direct visualization and instrumented retrieval of polyps and other suspicious lesions for further pathological evaluation. \cite{winter2024screening} Histopathological image analysis by pathologists is considered the gold standard for diagnosis of both colorectal cancer and its precursors, and for assessment of tumor staging and grading, surgical margin status and other relevant pathological features. \cite{washington2021diagnosis} \cite{hiareview}

Computational pathology (CoPath) is an interdisciplinary branch of pathology focusing on computational analysis and modelling of medical histopathology images, alleviating the workflow of pathologists. CoPath typically leverages artificial intelligence methods such as deep learning. The objective of CPath is to develop computer-aided diagnostics  systems that elicit transformational changes in disease diagnosis and treatment. Recent advancements in DL coupled with increasing ease of data flow from digital pathology are facilitating the integration of CPath into routine clinical workflow. \cite{cpath} \cite{dpa} In addition to enabling a more efficient pathology workflow, CPath provides pathologists with comprehensive and personalized tools to monitor complex diseases, leading to improved patient care. 

AI-based CoPath can process enormous quantities of digital pathology data to improve pathologic diagnosis, classification, prediction, and prognosis of diseases.\cite{aicpath} AI-based CPath generally utilizes DL models, including Convolutional Neural Networks\cite{cnn}, Graph Convolutional Networks \cite{gcn}, vision transformers\cite{vit}, and most recently, vision state space models.\cite{vmamba} The input data for training these models comes from digitized glass slides of tissue samples via whole-slide scanners, often up to a magnification of 400x. These resulting images, known as Whole Slide Images (WSI), are at a gigapixel scale and can be examined digitally by pathologists.\cite{acrobat} Currently, the majority of CoPath methods are performed on tissues stained with hematoxylin and eosin (H\&E).\cite{aidpath} \cite{aidpath1} 

In practice, using WSIs with deep learning methods involves numerous challenges. The first difficulty is the lack of imaging data in histopathology due to low incidence rates and a lack of publicly available data. Existing datasets are inherently much smaller than the natural image scene datasets used to train the state-of-the-art deep learning models used today. \cite{histopath} Applying histopathology datasets directly to these models will not produce ideal results. Secondly, extensive manual annotation is required on these datasets to create meaningful deep learning models. Also, adequate slide preparation is necessary for accurate model results. Artifacts such as blurriness, overstaining, understaining, air bubbles, and folded tissue can compromise results. \cite{aicpath} Moreover, unlike general image classification, granular patch-level annotations are crucial for WSI classification tasks due to their size. \cite{adp} To handle these, models must be computationally powerful and time-efficient. Cancer detection in particular introduces additional challenges such as variations within classes due to subtle differences in grading and subtypes across samples. \cite{cpath} Furthermore, many publicly available datasets feature either over-generalized slide-level annotations, or pixel/patch-level annotations too specific a few tissue types. To this end, it is more plausible to train on a dataset that is annotated according to all tissue types relevant to an organ, which strikes a balance in targeting a specific organ while general to all relevant disease possibilities.

Pathology-based dataset curation is expensive, time-consuming, and laborious. \cite{wahab2022semantic} Domain experts are required for annotations, and images are not easily shareable due to the privacy of patient data, so annotation crowdsourcing is generally not viable. \cite{wahab2022semantic}
Fortunately, this bottleneck can be mitigated through unsupervised learning, a machine learning technique that can leverage unannotated data to improve task performance. More specifically, self-supervised learning (SSL), a branch of unsupervised learning where the model generates supervisory signals based on training data, has found success in histopathological images. \cite{ciga} Experiments performed by Ciga et al. on the ILSVRC2012 digital histopathology dataset have shown that a certain contrastive SSL technique, SimCLR \cite{simclr}, is capable of achieving performance rivaling that of fully supervised training. 

Taking into account the limitations mentioned above, we propose ADPv2, a new dataset consisting of 20,004 image patches annotated according to a hierarchical taxonomy consisting of 32 types of colon tissue, extracted from 461 healthy colon biopsy slides. We leverage the recent self-supervised learning technique, Barlow Twins, to overcome the shortage in labelled training data, and utilize the recent VMamba model to achieve accurate multilabel classification of these tissue types. 

We furthermore demonstrate the utility of our learned multilabel model by analyzing our model's confidence scores on unseen images of polyp subtypes, and deriving useful insights to improve differentiation between closely related colorectal polyps. By evaluating the model’s predicted confidence scores on regions of interest (RoIs) extracted from four distinct pathological subtypes—Hyperplastic Polyps (HP), Sessile Serrated Lesions (SSL), Tubular Adenomas (TA), and Tubulovillous Adenomas (TVA)—we observe systematic shifts in the distribution of confidence scores relative to normal tissue. These shifts, characterized by decreased peak sharpness, leftward displacement, and broader spread, align with pathological expectations of two distinct colon cancer development pathways. This supports our hypothesis that deep models trained on healthy tissue can act as sensitive probes for detecting structural abnormalities in diseased tissue.

In short, our contributions are as follows:
\begin{itemize}\item Development of ADPv2 Dataset, a comprehensive dataset comprising 20,004 image patches extracted from healthy colon biopsy slides, each annotated with a 32-label hierarchical histological tissue taxonomy. Patch collection for our dataset is ongoing and the dataset will be continually updated. The latest version of our dataset is available for download at the following links 
    \begin{enumerate}
      \item \href{https://zenodo.org/records/15307021}{ADPv2 Dataset – Part 1}
      \item \href{https://zenodo.org/records/15312384}{ADPv2 Dataset – Part 2}
      \item \href{https://zenodo.org/records/15312792}{ADPv2 Dataset – Part 3}
    \end{enumerate}
The dataset accumulation is ongoing with new patches being continuously added.

\item{Development of multilabel tissue type classification model using VMamba architecture following a two-stage SSL pretraining and supervised fine-tuning procedure with strong classification performance on the ADPv2 dataset, demonstrating the merit of the model architecture and the quality of our data.}

\item{We show that, by comparing the statistical change in model's predicted confidence score on gland areas, we can confirm the two pathological pathways of colon cancer development and demonstrate the potential of future biomarker discovery for colon cancers.}\end{itemize}

\section{Related Works}

\subsection{Deep Learning in Computational Pathology}
Ongoing developments in advanced machine learning techniques have introduced novel DL frameworks for pathomics, the analysis of histopathological digital images with the aim of extracting quantitative features for clinical decision-making. \cite{Iv154} These techniques can uncover previously unknown relationships and features that provide biological insights and improve disease characterization. \cite{gupta2019emergence} In one study, Romo-Bucheli et al., \cite{romo2016automated} employ DL methods to learn granular features within the nuclei, discovering that nuclei tubule prominence, intensity, multicentricity, shape and texture, may predict breast cancer recurrence. Bejnordi and colleagues \cite{ehteshami2018using} highlighted rich insights which can be extracted from the stroma in breast tissue slides. Three CNN networks were trained in hierarchical fashion to segment stroma, identify tumor-stromal content, then derive a score for the likelihood of tumor malignancy. Colorectal cancer biomarker discovery has focused on analyzing genetic events in key genes to model tumor progression and observing differences in protein expression to differentiate between healthy and tumor tissues. \cite{martingarcia2024biomarker} \cite{10.1634/theoncologist.2009-0233} \cite{sole2014discovery} Some recent studies leveraged histology slides for deep learning-aided detection of microsatellite instability (MSI), a key biomarker for tumor mutation and immunogenicity. Gustav et al., \cite{gustav2024deep} use a transformer based deep learning model on colorectal tissue images, identifying both MSI and polymerase epsilon (POLE) mutations in tumors. Despite extensive research in colorectal biomarker discovery, to the best of our knowledge, there are only few studies that utilize deep learning approaches on histopathological slides from colorectal samples to derive biomarkers for differentiating between various types of colorectal polyps. \cite{wei2020evaluation} \cite{nasirmoin2021ai} \cite{kim2023adenocarcinoma} \cite{korbar2017polyp}

\subsection{Datasets for CoPath}
Although publicly available pathology datasets leveraging Whole Slide Images (WSIs) and patch-level annotations have grown significantly in recent years, crucial gaps still exist. Recent datasets primarily address multiclass or multilabel classification, incorporating region-level annotations, yet few feature extensive taxonomies (10+ labels). For instance, the BACH dataset \cite{bach2018} provides multiclass breast cancer annotations but only includes four classes. PANDA \cite{panda2020} features detailed prostate cancer annotations across several grades but does not meet the multilabel criterion. The Lizard \cite{lizard2021}, PanNuke \cite{pannuke2020}, and NCT-CRC-HE-100K \cite{nctcrc2019} datasets provide multiclass region-level annotations, but their label taxonomies remain limited to fewer than ten classes each, despite their diversity. ADPv1 \cite{adp} notably offers an extensive taxonomy (57 labels) , but covers various organs and integrates both normal and diseased tissues, not exclusively focusing on healthy samples of one organ type. Our proposed ADPv2 dataset with its extensive 32-label taxonomy annotated exclusively on healthy colon slides offers a novel approach to the task of colon disease diagnosis and polyp differentiation.

\subsection{Self-supervised Learning in CoPath}
Self-supervised learning (SSL) has become indispensable in computational pathology because exhaustive annotation of WSIs is prohibitively expensive. By pre-training on millions of unlabeled tissue tiles, SSL models acquire generic morphological features that can be fine-tuned with the limited expert labels available, consistently narrowing the performance gap between pathology-specific networks and those initialized from natural-image datasets such as ImageNet.

Early SSL success in CoPath came from contrastive objectives that pull two augmented views of the same tile together while pushing different tiles apart. SimCLR \cite{simclr} delivers strong features but demands very large batches to expose sufficient negatives; MoCo \cite{moco} sidesteps that requirement with a momentum-updated memory bank, trading batch size for extra bookkeeping. In colon-polyp studies, these contrastive schemes work well when stain-aware color jitter and multi-magnification cropping are carefully tuned, but their memory overhead limits their use on high-resolution slides. A second family removes explicit negatives and learns by feature prediction or distillation. BYOL \cite{byol} trains a “student” network to regress the embeddings of an exponential-moving-average “teacher,” while DINO \cite{dino} aligns the output distributions of two networks under diverse augmentations. Both methods cut GPU memory nearly in half relative to SimCLR and have shown stable convergence on imbalanced tissue datasets, though they can become sensitive to tile-sampling bias if the teacher lags too far behind the student.

More recently, redundancy-reduction objectives such as Barlow Twins \cite{barlow-twins} and VICReg \cite{vicreg} decorrelate features across views instead of contrasting samples. These objectives work with moderate batch sizes, require no memory bank, and encourage diverse texture cues that are particularly valuable for rare glandular patterns in gastrointestinal (GI) slides.

\subsection{Multilabel Classification}
Multilabel classification underpins many pathology tasks because a single tissue tile can express several histological attributes that sit at different depths of a diagnostic taxonomy. Two obstacles dominate. (i) Extreme label imbalance: rare lesions or gland patterns may appear in fewer than 1 \% of patches, so naive training is swamped by negatives. (ii) Strong label dependence and hierarchy: the presence of a high-level “Inflammatory cells” tag, for example, constrains the probabilities of its child classes (lymphocytes, eosinophils, etc.). Standard accuracy metrics overlook these subtleties, forcing researchers to rely on set-based scores like micro or macro F1.

The usual starting point is independent Binary-Cross-Entropy (BCE) with a sigmoid per label, but BCE amplifies the positive-negative skew: every patch contributes dozens of easy negative terms, drowning the rare positives in gradient noise. Re-weighting schemes seek to restore balance. Focal loss \cite{focal_loss} down-weights well-classified negatives; the asymmetric loss (ASL) of Ridnik et al.\cite{asl} goes further by giving separate, tunable exponents to positives and negatives and by introducing a hard-negative threshold, yielding consistent gains on long-tailed medical datasets. Class-balanced BCE and re-sampling strategies offer similar relief but at the cost of extra passes through the data or fragile weighting heuristics. However, none of these methods capture label correlations.

Dependencies can be introduced explicitly. Classifier chains\cite{class-chain} pass predicted labels down a sequence of binary heads so that each head conditions on the previous ones; performance rises, but inference grows linear in the label count and results depend on chain order. More scalable are graph-based or hierarchical networks that mirror the label tree in their architecture. HMCN\cite{mhcn} attaches shared low-level convolutional features to branch-specific classifiers, propagating information from coarse to fine labels. HiMulConE\cite{hmce} adds a supervised-contrastive objective\cite{supcon} that pulls embeddings of tiles sharing any ancestor closer together while enforcing a monotonicity constraint—child confidences can never exceed those of their ancestors. Similar hierarchy-aware ideas now appear in graph-neural-network heads and conditional probability models for WSIs.

\section{Dataset}
\begin{figure}
    \centering
    \includegraphics[width=\linewidth]{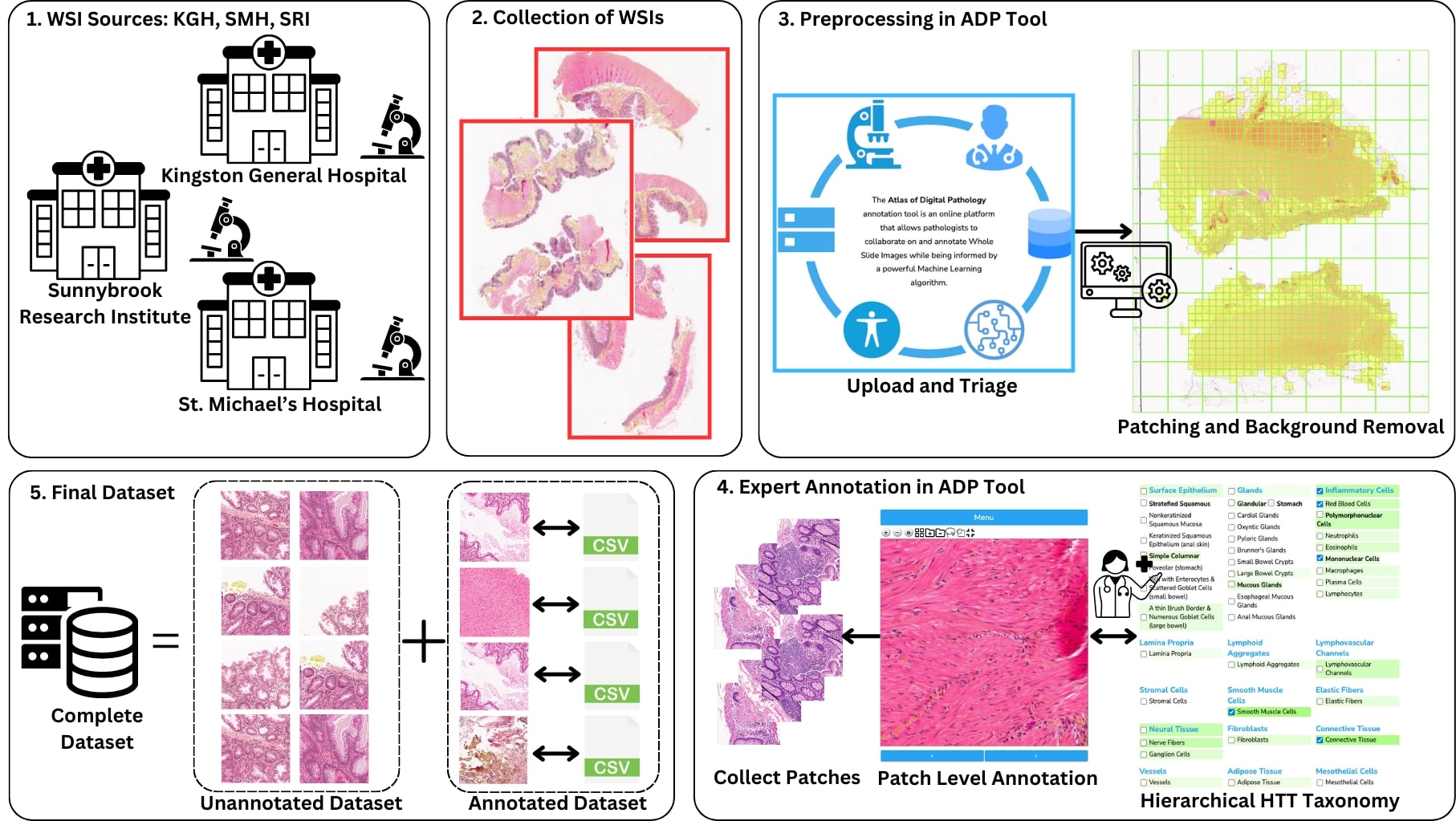}
    \caption{High-level overview of our WSI curation workflow. First, biopsy slides are prepared and scanned from numerous hospitals. Afterwards, we collect and preprocess these slides by uploading them to our ADP tool for automatic patching and background filtering. Then, our expert in-house annotator annotates these image patches according to our hierarchical taxonomy. Lastly, we compile these patches to form our dataset, which consists of both annotated and unannotated. These are later used in our training procedures.  
    \label{workflow}}
\end{figure}

In this section, we detail the process of developing our digital pathology database, ADPv2. Figure \ref{workflow} illustrates our data collection procedure. We first collect WSIs from multiple hospitals, preprocess the WSIs using our online ADP annotation platform to extract patches for annotations. Pathologists inspect and select \textbf{all} HTTs associated with each patch, creating a \textbf{multilabel} annotation for each patch. The final complete dataset is split into an annotated split and an unannotated split, both of which will be presented to the public for research and development use. 

\subsection{WSI Collection}

The 461 whole slide images (WSIs) used in this project originate from multiple sources, including "The Cancer Genome Atlas (TCGA)" provided by the National Cancer Institute (NCI); Kingston General Hospital (KGH; Canada), St. Michael's Hospital (SMH; Canada), and Sunnybrook Research Institute (SRI; Canada). Each slide used in our project is derived from healthy colon biopsies. KGH slides are acquired using a TissueScope LE brighfield Scanner while SMH and SRI slides are acquired from TissueScope IQ brighfield Scanner. The staining techniques, image magnifications, and resolutions vary across institutions, as shown in Table \ref{tab:WSI_info}. 

\begin{table}[ht]
\centering
\scriptsize
\resizebox{\linewidth}{!}{%
\begin{tabular}{l c c c c l}
\hline
\textbf{Institution} & \textbf{Slide Count} & \textbf{Magnification} & \textbf{Resolution} & \textbf{Patch Size} & \textbf{Staining} \\
\hline
Kingston General Hospital (KGH) & 200 & 20x & 0.4 mpp & 1360 x 1360 & Hematoxylin phloxine saffron (HPS) \\
The Cancer Genome Atlas (TCGA) & 110 & 20x & 0.5 mpp & 1088 x 1088 & Hematoxylin and eosin (H\&E) \\
St. Michael's Hospital (SMH)& 131 & 50x & 0.2 mpp & 2720 x 2720 & Hematoxylin and eosin (H\&E) \\
Sunnybrook Research Institute (SRI) & 20 & 40x & 0.25 mpp & 2176 x 2176 & Hematoxylin and eosin (H\&E) \\
\hline
\end{tabular}%
}
\caption{Properties of the WSIs used in our training, including institution, slide count, magnification, resolution, patch size, and staining method.}
\label{tab:WSI_info}
\end{table}

\subsection{Patch Extraction and Preprocessing}
We evenly extract non-overlapping image patches of fixed size 544 x 544 microns from each of our slides. To account for variations in image resolutions of WSIs, we adjust the digital resolution of the patches according to their microns per pixel (mpp) such that each image patch represents the same physical patch of 544 x 544 microns. Namely, we compute pixel resolution as follows: \[Image\ Resolution = \frac{544\mu m}{mpp}\]

Refer to table \ref{tab:WSI_info} for specific pixel resolutions of each type of slide. 
The extracted patches are then passed into a background detection algorithm that filters out non-tissue patches by comparing the pixel colors and contrast of the patch. As a result, we extracted a total of 125,808 non-background image patches. Among these patches, 20,004 image patches are selected by our annotator for labeling. Only patches deemed the most informative and containing sufficient numbers of tissue types from our taxonomy were selected. 

\subsection{Annotation Process}
The annotation of this dataset was conducted on colon polyp WSIs at patch level using the Atlas of Digital Pathology annotation tool. The annotator was trained under close supervision of an expert pathologist to specifically recognize gastrointestinal (GI) tract tissues under different staining conditions on WSIs. The annotation criteria were designed to be highly precise, ensuring maximum specificity. The detection of even a single cell within a patch warranted its labeling, regardless of the number or quantity of cells present. All annotated patches are subsequently evaluated by the expert pathologist for accuracy. 

\begin{table}[h!]
\centering
\small 
\resizebox{\textwidth}{!}{%
\begin{tabular}{|c|c|c|c|c|}
\hline
\multirow{2}{*}{\#} 
 & \multirow{2}{*}{Level 1} 
 & \multirow{2}{*}{Level 2} 
 & \multirow{2}{*}{Level 3}
 & \multirow{2}{*}{Count} \\
 &  &  &  &  \\
\hline
1 & \multirow{5}{*}{\textcolor{green}{Surface Epithelium (SE)}} 
  & \multirow{2}{*}{\textcolor{red}{Stratified Squamous (SS)}} 
  & \textcolor{red}{Nonkeratinized squamous mucosa (NSM)} 
  & - \\
\cline{4-5}
2 &  &  & \textcolor{red}{Keratinized squamous epithelium (anal skin) (KSE)} & - \\
\cline{3-5}
3 &  & \multirow{3}{*}{Simple Columnar (SCL)} 
  & \textcolor{red}{Foveolar (stomach) (F)}
  & - \\
\cline{4-5}
4 &  &  & \multirow{2}{*}{Small Bowel and Large Bowel (VESGC + BBNGC)} & \multirow{2}{*}{2729} \\
5 &  &  & & \\
\cline{2-5}
6 & \multirow{4}{*}{\textcolor{green}{Glands (GD)}} 
  & \multirow{2}{*}{Glandular (GL)} 
  & \multirow{2}{*}{Small Bowel and Large Bowel Crypts (SBC + LBC)} & \multirow{2}{*}{7493} \\
7 &  &  & & \\
\cline{3-5}
8 &  & \multirow{2}{*}{\textcolor{red}{Mucous Glands (MG)}} & \textcolor{red}{Esophageal mucous glands (EMG)} & - \\
\cline{4-5}
9 &  &  & \textcolor{red}{Anal mucous glands (AMG)} & - \\
\cline{2-5}
10 & -- & -- & \textcolor{green}{Lamina Propria (LP)} & 7181 \\
\cline{2-5}
11 & -- & -- & \textcolor{green}{Lymphoid Aggregates (LA)} & 1382 \\
\cline{2-5}
12 & -- & -- & \textcolor{green}{Lymphovascular Channels (LC)} & 16418 \\
\cline{2-5}
13 & -- & -- & \textcolor{green}{Smooth Muscle Cells (SMC)} & 13840 \\
\cline{2-5}
14 & -- & -- & \textcolor{green}{Elastic Fibers (EF)} & 1706 \\
\cline{2-5}
15 & \multirow{2}{*}{\textcolor{green}{Neural Tissue (NT)}} 
   & \multirow{2}{*}{--} 
   & Nerve Fiber (NF) & 4159 \\
\cline{4-5}
16 &  &  & Ganglion Cells (GC) & 2644 \\
\cline{2-5}
17 & -- & -- & Connective Tissue (CT) & 19277 \\
\cline{2-5}
18 & \multirow{6}{*}{Inflammatory cells and other blood cells(IC)}& \multirow{1}{*}{Polymorphonuclear Cells (POC)} 
   & Neutrophils (NTR) & 2391 \\
\cline{4-5}
19 &  &  & \textcolor{green}{Eosinophils (ES)} & 6799 \\
\cline{3-5}
20 &  & \multirow{3}{*}{Mononuclear Cells (MOC)} & \textcolor{green}{Macrophages (MA)} & 1246 \\
\cline{4-5}
21 &  &  & \textcolor{green}{Plasma Cells (PC)} & 7257 \\
\cline{4-5}
22 &  &  & \textcolor{green}{Lymphocytes (LY)} & 8142 \\
\cline{3-5}
23 &  & -- & \textcolor{green}{Red Blood Cells (RBC)} & 16543 \\
\cline{2-5}
24 & -- & -- & \textcolor{green}{Vessels (V)} & 1735 \\
\cline{2-5}
25 & -- & -- & \textcolor{green}{Adipose Tissue (AT)} & 3621 \\
\cline{2-5}
26 & -- & -- & \textcolor{red}{Mesothelial Cells (MC)} & - \\ 
\hline
\end{tabular}%
}
\caption{Hierarchical taxonomy of histological tissue types in ADPv2. Red indicates HTTs without any presence in our current dataset. Green indicates the HTTs included in our multilabel representation model training.}
\label{htt}
\end{table}

All annotations follow a predefined hierarchical taxonomy of Histological Tissue Types (HTTs), as shown in Table \ref{htt}. Although our dataset only contains patches extracted from colon polyp WSIs, the taxonomy includes the entire GI tract tissues for possible future dataset expansion. Developed by expert pathologists, this taxonomy carefully selected 32 HTTs belonging to three levels of hierarchy based on their critical role in both normal physiology and pathological transformations. These HTTs generally represent the key sites where precancerous and cancerous changes occur. For instance, the HTTs under the \textit{Surface Epithelium} branch are often examined for the shift from stratified squamous to columnar epithelium, which is a well-documented precancerous change. Similarly, glandular structures (under the \textit{Glands} branch), including crypts, are common sites of dysplasia and neoplastic transformation. Inflammatory cells were included due to their role in the tumor microenvironment and chronic inflammation-driven carcinogenesis. Additionally, connective tissue, stromal, neural, and vascular components were incorporated due to their role in tumor progression, as remodelling, perineural invasion, and lymphovascular invasion are key markers of cancer aggressiveness. 

\subsection{Dataset Statistics and HTT Interaction}

\begin{figure}[!ht]
    \centering
    \includegraphics[width=0.75\linewidth]{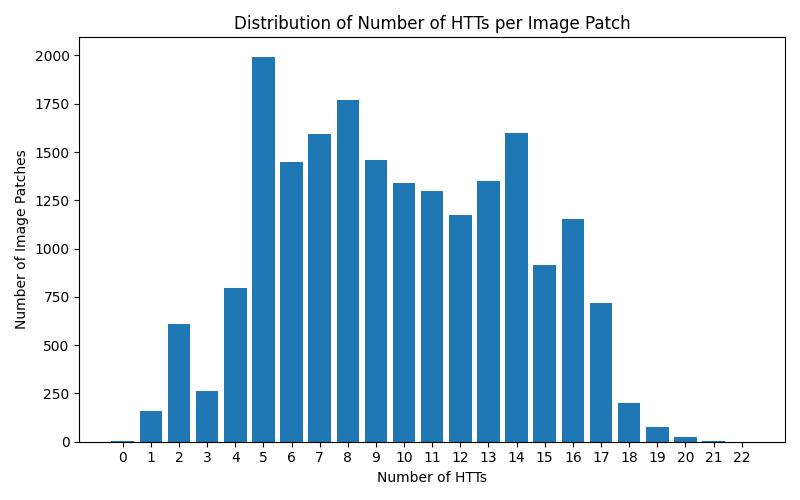}
    \caption{Distribution of the number of HTTs found per annotated image patch in the ADPv2 dataset. Annotated patches are associated with 1 to 22 HTT labels, with a mean value of 10 HTT labels. 90\% of the patches have more than 9 labels.}
    \label{label_hist}
\end{figure}

As described in Section 3.2, our database comprises a total of 20,004 multilabel annotated image patches.  Each image patch is associated with at least one HTT label to a highest of 22 labels, with an average of 10 HTTs being associated with each annotated patch in our dataset, as shown in Figure \ref{label_hist}. It can be observed that 90\% of the image patches in the dataset have more than 9 labels, and 75\% of them contain between five and 15 labels. This pattern reflects the inherent heterogeneity of gastrointestinal tissues, where multiple structures coexist within a given area. The layered organization of the colon, consisting of the mucosa, submucosa, muscularis, and serosa, means that a single patch often captures portions of multiple layers. For example, a single patch may contain epithelial components such as simple columnar epithelium and glandular structures alongside connective tissue elements, including the lamina propria and smooth muscle cells. Additionally, patches frequently include vascular structures, which naturally contain inflammatory cells and elastic fibres. Since these HTTs are structurally and functionally interrelated, their close proximity within the tissue contributes to a higher HTT count per patch. Furthermore, patch selection was intentionally designed to prioritize histologically rich and informative regions. Instead of choosing entirely homogeneous areas, patches were selected based on the presence of multiple HTTs to maximize information capture. 

\begin{figure}[htbp]
  \centering
  \begin{subfigure}[t]{0.32\textwidth}
      \includegraphics[width=\linewidth]{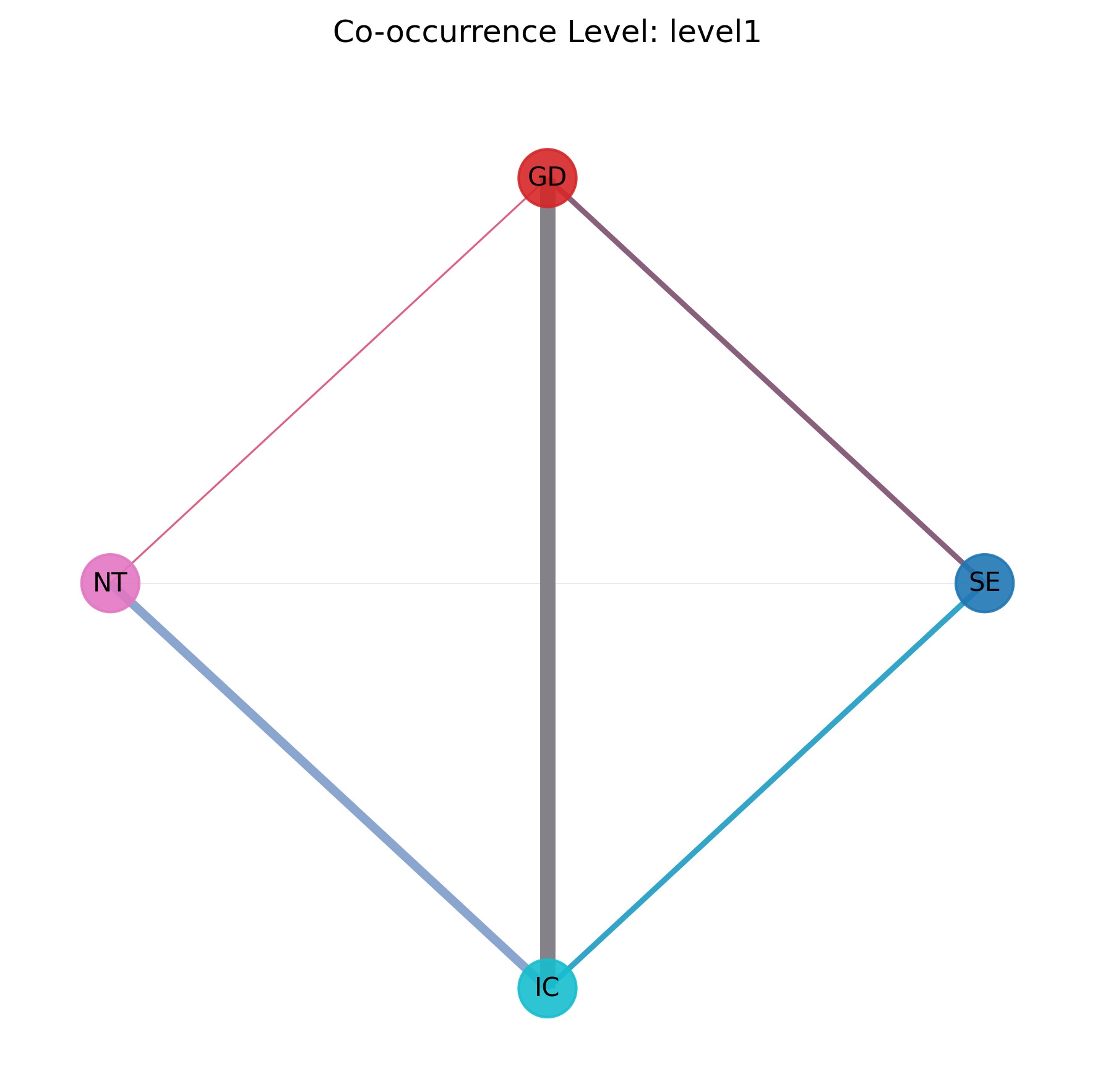}
      \caption{Level-1}
      \label{fig:co_occur_l1}
  \end{subfigure}\hfill
  \begin{subfigure}[t]{0.32\textwidth}
      \includegraphics[width=\linewidth]{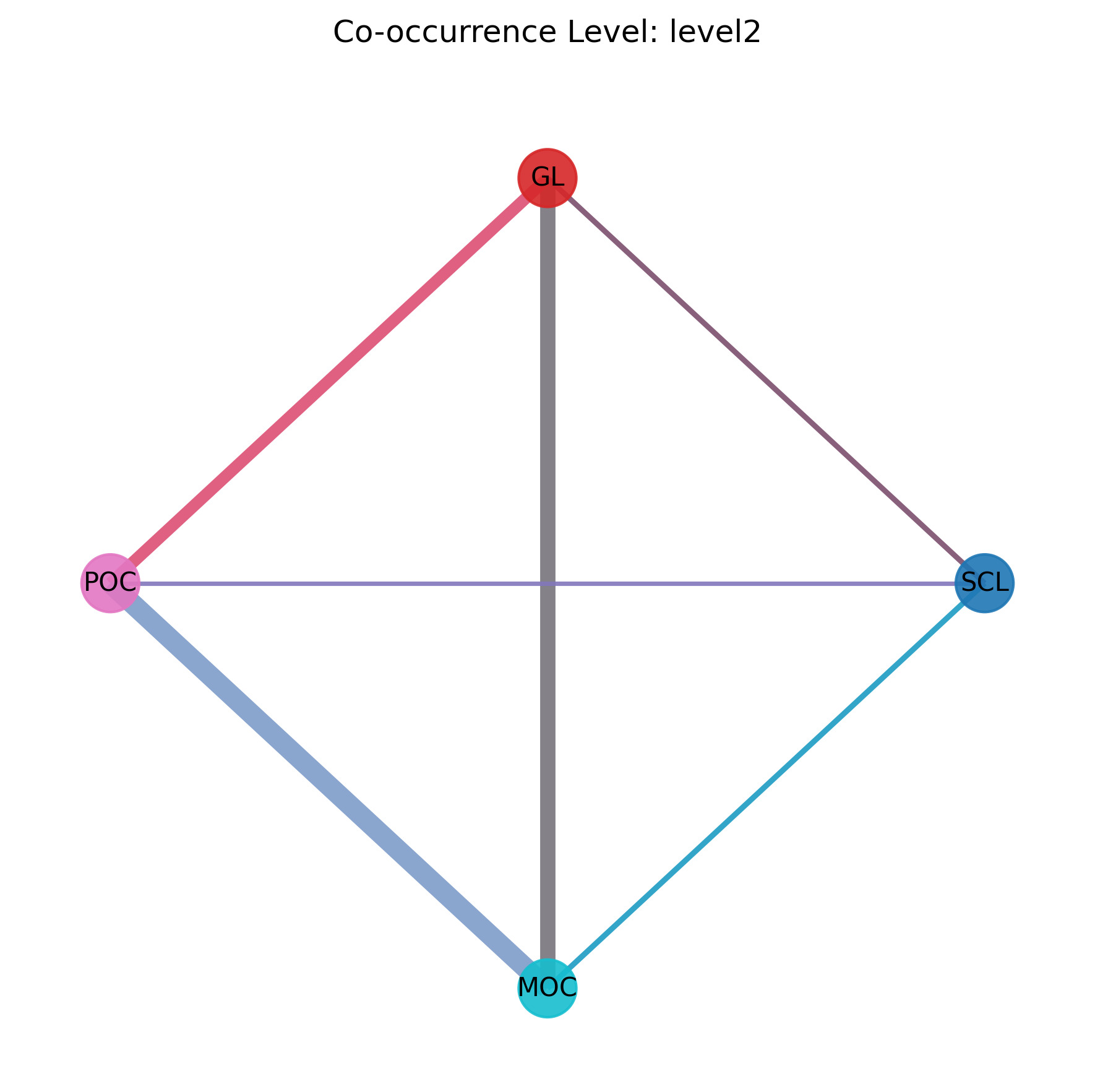}
      \caption{Level-2}
      \label{fig:co_occur_l2}
  \end{subfigure}\hfill
  \begin{subfigure}[t]{0.32\textwidth}
      \includegraphics[width=\linewidth]{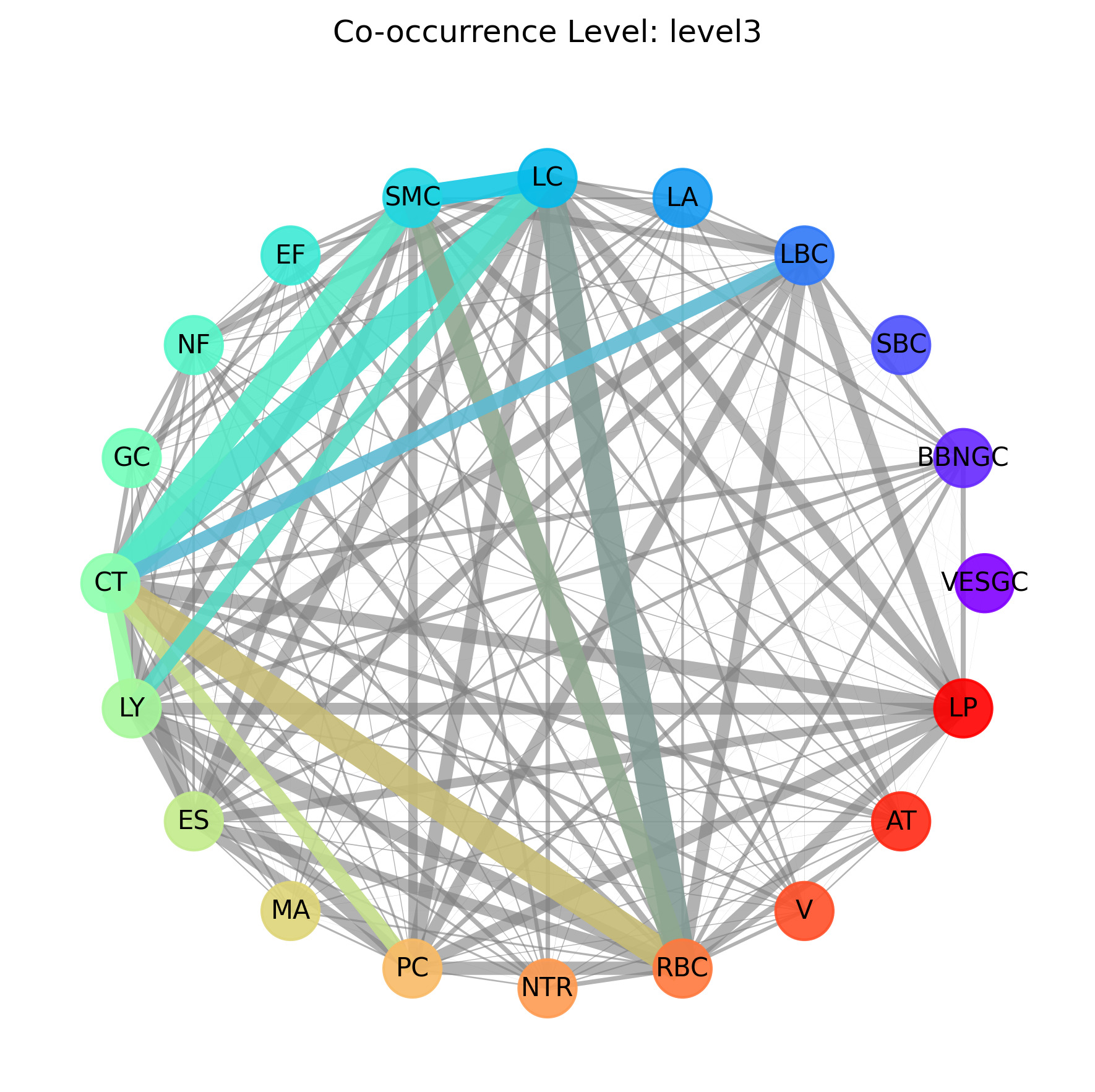}
      \caption{Level-3}
      \label{fig:co_occur_l3}
  \end{subfigure}

  \caption{Co-occurrence network for each level of HTTs in the taxonomy. The varying thickness of the edges in the network graph indicates differences in co-occurrence strength, suggesting that some HTTs are more closely associated than others.}
  \label{fig:co_occur_all}
\end{figure}

HTTs exhibit a network of co-existence by nature, with certain types frequently co-occurring. As shown in Figure \ref{fig:co_occur_all}, we visualize the co-existence network on the HTTs in the dataset. In the figure, the varying thickness of the edges in the network graph indicates differences in co-occurrence strength, suggesting that some HTTs are more closely associated than others. Level-1 (Figure \ref{fig:co_occur_l1}) shows four broad compartments—surface epithelium, glands, neural tissue and inflammatory cells. The thickest edge links glands to inflammatory cells, reflecting the well-known propensity of colonic glands to attract inflammatory infiltrates when epithelial integrity is perturbed (e.g., glandular distortion, crypt abscesses). A similarly prominent SE–IC edge mirrors the fact that the luminal epithelium is the first barrier breached by pathogens or mechanical injury, again driving immune cell recruitment. In comparison, NT is more isolated, consistent with the relative scarcity of enteric neurons in routine mucosal biopsies and their limited exposure to inflammatory exudates. Level-2 (Figure \ref{fig:co_occur_l2}) differentiates epithelium into simple columnar versus glandular phenotypes and splits the immune compartment into polymorphonuclear (POC, acute) and mononuclear (MOC, chronic) cells. The GL–MOC and POC–MOC edges dominate the graph: gland-associated chronic inflammation (e.g., lymphoplasmacytic infiltrates in long-standing colitis) explains the first, while the second captures the typical succession of neutrophil-led acute responses that transition into mononuclear-cell predominance during tissue repair. A thinner SCL–POC connection suggests that the surface epithelium is more often attacked by neutrophils during acute flares, whereas deeper glands harbour mixed inflammatory infiltrates. Level-3 (Figure \ref{fig:co_occur_l3}) explodes the network into 16 fine-grained HTTs, revealing a dense mesh of anatomical and pathological relationships. Notable thick edges include LC–SMC (lymphovascular channels enwrapped by muscularis mucosa), CT–LY (lymphoid aggregates embedded in connective stroma), and RBC-V / RBC-NTR (erythrocytes pooling within damaged vessels and neutrophil-rich exudates). Strong SMC–EF and V–AT links echo the structural coupling of smooth muscle, elastic fibres, vessels and perivascular adipose in the submucosa, while the cluster of NTR–ES–MA–PC interactions typifies mixed acute-on-chronic inflammation where neutrophils, eosinophils, macrophages and plasma cells co-localise around sites of persistent injury. Collectively, the graphs recapitulate the expected micro-architectural hierarchy of the colonic wall and align with classical pathophysiological sequences—from epithelial damage to gland-centred chronicity and, finally, to the complex multicellular milieu observed in advanced lesions or reparative fibrosis.

\subsection{Visualization of HTTs in our taxonomy}
\begin{figure}[h]
\centering
\begin{subfigure}{0.22\textwidth}
\centering
\includegraphics[width=\linewidth]{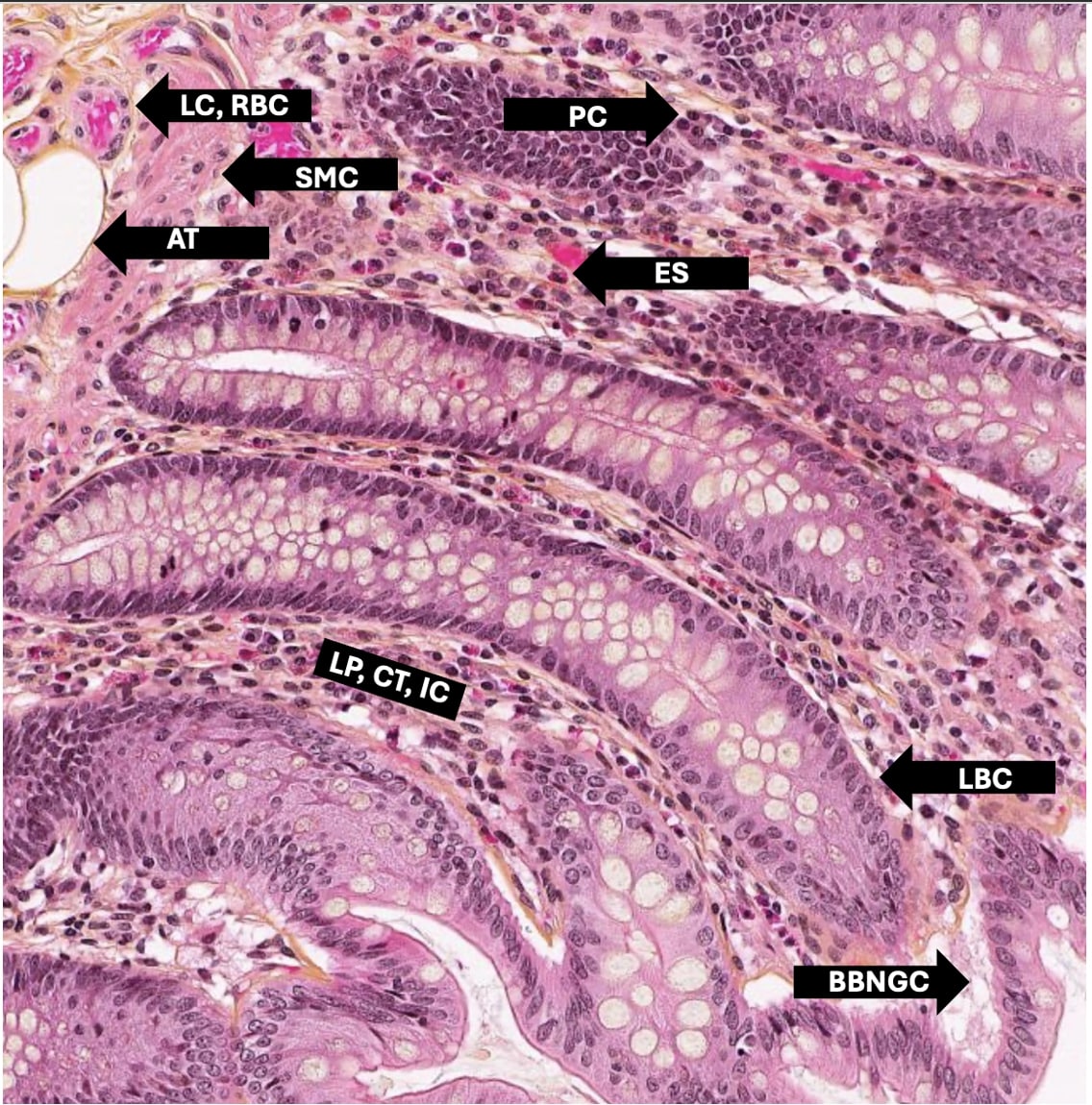}
\end{subfigure}
\begin{subfigure}{0.22\textwidth}
\centering
\includegraphics[width=\linewidth]{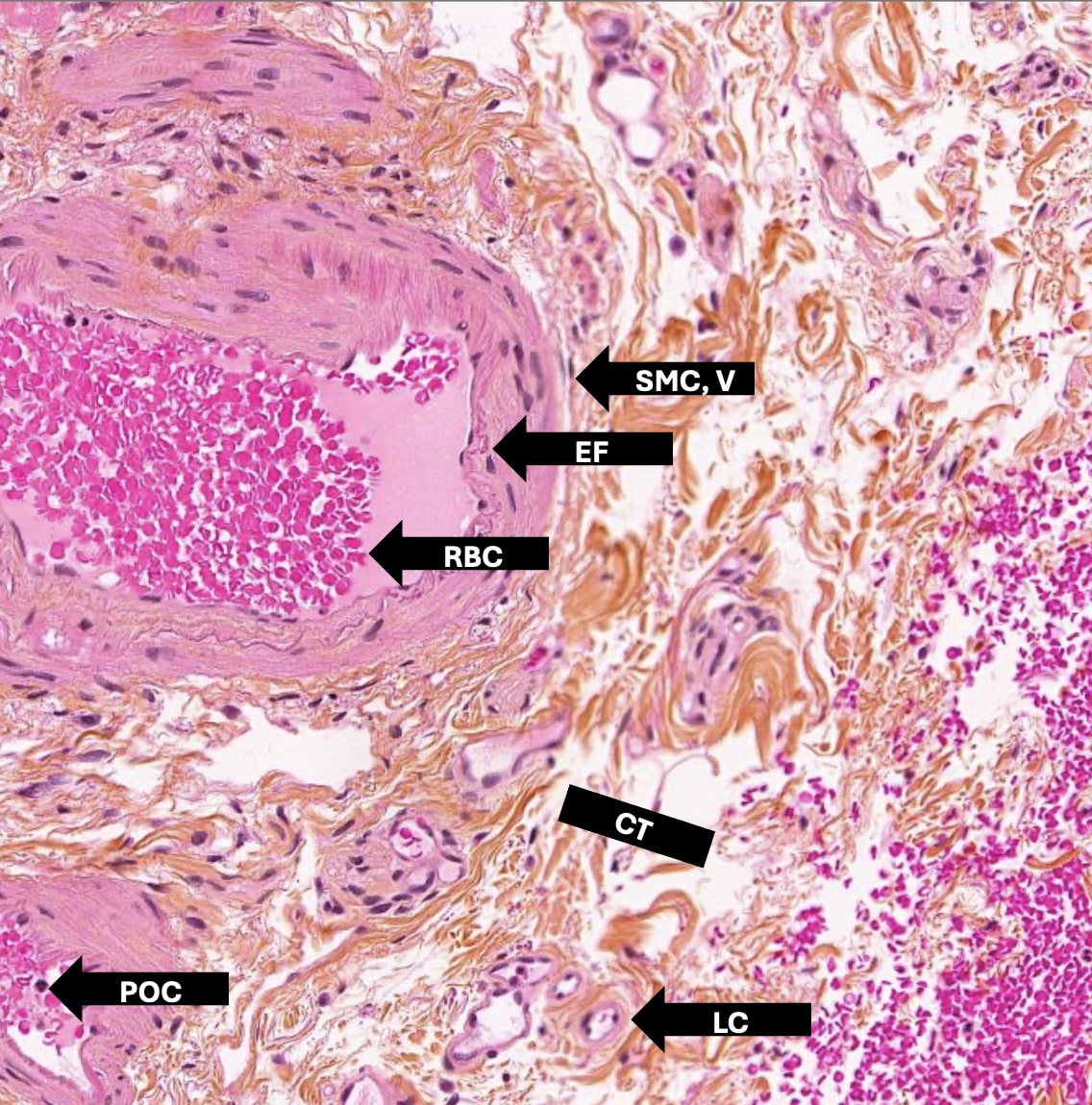}  
\end{subfigure}
\begin{subfigure}{0.22\textwidth}
\centering
\includegraphics[width=\linewidth]{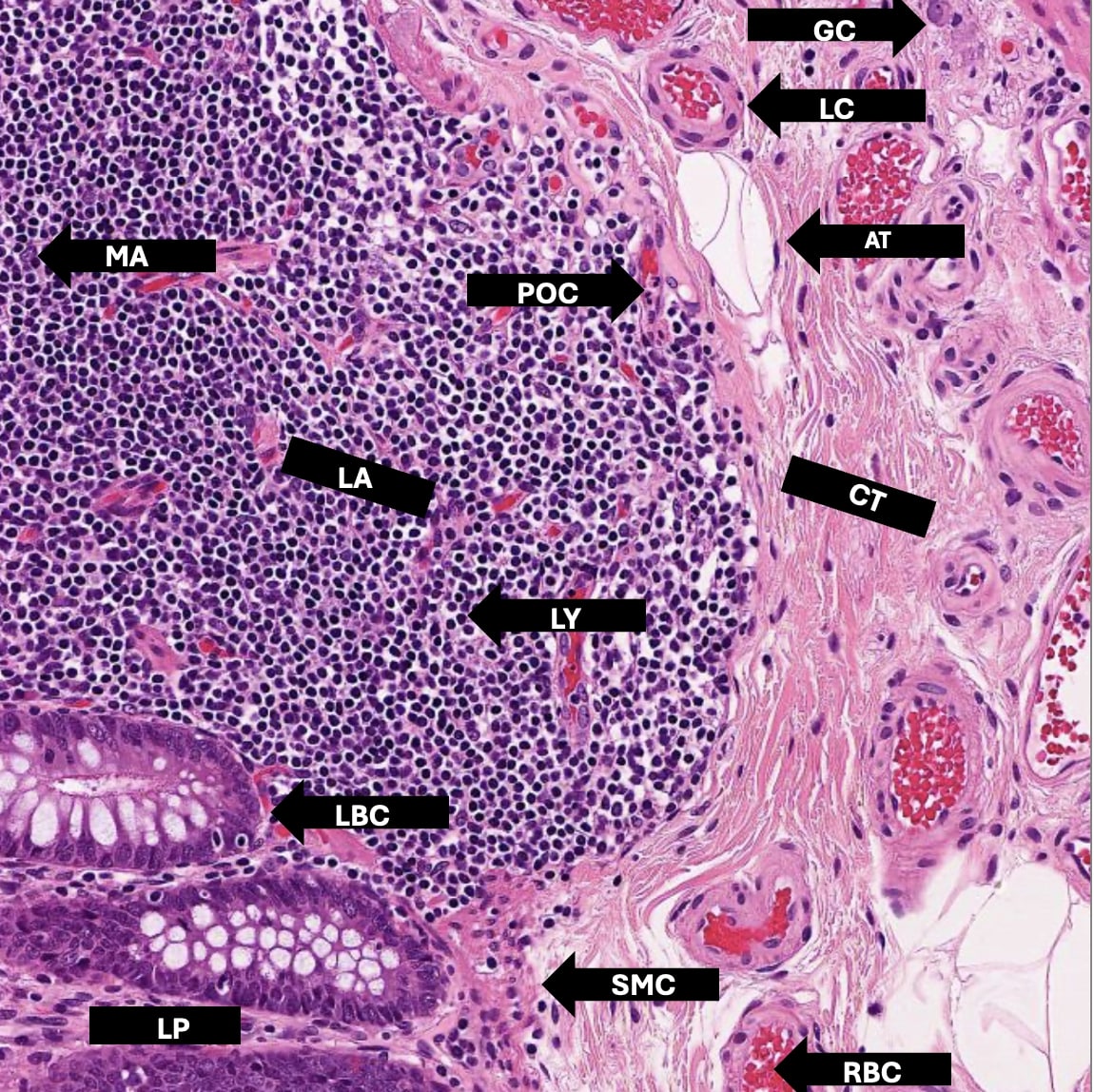}
\end{subfigure}
\begin{subfigure}{0.22\textwidth}
\centering
\includegraphics[width=\linewidth]{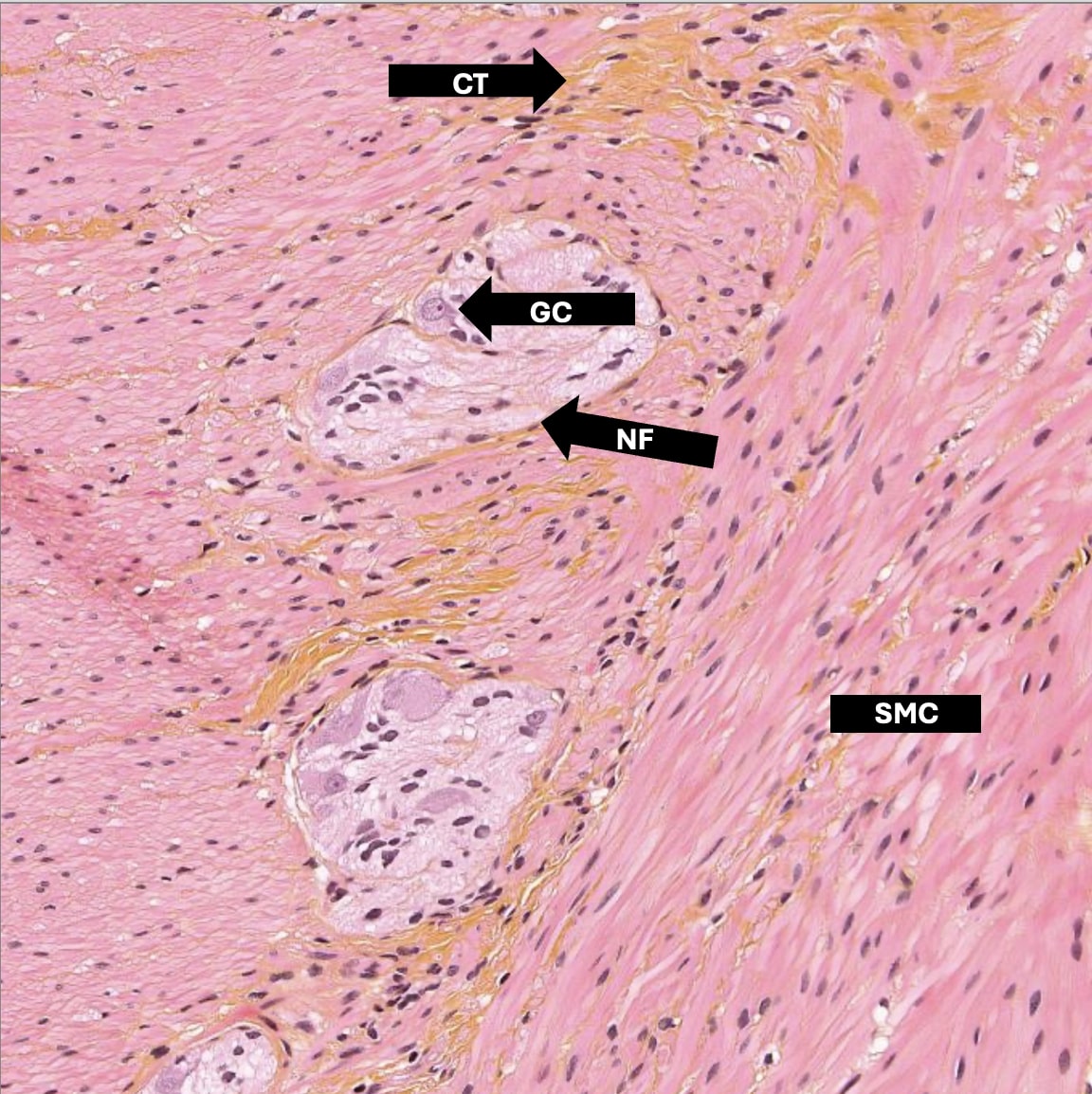}
\end{subfigure}
\caption{Example patches annotated with arrows indicating the regions containing HTTs. Each patch demonstrates the morphological complexity and structural diversity of HTT regions, often containing multiple distinct HTTs within a heterogeneous tissue environment. For tissue type abbreviations, please refer to Table 2.}
\label{fig:image_grid}
\end{figure}

\begin{figure}[h]
\centering
\begin{subfigure}{0.4\textwidth}
\centering
\includegraphics[width=\textwidth]{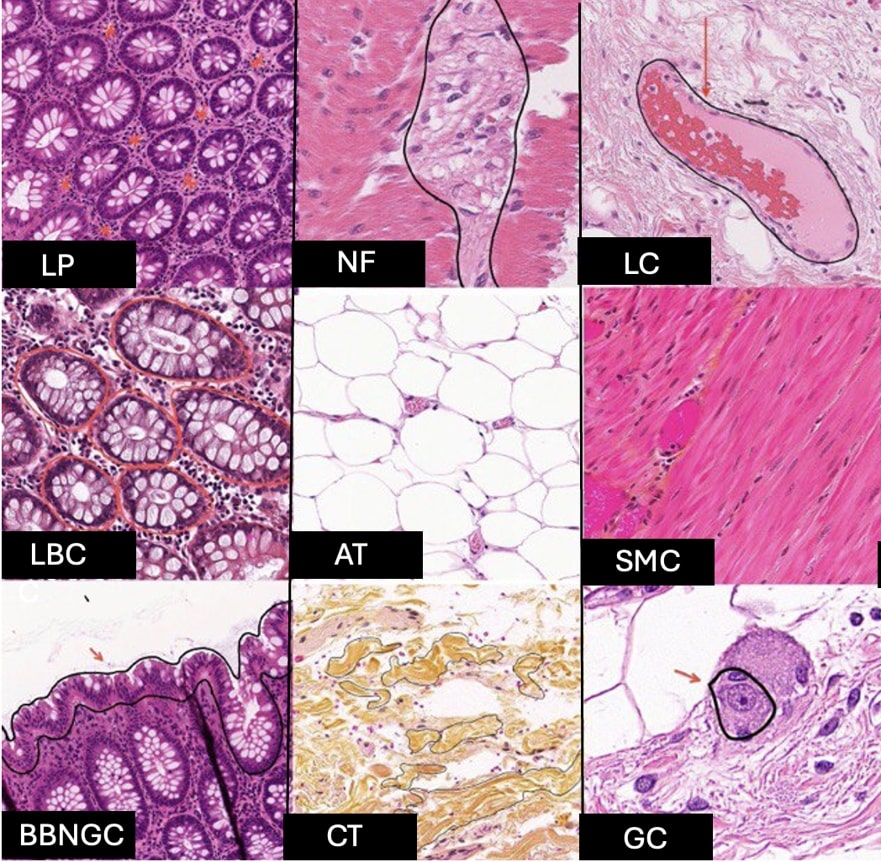}
\end{subfigure}
\hspace{0.09 cm}
\begin{subfigure}{0.4\textwidth}
\centering
\includegraphics[width=\textwidth]{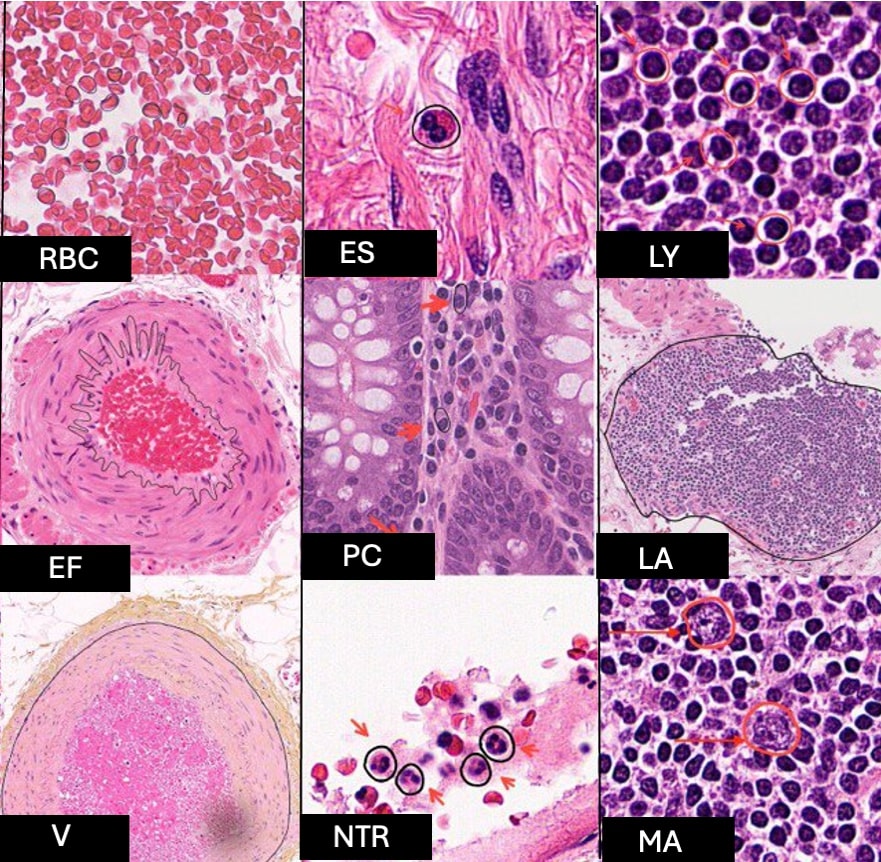}
\end{subfigure}

\caption{Representative examples of Level 3 HTTs. Each panel displays a cropped region centered on a specific HTT, with the relevant structure outlined or highlighted. These visualizations provide a reference for the diverse morphological patterns captured at this classification level. For definitions of each HTT abbreviation, refer to Table 2.}
    \label{fig:htts_examples}   
\end{figure}

We visualize four example patches in \textbf{Figure \ref{fig:image_grid}}. The regions corresponding to HTTs in each patch is highlighted with black arrows. As observed, the HTTs on these patches exhibit specific and identifiable morphological characteristics. Notably, the patches tend to be multiplex, often containing multiple HTTs with diverse structural features. Specifically, the surface epithelium or brush border with numerous goblet cells appears as a monolayer of goblet cells, while large bowel crypts present as round or oval structures arranged in clusters in cross-sectional cuts. Embedded between these crypts is the lamina propria, which mainly consists of loose connective tissue and various inflammatory cells. Among these inflammatory cells, eosinophils are easily distinguishable due to their red granules and multilobed nuclei. Additionally, plasma cells, recognized by their round, dense, and eccentrically positioned nuclei, are also present within the lamina propria. Within the lamina propria, small lymphovascular channels—thinner than large blood vessels and lacking elastic fibers—can be detected, often containing numerous red blood cells. The smooth muscle cells of the muscularis mucosae, stained pink, can be abundant near the bases of the crypts, at the junction with the submucosa. Moreover, this patch also contains hollow-appearing white lipid vacuoles of adipose tissue, further emphasizing the complexity and heterogeneity of the visualized regions. To provide a more detailed reference for these HTTs, \textbf{Figure \ref{fig:htts_examples}} presents representative examples of Level 3 HTTs. Each panel in the figure displays a cropped region centered on a specific HTT, with the relevant structure outlined or highlighted. These visualizations serve as examples of the diverse morphological patterns captured at this classification level. 
Further details can be found in Appendix, where Table \ref{visual_htt} provides a detailed description of all HTTs.

\section{Method}

\begin{figure}
    \centering
    \includegraphics[width=\linewidth]{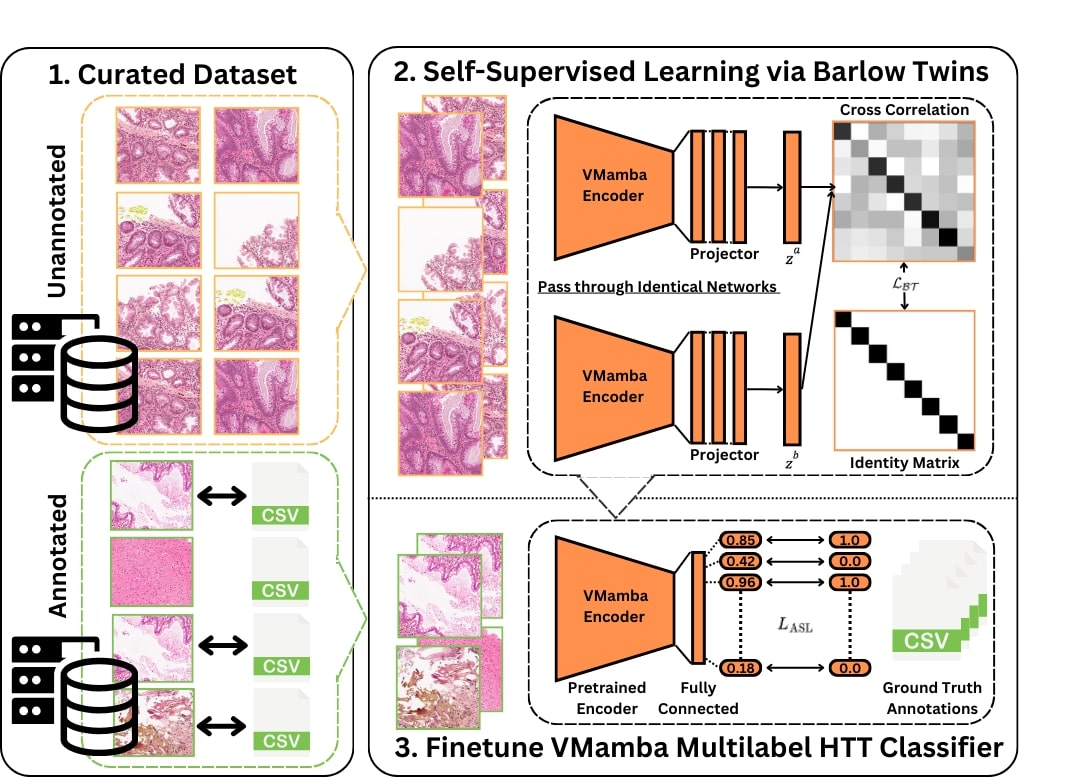}
    \caption{Training pipeline of the VMamba model. First, WSIs are split into fixed size 544 x 544 micron patches. Unlabelled patches are used for Barlow Twin SSL pretraining. Once pretrained weights are obtained, labelled patches are then used for downstream finetuning of the VMamba encoder for multilabel classification. \label{overview}}
\end{figure}

In this section, we provide an overview of our multilabel HTT representation learning pipeline (Figure \ref{overview}) to efficiently train a model and how we use the model's predicted confidence scores to uncover statistical patterns that confirms the two pathological development pathways of colon cancers (Figure \ref{workflow1}). 

\subsection{Multilabel Representation Learning}

Our multilabel representation learning procedure follows three steps: 1. choice of the representation learning encoder, 2. self-supervised pretraining on unlabeled image patches, and 3. model fine-tuning on the multilabel HTT classification task using labeled ADPv2 patches. The learned model is further used for potential biomarker discovery, which will be detailed in Section 4.2.

\subsubsection{Encoder for Representation Learning} 
We adopt VMamba \cite{vmamba} as the feature encoder for HTT classification. VMamba is a Selective State-Space Model that scans image tokens sequentially, storing their latent “state” in a lightweight gating mechanism. This design delivers transformer-level accuracy while keeping time and memory strictly linear in the number of tokens—an important property when a single pathology patch may contain many thousands of 16 × 16 pixel tokens.

Linear scaling lets us present the network with a large field of view (multiple millimetres of tissue) without down-sampling, so VMamba simultaneously sees architectural patterns (e.g., crypt orientation, glandular shape and spacing) and cellular details (e.g., nuclear atypia) that often co-occur in colon polyps. The recurrent state captures short-range context naturally, while the long convolutional kernel implicit in the SSM propagates information across the entire patch, yielding robust representations for both coarse and fine labels in our hierarchy.

Clinically, this token-by-token sweep echoes how pathologists move a microscope stage: first surveying broad regions, then zooming in on suspicious micro-structures. That behavioural analogy, together with VMamba’s favourable compute profile, makes it a more apt backbone for ADPv2 than quadratic Vision Transformers \cite{vit} or pure CNN alternatives.

\subsubsection{Self-supervised Learning of Tissues using Barlow Twins}
In the pre-training stage, we select Barlow Twins \cite{barlow-twins} as our self-supervised learning method as it efficiently reduces feature redundancy—the presence of repetitive or duplicate information in the learned representation of the data. By reducing redundancy, it extracts diverse and meaningful features crucial for accurately capturing the subtle variations in pathology images, which is essential for tasks such as cancer diagnosis. Barlow Twins utilizes an objective function that measures the cross-correlation between the output features computed from two distorted versions of each image sample and tries to make this matrix as close to the identity as possible, reducing redundancy between components of these vectors. The Barlow Twins objective function is given as:

\begin{equation}
\label{eq:barlow_twins}
\mathcal{L}_{BT} \;\triangleq\;
\underbrace{\sum_{i} \bigl(1 - C_{ii}\bigr)^{2}}_{\text{invariance term}}
\;+\;
\lambda\,
\underbrace{\sum_{i} \sum_{j \neq i} \bigl(C_{ij}\bigr)^{2}}_{\text{redundancy reduction term}}
\end{equation}

Where the $\mathcal{\lambda}$ is a positive constant trading off the importance of
the first and second terms of the loss, and $\mathcal{C}$ is the
cross-correlation matrix computed between the two outputs of
the network along the batch dimension:

\begin{equation}
\label{eq:correlation}
c_{ij} \;\triangleq\;
\frac{\displaystyle \sum_{b} \bigl(z_{b,i}^{A}\bigr)\bigl(z_{b,j}^{B}\bigr)}%
     {\sqrt{\displaystyle \sum_{b} \bigl(z_{b,i}^{A}\bigr)^{2}}\,
      \sqrt{\displaystyle \sum_{b} \bigl(z_{b,j}^{B}\bigr)^{2}}}
\end{equation}

During training we follow the standard Barlow Twins data augmentation settings, with an extra RandStainNA \cite{randstainna} prior to color normalization. RandStainNA is a hybrid framework combining stain normalization and stain augmentation while incorporating variations in color space to produce realistic stains. Using RandStainNA, we can exclude the default color jitter used in the standard Barlow Twins implementation while gaining increased performance. Training setup details can be found in Table \ref{tab:hyperparameters_selected} and Table \ref{tab:general_stats} in the appendix.

\subsubsection{Finetuning Model on ADPv2 for Multilabel HTT Classification}
After pre-training, we finetune the pretrained encoder for multilabel HTT classification. During finetuning, we remove the projection head of the pretrained VMamba encoder and add a classification
head of a single linear layer.
Prior to training, we discard all images originating from the 110 TCGA slides due to their heterogeneity in staining and inconsistent quality. We split the dataset into training, validation and test using an 80-10-10 ratio. To prevent data leakage, we create dataset splits so that no two patches from different splits come from the same slide. The training split covers 364 slides and the testing split covers 86 slides. For data augmentation, we applied RandomResizedCrop, RandomResizedHorizontalFlip, RandomRotation, RandomGrayscale, Solarization, RandStainNA, and color normalization.

Before training, we perform label pruning to address the dataset imbalance issue and reduce label noise. First, we exclude all HTTs that are non-colon or occur very sparsely in our dataset, which are denoted in a red font in table \ref{htt}. Then, we remove connective tissues due to their extreme abundance (\textasciitilde95\%) in the image patches, making the model extremely biased in predicting this class. Furthermore, we discard HTTs with noisy annotations due to the challenge for pathologist annotation. We remove inflammatory cells, polymorphonuclear cells, and mononuclear cells from training as these are deemed "low-confidence" labels, which are selected in cases where the presence of the children HTTs is ambiguous, thereby possibly introducing inaccuracies in our dataset. Finally, we merge surface epithelium and glands as a single class GD, because they are the same biological structure cut at different angles. All pruned HTTs are denoted in standard black font color. As a result, we reduced the total number of HTTs for training from 32 labels to 14 labels. These 14 labels are denoted in green in table \ref{htt}.

We use ASL\cite{benbaruch2021asymmetriclossmultilabelclassification} as our objective function, which is described in the following equation:
\begin{equation}
L_{\text{ASL}} = \sum_{k=1}^{K} \left[-y_k (1 - p_k)^{\gamma^+} \log(p_k) - (1 - y_k) (p_{m_k})^{\gamma^-} \log(1 - p_{m_k})\right]
\end{equation}
where $K$ is the number of labels, $y_k$ is the ground truth label for the $k$-th class, $p_k$ is the predicted probability for the $k$-th label, $p_{m_k} = \max(p_k - m, 0)$ is the shifted probability with the margin $m$ applied, and $\gamma^+$ and $\gamma^-$ are tunable focusing parameters for the positive and negative samples, respectively. This objective function enhances multilabel classification by applying different penalties to false positives and false negatives, allowing the model to focus more on correctly identifying rare but critical labels. This approach addresses label imbalance and asymmetry in error importance, leading to improved performance in scenarios where certain misclassifications carry higher consequences, especially in digital pathology. All hyperparameter settings can be found in Table \ref{tab:hyperparameters_selected} in the appendix.

\begin{figure}
    \centering
    \includegraphics[width=1.1\linewidth]{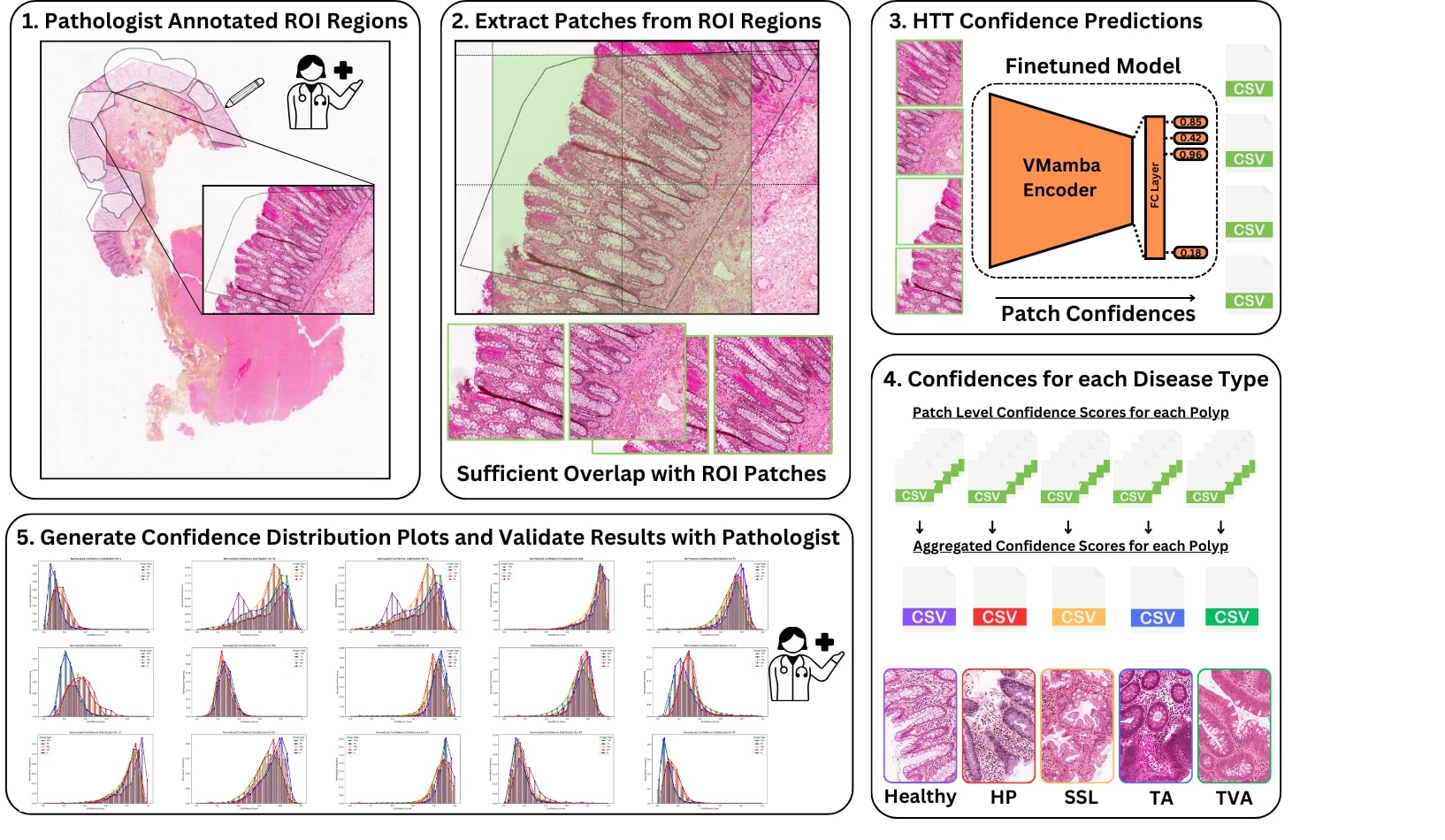}
    \caption{RoI Confidence Score Distribution Analysis Workflow. Pathologists annotate RoIs that are important to diseases on each Slide. Patches are extracted from RoI regions. We filter out patches with insufficient overlaps with the RoI annotations. Then, we use our VMamba model trained on our ADPv2 dataset to get per-HTT confidence scores for each RoI patch. We group the confidence scores according to the disease types and plot the per-HTT distribution histograms. We quantify the distribution shifts using two-tailed t-test. Pathologists analyze the plots and t-statistics with medical interpretations.}
    \label{workflow1}
\end{figure}

\subsection{Confidence Score Distribution Analysis}

Tissues affected by disease change in morphology, structure, and even color. These changes may be subtle but will affect the model's output confidence scores if the model has only seen healthy tissue patches during training. \cite{on_trans} Therefore, the distribution of the model’s predicted confidence scores on patches from the RoIs of diseased slides will shift from the distribution of normal slides on certain HTTs. The shift pattern on different sets of HTTs can be potentially used as a marker for different diseases. 

Figure \ref{workflow1} shows an overall illustration of the confidence score distribution analysis workflow. To find the distribution shift, we first additionally collected colon polyp WSIs belonging to four different diseases: Hyperplastic (HP), Sessile serrated lesion (SSL), Tubular Adenomas (TA) and Tubulovillous adenoma (TVA). All slides are annotated by pathologists at RoI level to have areas of disease identified. For our healthy slides in hand, pathologists simply annotate areas where these diseases are most likely to develop as RoIs. We extracted non-overlapping patches corresponding to a FOV of 544 x 544 microns from the RoI annotation areas on all slides (both healthy and diseased). The statistics of the extracted patches are shown in Table \ref{tab:slide_sets}. Then, we applied our VMamba model to all the RoI patches and collected the multilabel confidence scores for all HTTs for each patch, giving us a 2D tensor of shape [$N, C$], where $N$ is the total number of image patches extracted from all WSI RoIs and $C$ is the number of HTTs. We divide this collection of confidence scores into $M$ non-overlapping subsets [$N_1$, C] ... [$N_M$, C], depending on the group of the diseases to which the patch belongs. After that, we visualize the distribution shifts using normalized histograms. We generate a histogram for each HTT using the $M$ subsets of predicted confidence scores, giving us a total of $C$ histogram plots having $M$ distribution histograms on each plot (with $M-1$ diseased distributions and one healthy distribution). Furthermore, we quantify the distribution shifts using Welch’s two-sample t-test (two-tailed). $\alpha$ was set at $0.05$ for significance for the tests described. Since the confidence scores are heavily right-skewed, we apply logit-transformation to all confidence values to improve normality. The logit-transform is defined as \( y = \operatorname{logit}(p)\;=\;
   \ln\!\left(\frac{p+\varepsilon}{1-(p+\varepsilon)}\right) \), where $p$ is the confidence score and $\varepsilon=1^{-6}$ is added to avoid infinity when $p$ is exactly 0 or 1. We compute p-values with Holm-Bonferroni's correction method to control the family-wise error rate (adjustment for multiple comparisons).

   When grouping diseases for confidence score analysis, we treat SSL and HP as the first group, TA and TVA as the second group, and healthy slides as the last group. Finally, pathologists interpret the distribution shifts from the medical perspective to confirm the two cancer development pathways.

\section{Experiments and Results}
\subsection{Multilabel Representation Learning Results}
In this section, we detail both quantitative and qualitative results achieved by our VMamba model. For quantitative results, we show the model's general multilabel classification performance and per-class performance using multiple metrics on the test split of our ADPv2 dataset. For qualitative results, we first visualize the spreads in the multilabel representations learned by our model in the latent space using t-SNE on 500 randomly sampled image patches from our ADPv2 dataset. Then, we provide heatmaps on in-house test slides to check the quality of HTT identification on WSIs. Finally, we select the most representative image patch for each HTT in our dataset and apply GradCAM on the model's second VSSM block to visualize the model's pixel-level attentions to make predictions.

\subsubsection{Quantitative Results}

\begin{table}[ht]
\centering
\scriptsize
\caption{Performance Metrics by HTT with Best (green) and Worst (red) Values Highlighted}
\label{tab:metrics_highlight}
\begin{tabular}{lrrrrrrrr}
\hline
\textbf{HTT} & \textbf{TPR} & \textbf{FPR} & \textbf{TNR} & \textbf{FNR} & \textbf{Accuracy} & \textbf{Precision} & \textbf{F1} & \textbf{AUC} \\
\hline
SE  & 0.944            & 0.082           & 0.918           & 0.056            & 0.922            & 0.670            & 0.784            & 0.973 \\
GD  & 0.981            & 0.057           & 0.943           & 0.019            & 0.959            & 0.920            & 0.950            & \cellcolor{green!20}0.986 \\
LA  & 0.889            & \cellcolor{green!20}0.021 & \cellcolor{green!20}0.979 & 0.111            & \cellcolor{green!20}0.972 & 0.780            & 0.831            & 0.977 \\
LP  & 0.984            & 0.073           & 0.927           & 0.016            & 0.948            & 0.889            & 0.934            & 0.976 \\
LC  & 0.995            & 0.913           & 0.087           & 0.005            & 0.867            & 0.869            & 0.928            & \cellcolor{red!20}0.843 \\
SMC & 0.969            & 0.184           & 0.816           & 0.031            & 0.926            & \cellcolor{green!20}0.931 & 0.950            & 0.968 \\
EF  & 0.831            & 0.034           & 0.966           & 0.169            & 0.950            & 0.766            & 0.797            & 0.958 \\
NT  & 0.871            & 0.196           & 0.804           & 0.129            & 0.820            & 0.584            & 0.699            & 0.918 \\
ES  & 0.948            & 0.459           & 0.541           & 0.052            & \cellcolor{red!20}0.697 & 0.562            & 0.706            & 0.880 \\
MA  & \cellcolor{red!20}0.716 & 0.055           & 0.945           & \cellcolor{red!20}0.284 & 0.925            & \cellcolor{red!20}0.551 & \cellcolor{red!20}0.623 & 0.927 \\
PC  & 0.869            & 0.228           & 0.772           & 0.131            & 0.810            & 0.704            & 0.778            & 0.909 \\
LY  & 0.915            & 0.331           & 0.669           & 0.085            & 0.780            & 0.693            & 0.789            & 0.892 \\
RBC & \cellcolor{green!20}0.999 & \cellcolor{red!20}0.961 & \cellcolor{red!20}0.039 & \cellcolor{green!20}0.001 & 0.913            & 0.914            & \cellcolor{green!20}0.954 & 0.887 \\
V   & 0.847            & 0.035           & 0.965           & 0.153            & 0.950            & 0.770            & 0.807            & 0.959 \\
AT  & 0.904            & 0.079           & 0.921           & 0.096            & 0.917            & 0.729            & 0.807            & 0.967 \\
\hline
\end{tabular}
\end{table}

Our finetuned VMamba model achieves a test mAP of 0.879. As shown in Table \ref{tab:metrics_highlight},  for each HTT used in training, we compute the True Positive Rate (TPR), True Negative Rate (TNR), False Positive Rate (FPR), False Negative Rate (FNR), Accuracy (ACC), F1-score, Precision, and Recall. Notably, the model performs strongly on most HTTs, importantly on HTTs pathologists consider diagnostically relevant. We also find the worst performances primarily stem from four HTTs: LC, ES, MA, and RBC. For LC and RBC, both HTTs are present in a large majority of our dataset, with LC occurring in 16418 image patches and RBC occurring in 16543 image patches, meaning these HTTs are found in about 82\% of all image patches in our dataset. Furthermore, LC and RBC exhibit similarities in their performances: exceptionally high TPR and FPR, and abnormally low TNR and FNR. One interpretation for this is the lack of negative samples makes it difficult for the model to discern these HTTs properly. In addition, our objective function incentivizes predicting positive labels correctly via asymmetric weighting of positive losses, thereby neglecting to 'learn' negative samples when there exist so few. Conversely, for macrophages, we observe low TPR, high FNR, low precision, and low F1, indicative that the model cannot recognize positive samples. This can be attributed to macrophages' low positive occurrence of positives in our dataset—at just 1020 samples, or 6\% of the dataset. Lastly, EF has a high FPR, low TNR, and low accuracy, indicating the model is quite poor at correctly identifying negatives. This can be attributed to the general difficulty of identifying this HTT in our images, as is the case with neutrophils (refer to section 4.1.2) which is the sibling of EF under POC. Thus, EF is susceptible to mislabelling by the pathologists themselves and as a result, the 'real' FPR is likely to be slightly lower than reported.

\subsubsection{Qualitative Results}
\begin{figure}[ht]
    \centering
    \includegraphics[width=0.5\textwidth]{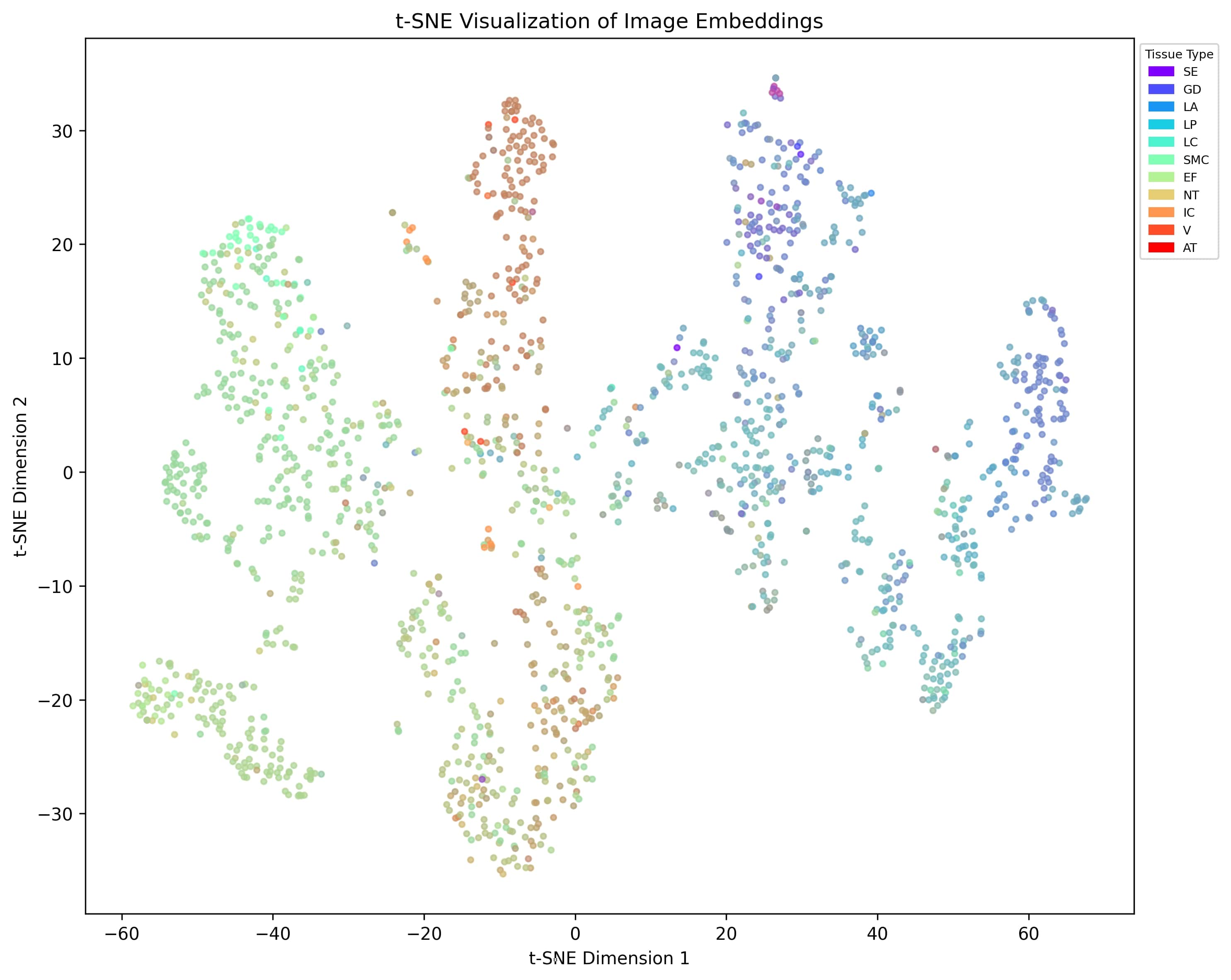}
    \caption{Visualization of image embeddings using t-SNE. Each HTT is denoted by a single color; The color for each multilabel point is obtained by mixing the colors of its individual HTTs. For visual clarity, we assign colors only to level-1 and level-2 labels}
    \label{fig:tsne_plot}
\end{figure}
In this section, we first visualize the representation of our VMamba model learned on ADPv2 using t-SNE, then we show the heatmaps on unseen slides using the model's predictions. Since our dataset is multilabeled, we use the color blending technique to color-label the t-SNE plot, by assigning each HTT a color and blending the colors into one if a sample has multiple labels associated with it. For visualization clarity, we only consider the parent HTTs for color blending, reducing the total number of colors to blend from 16 to 11. As shown in Figure \ref{fig:tsne_plot}, we observe clear clustering effect on the blended colors. It suggests that the VMamba model effectively captures high-level morphological differences—e.g., separating glandular crypt structures from surface epithelium or adipose tissue. Where colors blend, the patches exhibit multiple co-occurring tissue/cell types, reflecting the biological reality that colon subregions consist of overlapping structures. For instance, smooth muscle cells and elastic fibers coexist as key components of blood vessel walls in the submucosa. From a histopathological point of view, this clustering and partial overlap align with the preserved colon wall architecture, e.g., epithelial lining is found at the mucosal surface, and lymphatic and vascular channels can be seen within the submucosa.

\begin{figure}[h]
\centering
\begin{subfigure}{0.18\textwidth}
\centering
\includegraphics[width=\textwidth]{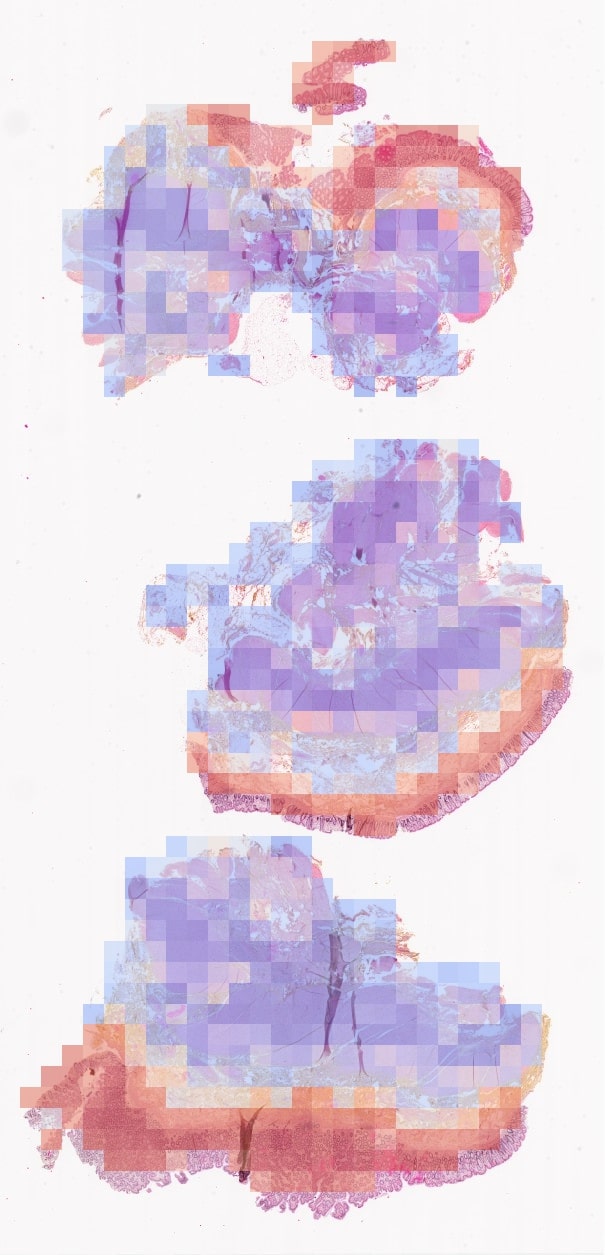}
\caption{Normal - GD}
\end{subfigure}
\begin{subfigure}{0.18\textwidth}
\centering
\includegraphics[width=\textwidth]{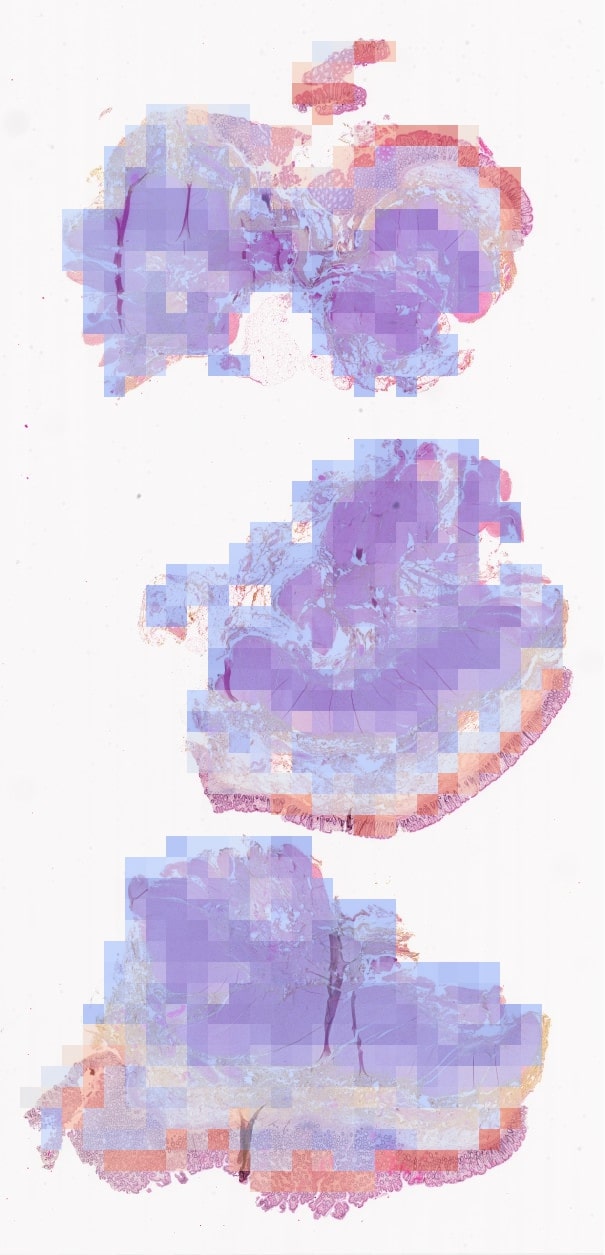}
\caption{Normal - SE}
\end{subfigure}
\begin{subfigure}{0.18\textwidth}
\centering
\includegraphics[width=\textwidth]{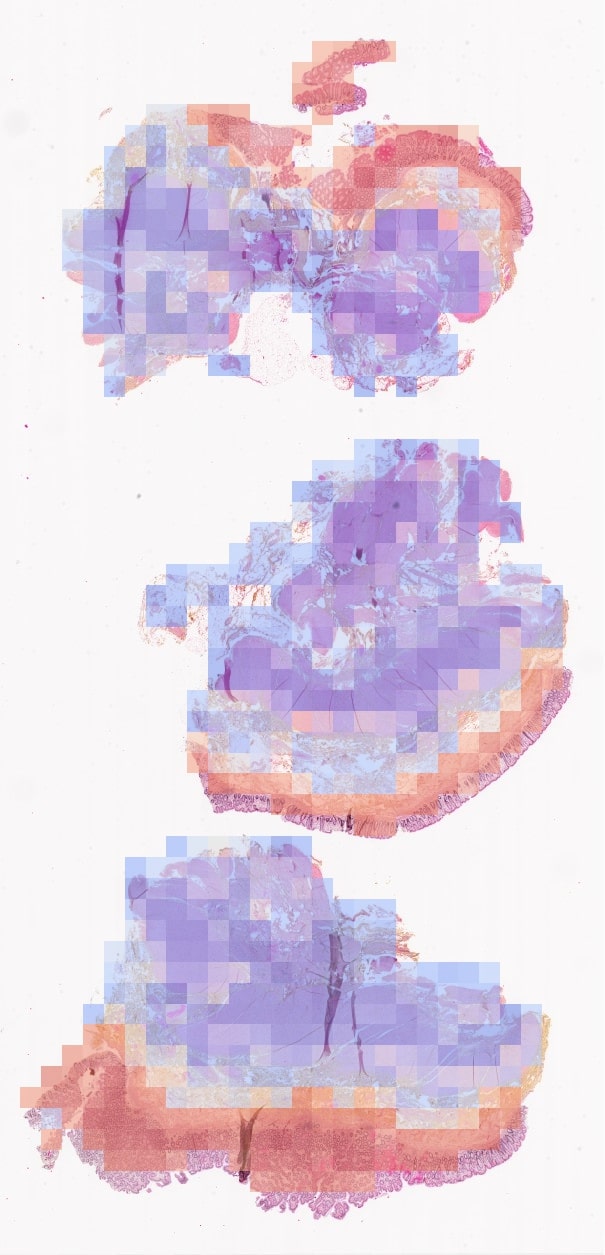}
\caption{Normal - LP}
\end{subfigure}
\begin{subfigure}{0.18\textwidth}
\centering
\includegraphics[width=\textwidth]{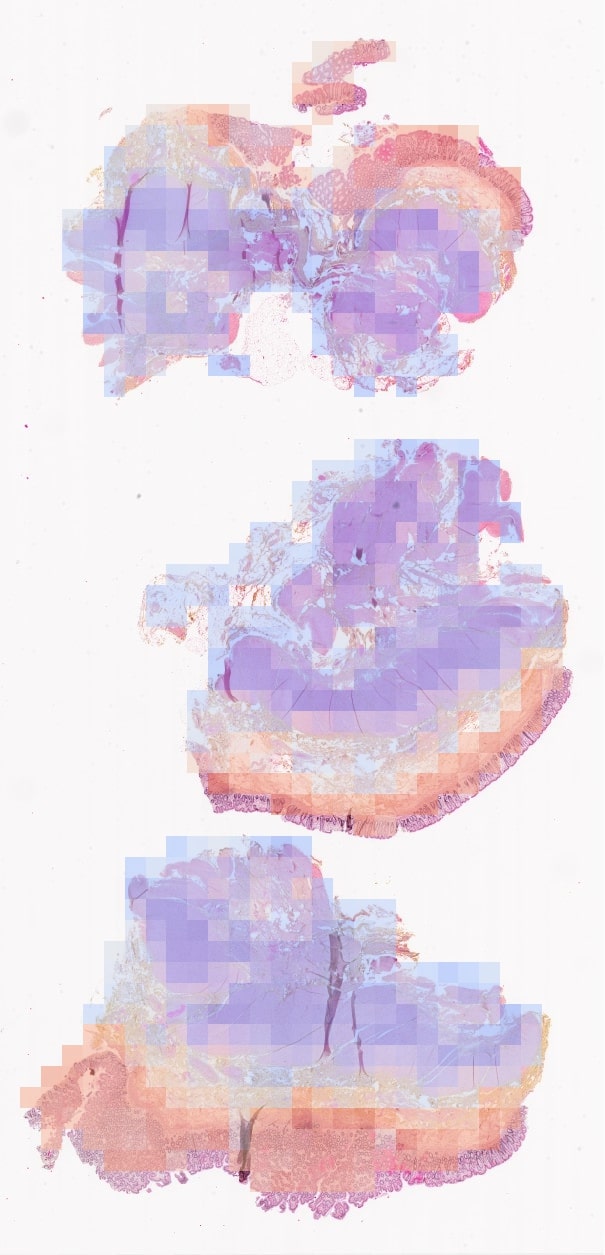}
\caption{Normal - LY}
\end{subfigure}
\begin{subfigure}{0.18\textwidth}
\centering
\includegraphics[width=\textwidth]{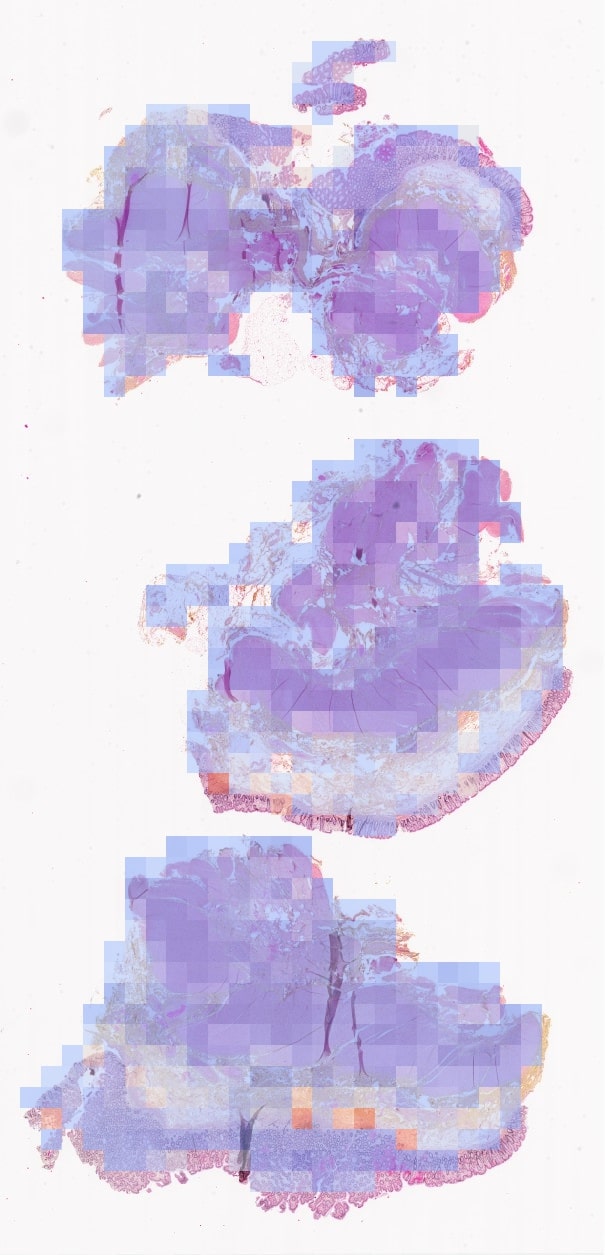}
\caption{Normal - LA}
\end{subfigure}
\begin{subfigure}{0.18\textwidth}
\centering
\includegraphics[width=\textwidth]{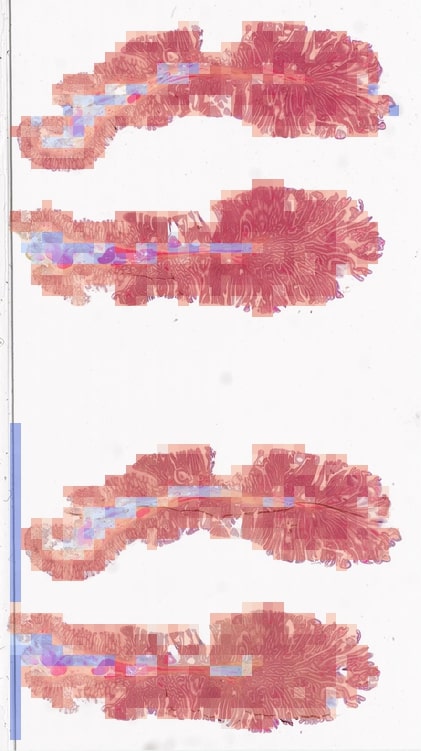}
\caption{TVA - GD}
\end{subfigure}
\begin{subfigure}{0.18\textwidth}
\centering
\includegraphics[width=\textwidth]{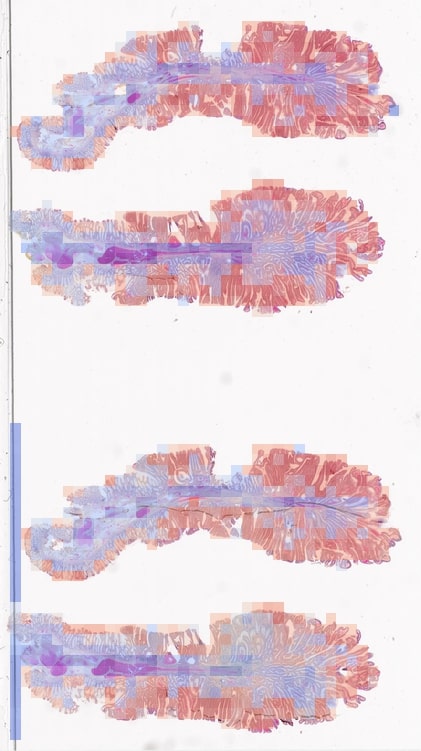}
\caption{TVA - SE}
\end{subfigure}
\begin{subfigure}{0.18\textwidth}
\centering
\includegraphics[width=\textwidth]{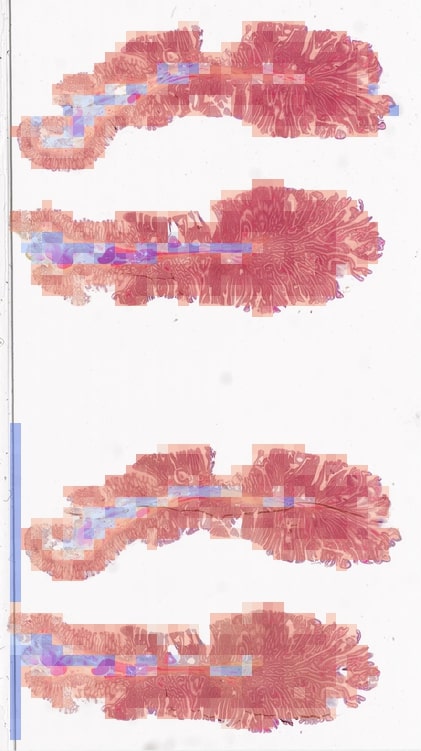}
\caption{TVA - LP}
\end{subfigure}
\begin{subfigure}{0.18\textwidth}
\centering
\includegraphics[width=\textwidth]{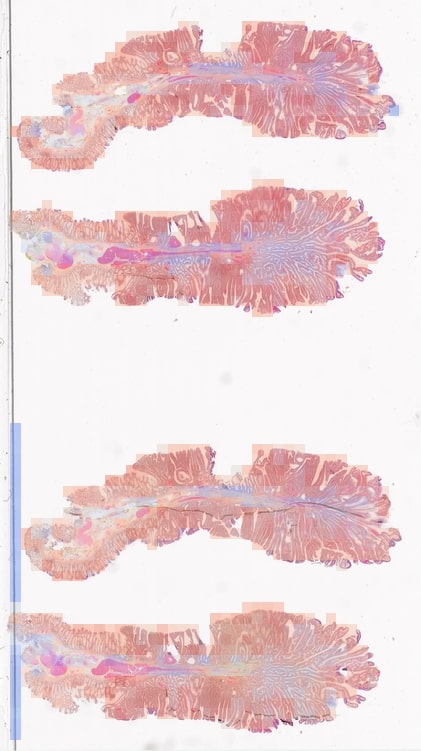}
\caption{TVA - LY}
\end{subfigure}
\begin{subfigure}{0.18\textwidth}
\centering
\includegraphics[width=\textwidth]{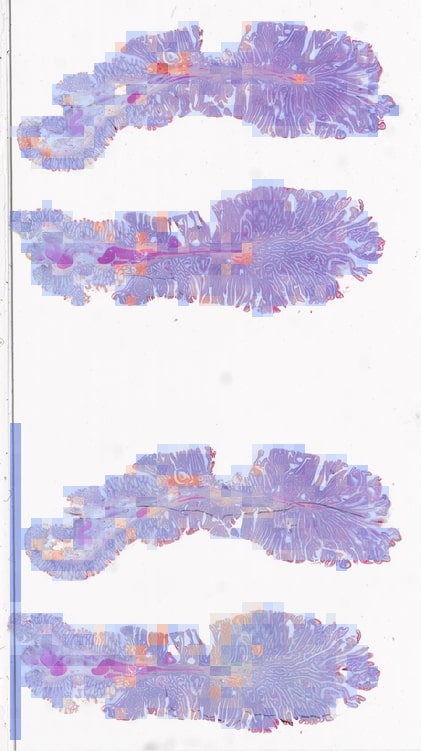}
\caption{TVA - LA}
\end{subfigure}
\caption{The heatmaps for diagnostically relevant HTTs on a healthy slide and a tubulovillous adenoma (TVA) slide. The model's prediction accurately indicates corresponding HTTs on healthy and diseased slides.}
\label{fig:heatmaps}
\end{figure}

We also visualize the model's prediction results for each HTT as heatmaps on slides in the test split. Each slide is tiled into patches corresponding to a FOV of 544x544 microns at 20x level. The patches are resized to 512x512 pixels and fed into our VMamba model to get multilabel confidence scores. For each HTT, we visualize the patch confidence scores as an overlay on the original slides. Figure \ref{fig:heatmaps} shows the heatmaps for a healthy slide and a tubulovillous adenoma (TVA) slide on five diagnostically relevant HTTs. On tubulovillous adenoma slides, the detection activates mostly on glands and lamina propria, matching key diagnostic architectural features of tubulovillous adenomas, with increased crowding of dysplastic glands and presence of both tubular and villous components (the latter representing >25\% but <75\% of the TVA). The observed detection of immune cells, such as lymphocytes and plasma cells, within the adenoma's mucosa, could be related to local responses to dysplastic changes, especially if increased, but such cells also seen within the lamina propria of the normal colorectal mucosa.

\begin{figure}[!ht]
\centering
\resizebox{\linewidth}{!}{%
\begin{tabular}{ccccccc}

\begin{minipage}{0.13\linewidth}
\centering
\includegraphics[width=\linewidth]{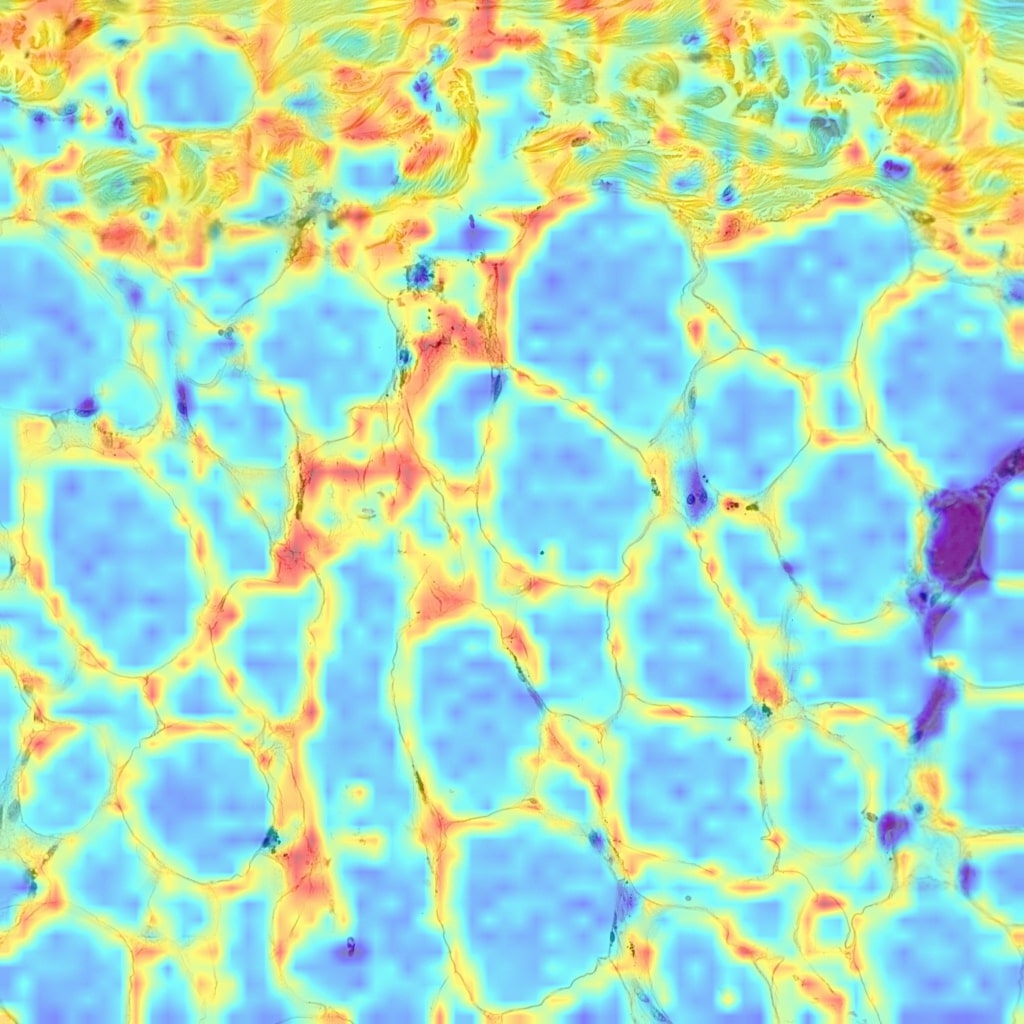}\\
AT
\end{minipage} &
\begin{minipage}{0.13\linewidth}
\centering
\includegraphics[width=\linewidth]{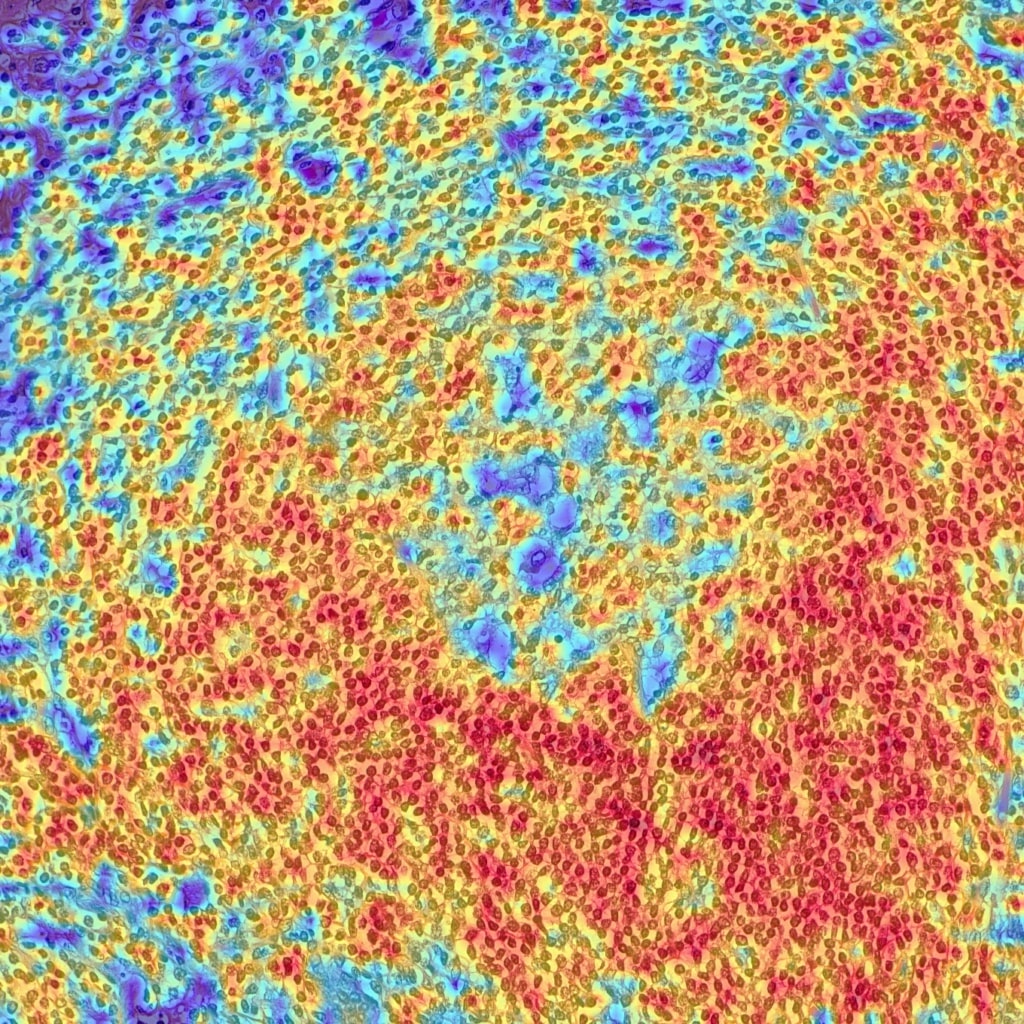}\\
LA
\end{minipage} &
\begin{minipage}{0.13\linewidth}
\centering
\includegraphics[width=\linewidth]{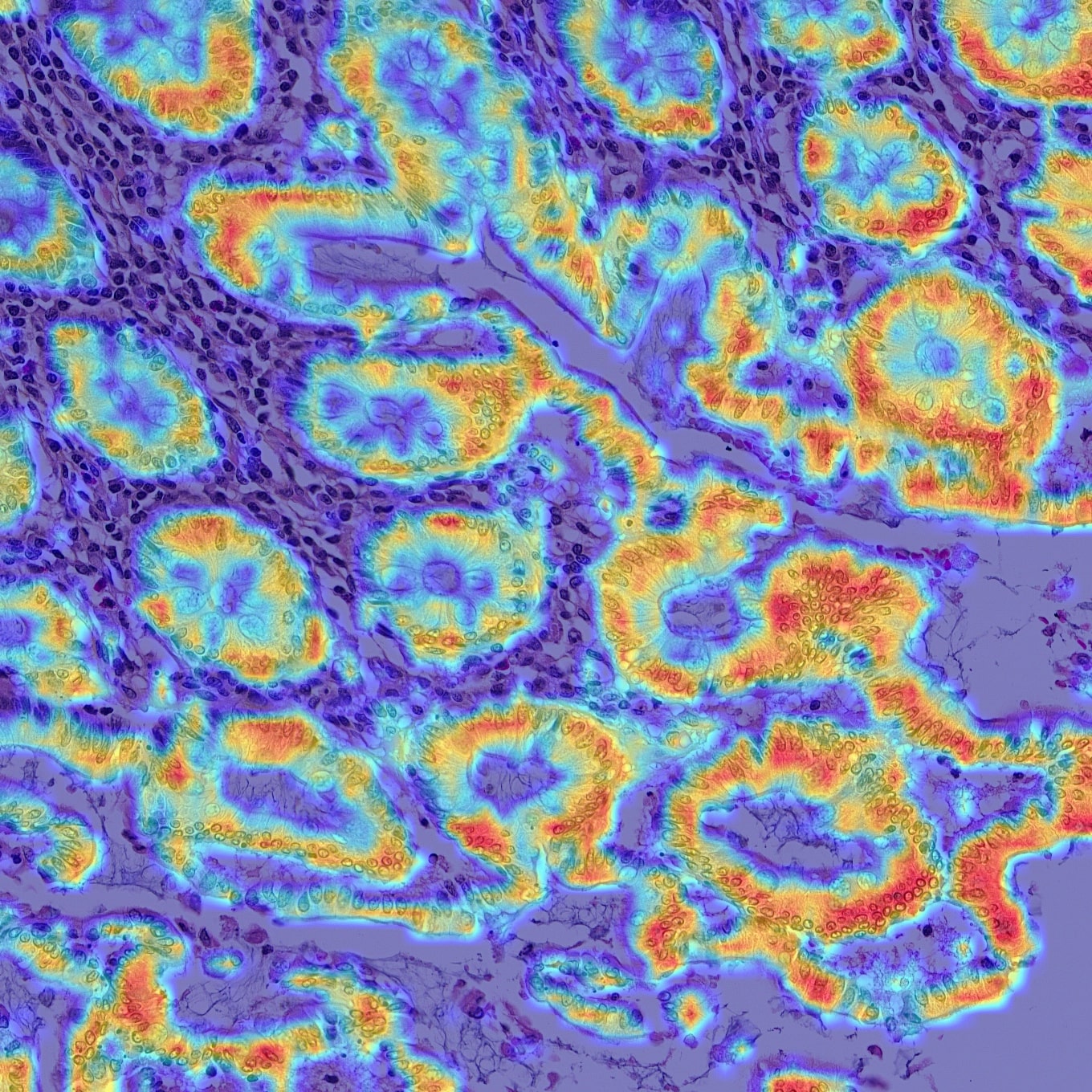}\\
GD
\end{minipage} &
\begin{minipage}{0.13\linewidth}
\centering
\includegraphics[width=\linewidth]{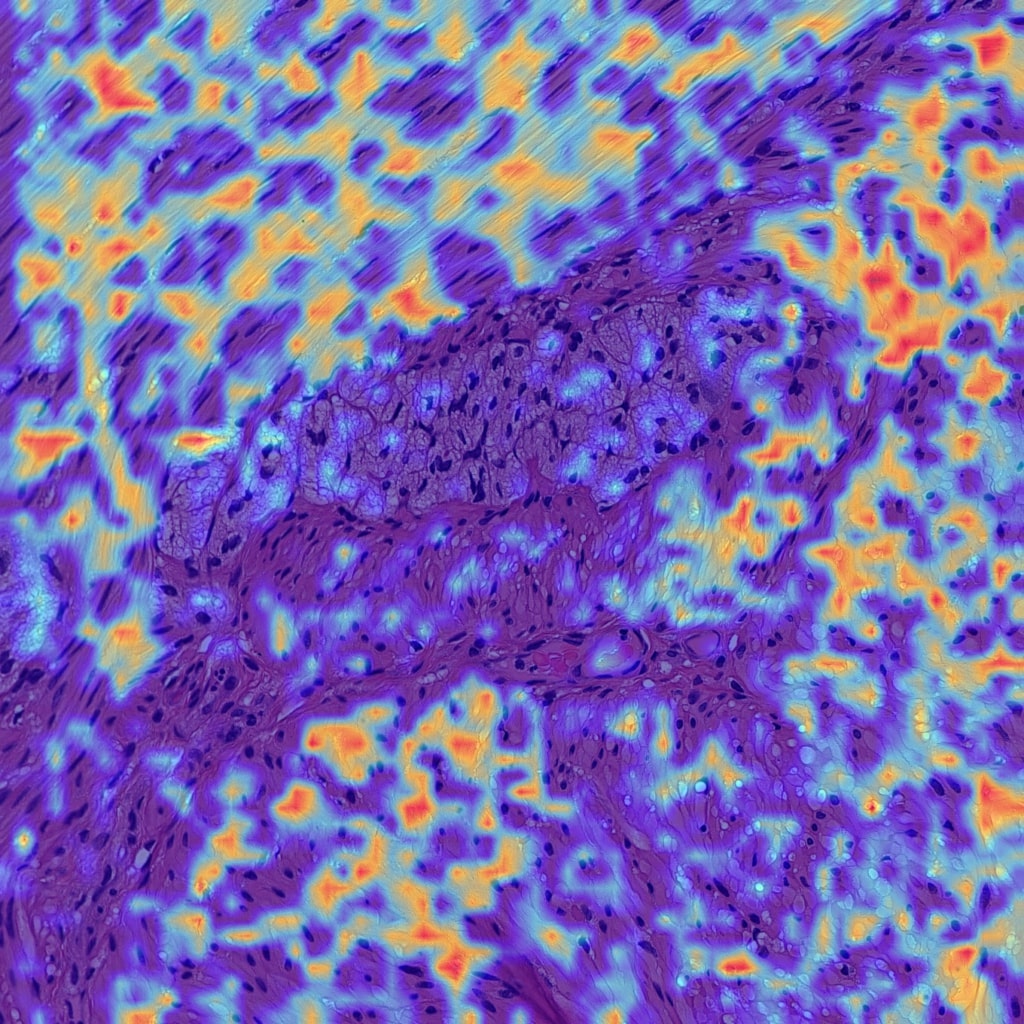}\\
NT
\end{minipage} &
\begin{minipage}{0.13\linewidth}
\centering
\includegraphics[width=\linewidth]{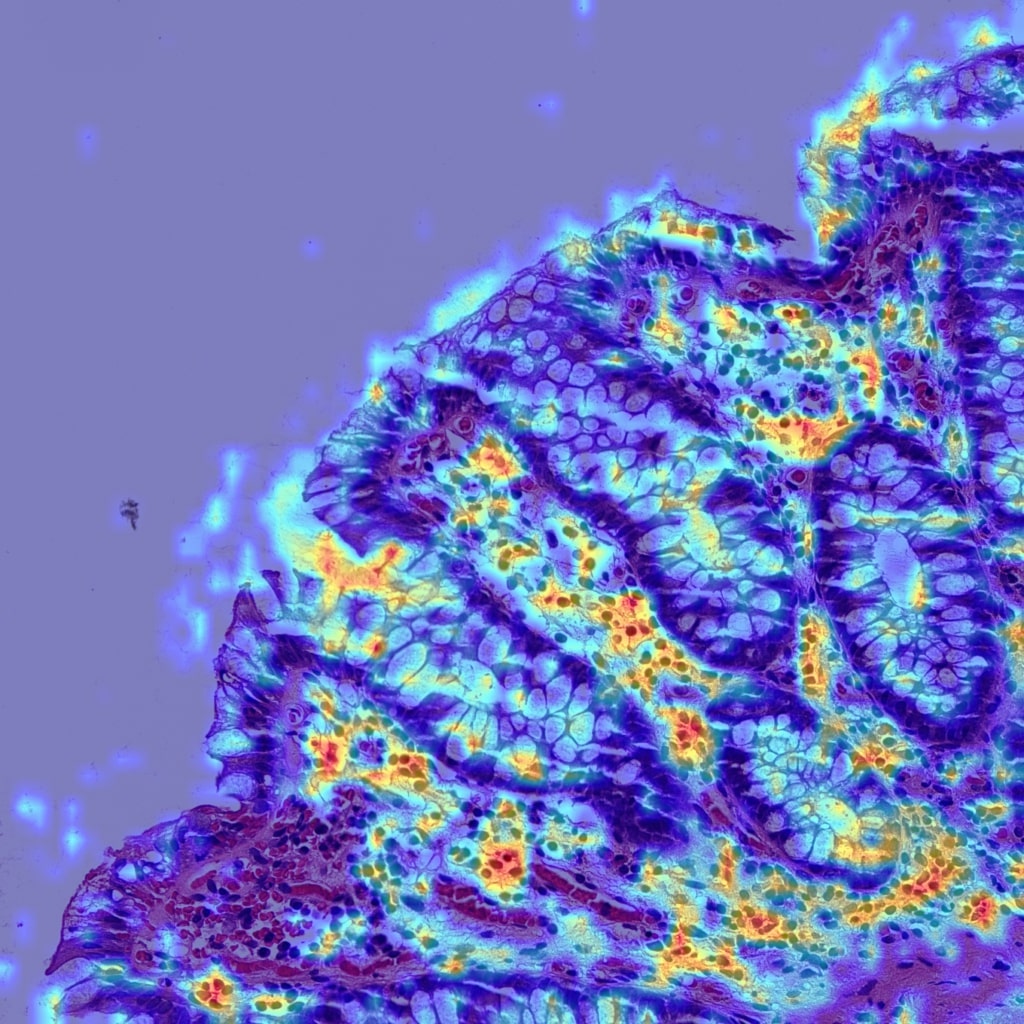}\\
ES
\end{minipage} &
\begin{minipage}{0.13\linewidth}
\centering
\includegraphics[width=\linewidth]{ 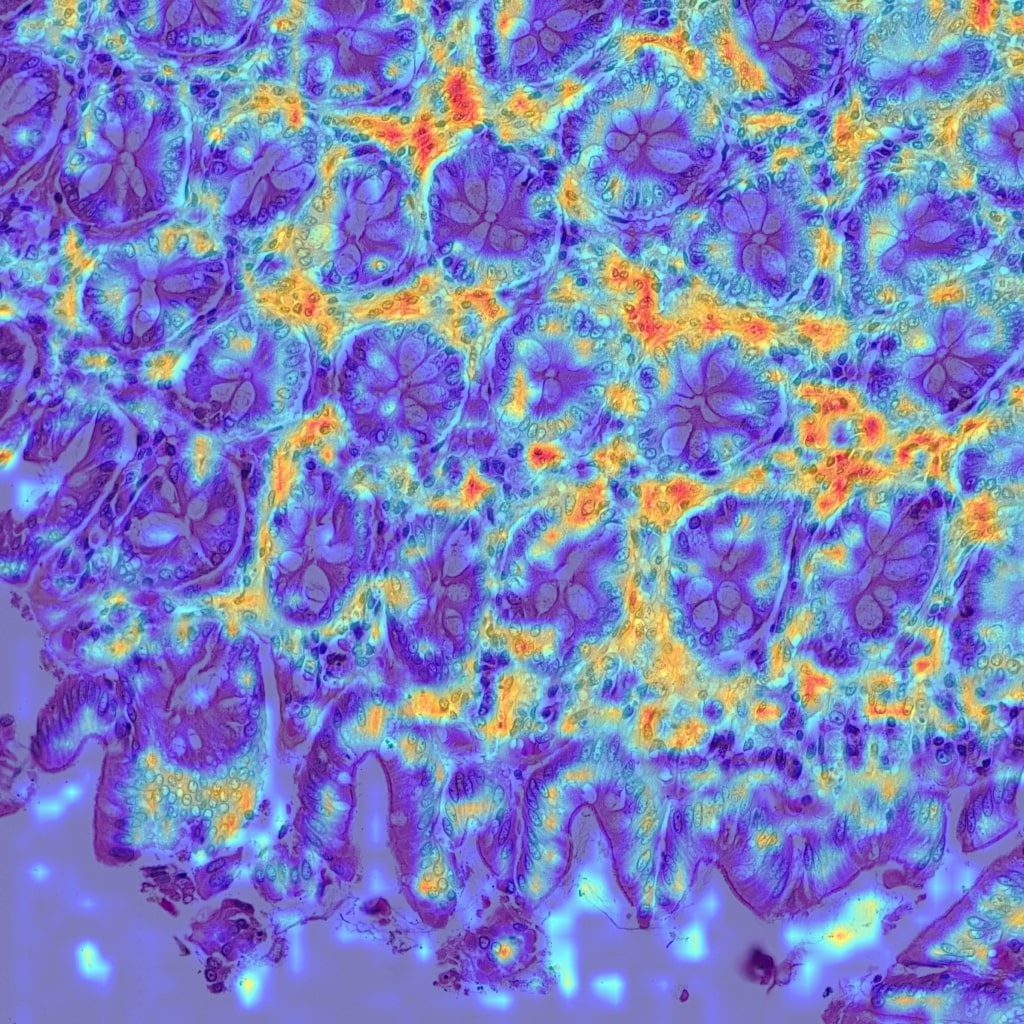}\\
LP
\end{minipage} &
\begin{minipage}{0.13\linewidth}
\centering
\includegraphics[width=\linewidth]{ 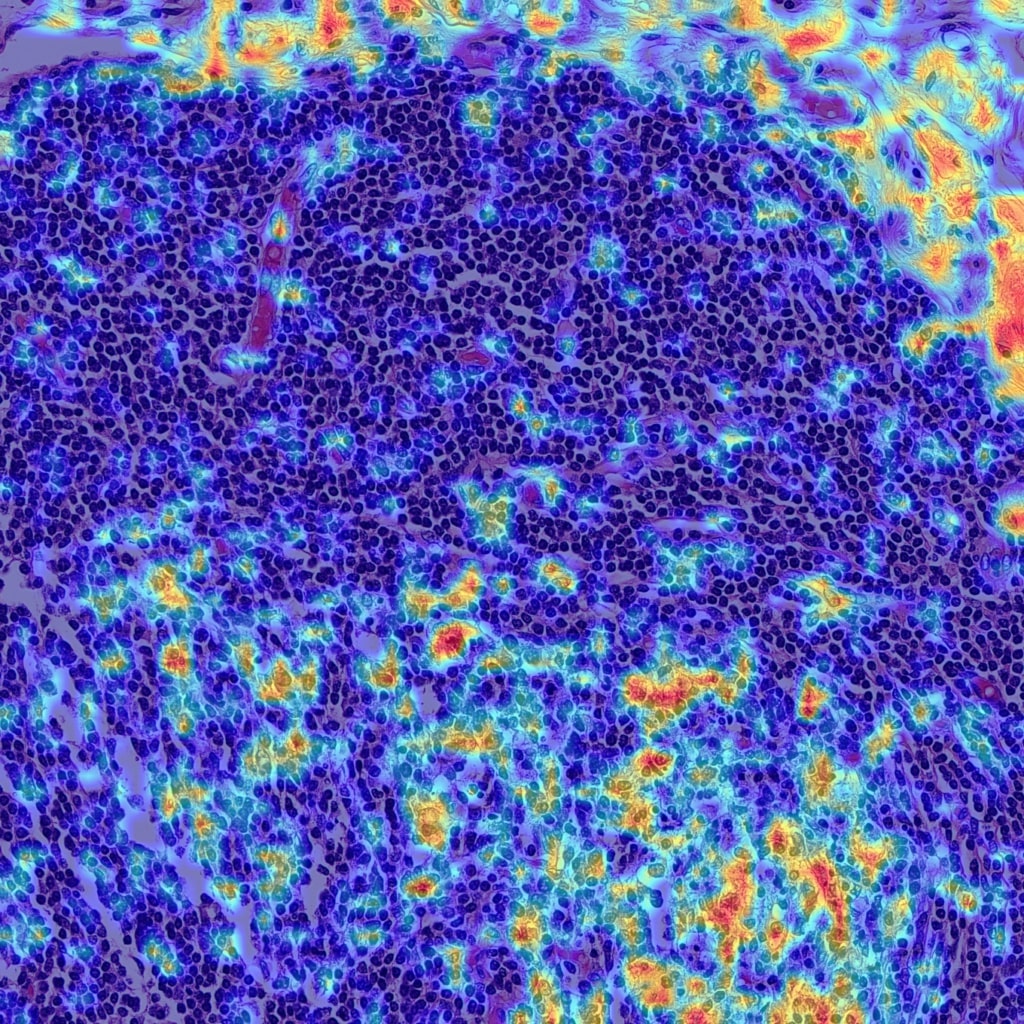}\\
MA
\end{minipage} \\

\begin{minipage}{0.13\linewidth}
\centering
\includegraphics[width=\linewidth]{ 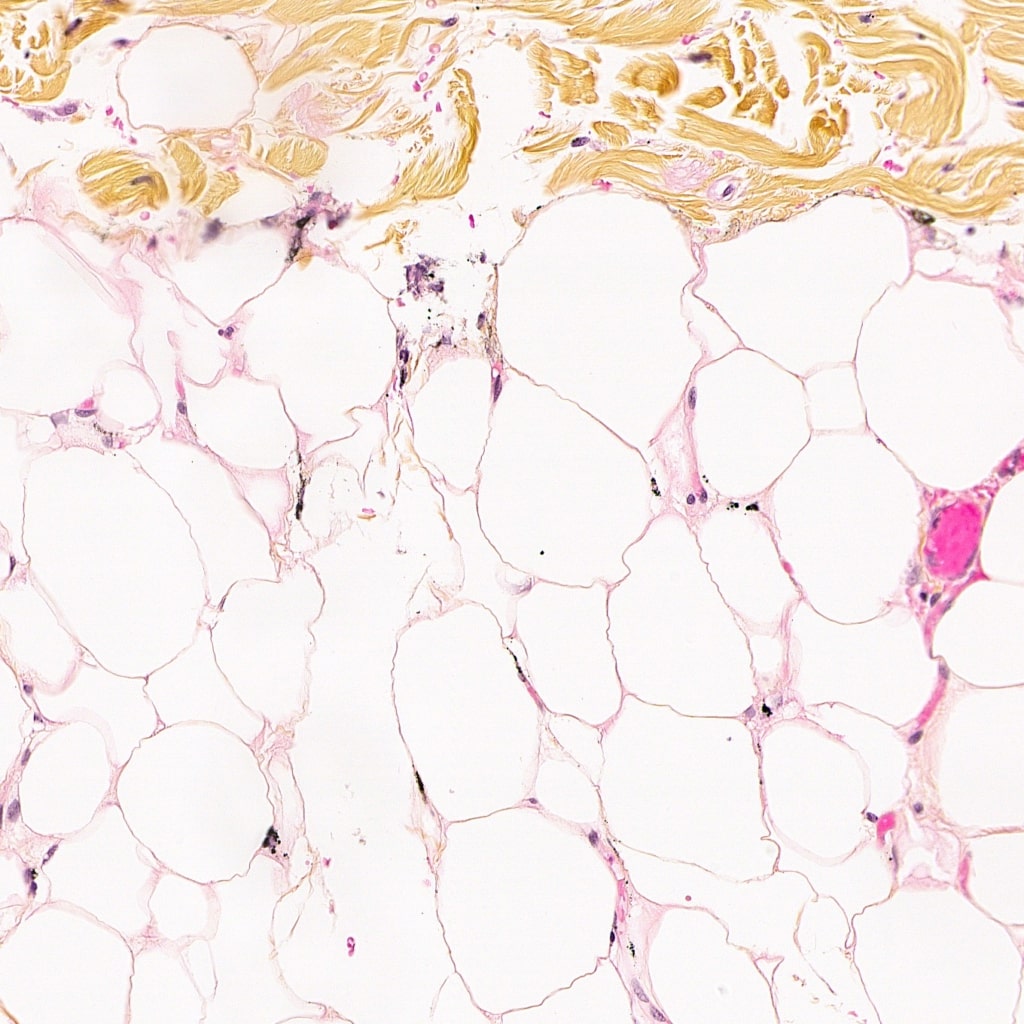}\\
AT (orig)
\end{minipage} &
\begin{minipage}{0.13\linewidth}
\centering
\includegraphics[width=\linewidth]{ 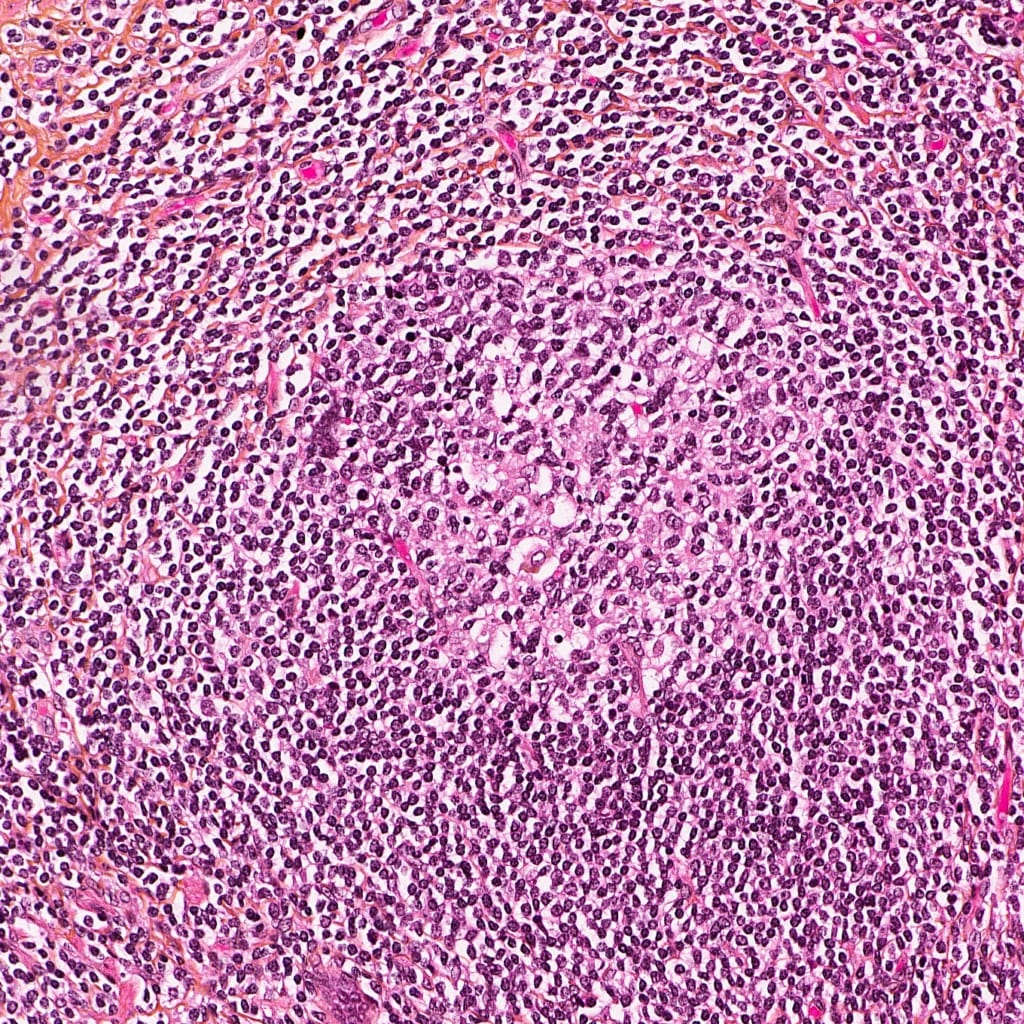}\\
LA (orig)
\end{minipage} &
\begin{minipage}{0.13\linewidth}
\centering
\includegraphics[width=\linewidth]{ 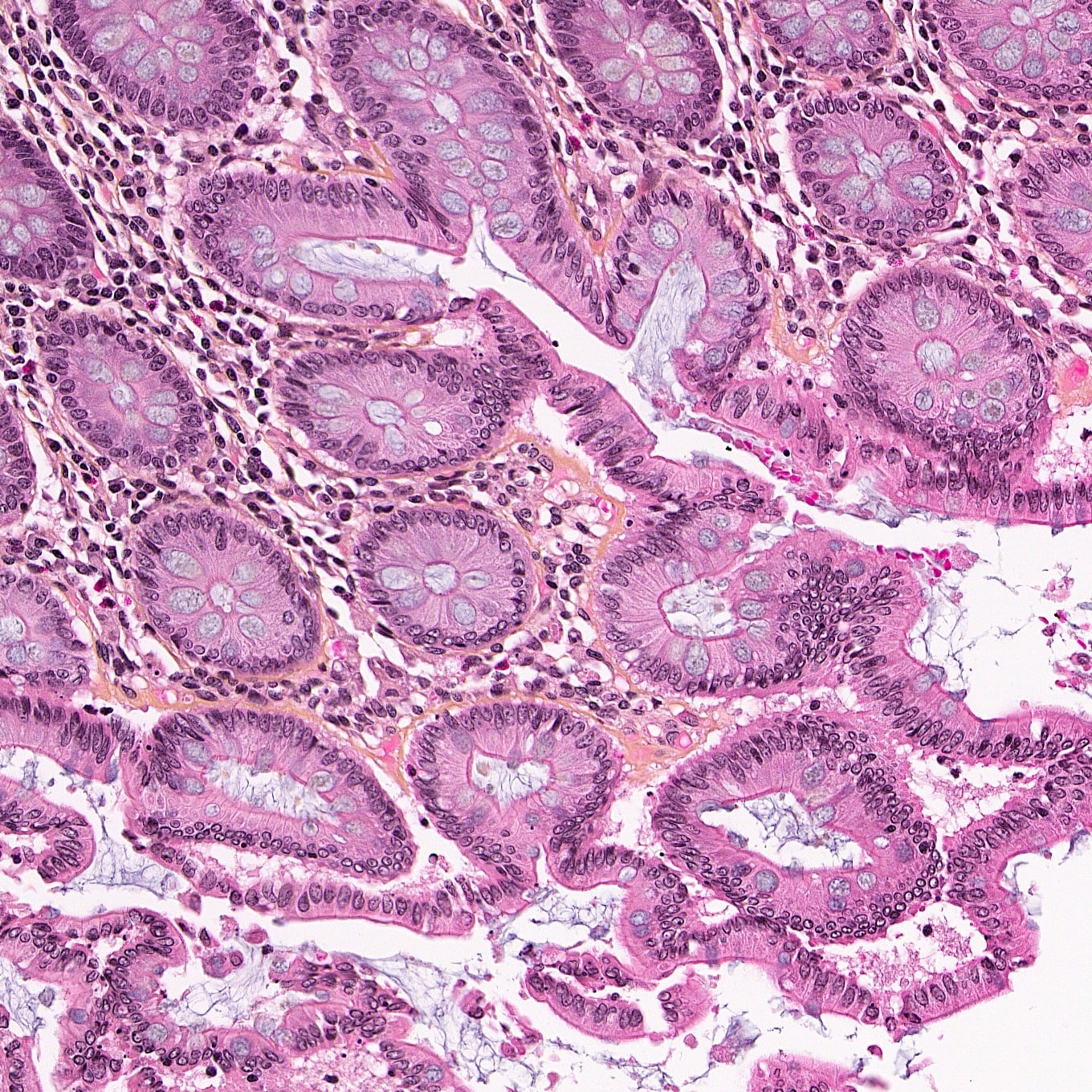}\\
GD (orig)
\end{minipage} &
\begin{minipage}{0.13\linewidth}
\centering
\includegraphics[width=\linewidth]{ 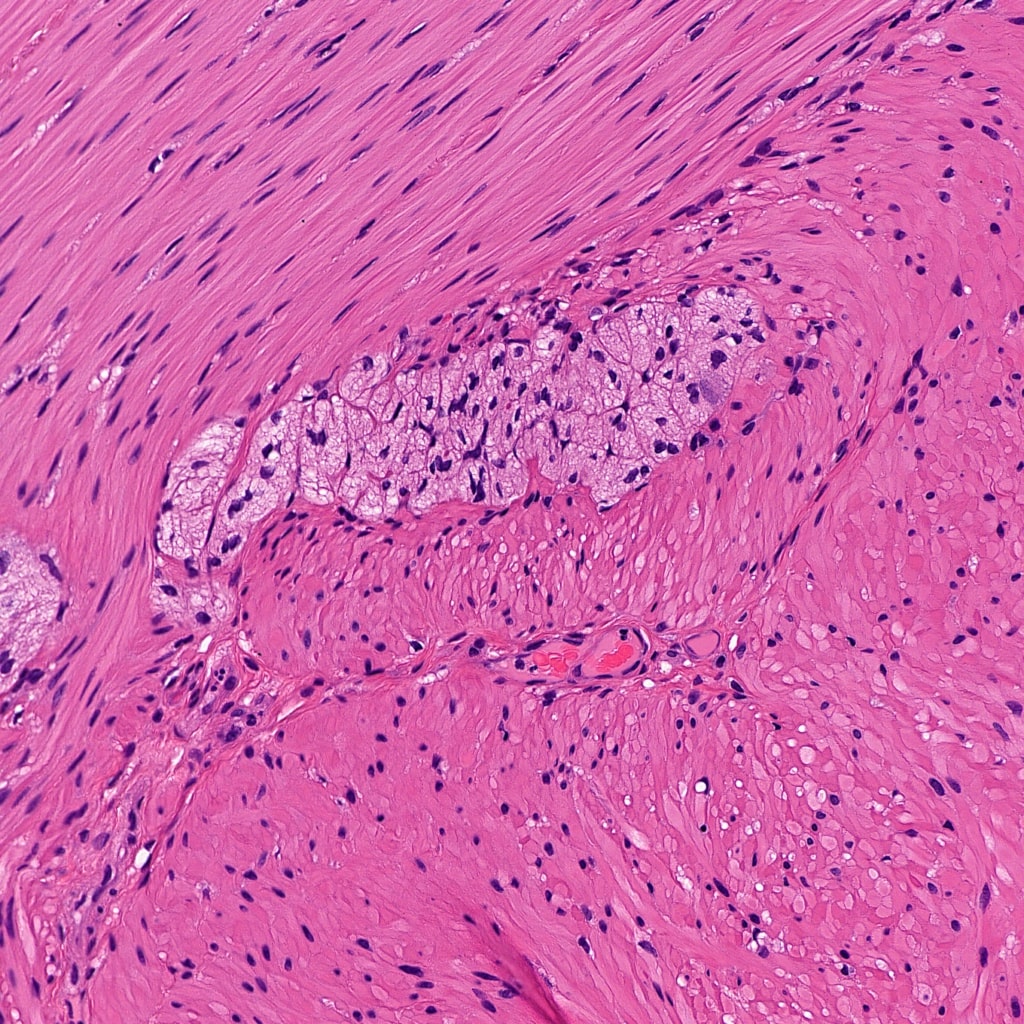}\\
NT (orig)
\end{minipage} &
\begin{minipage}{0.13\linewidth}
\centering
\includegraphics[width=\linewidth]{ 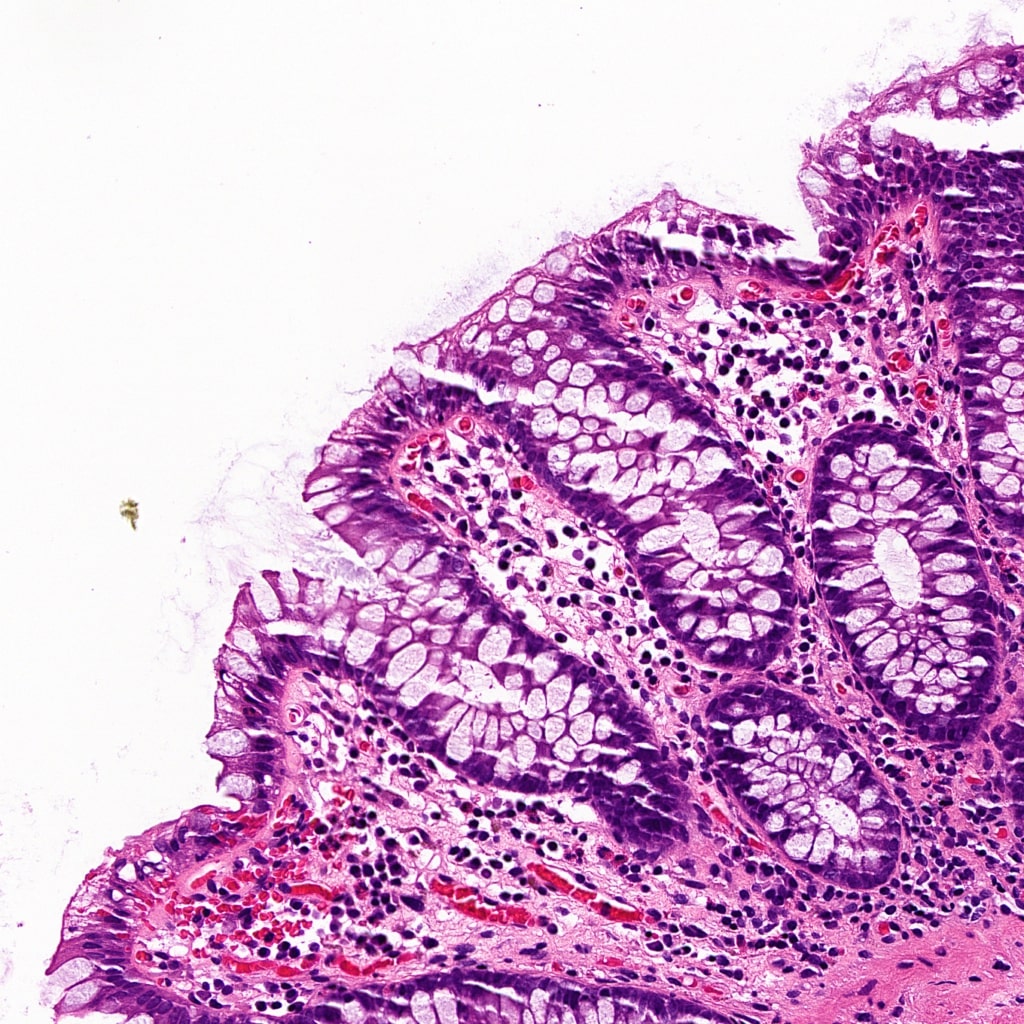}\\
ES (orig)
\end{minipage} &
\begin{minipage}{0.13\linewidth}
\centering
\includegraphics[width=\linewidth]{ 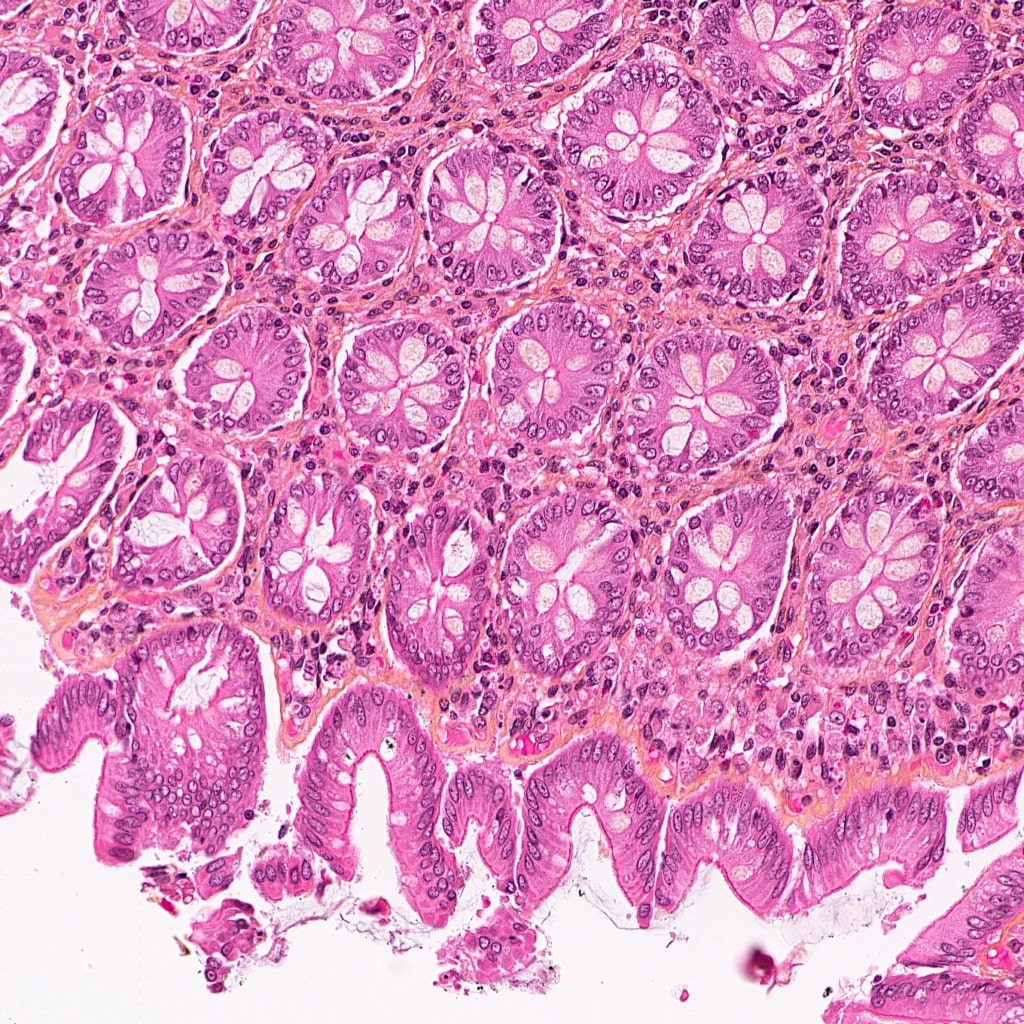}\\
LP (orig)
\end{minipage} &
\begin{minipage}{0.13\linewidth}
\centering
\includegraphics[width=\linewidth]{ 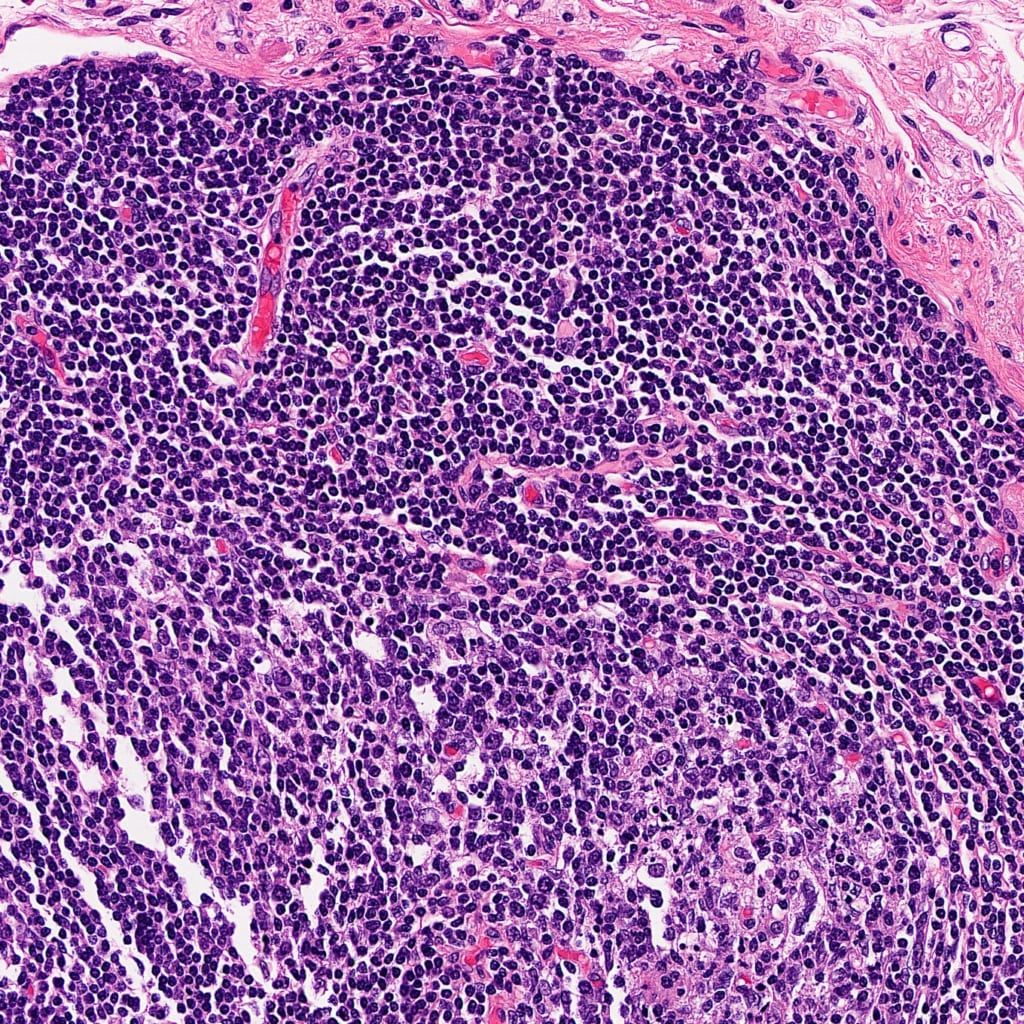}\\
MA (orig)
\end{minipage} \\

\begin{minipage}{0.13\linewidth}
\centering
\includegraphics[width=\linewidth]{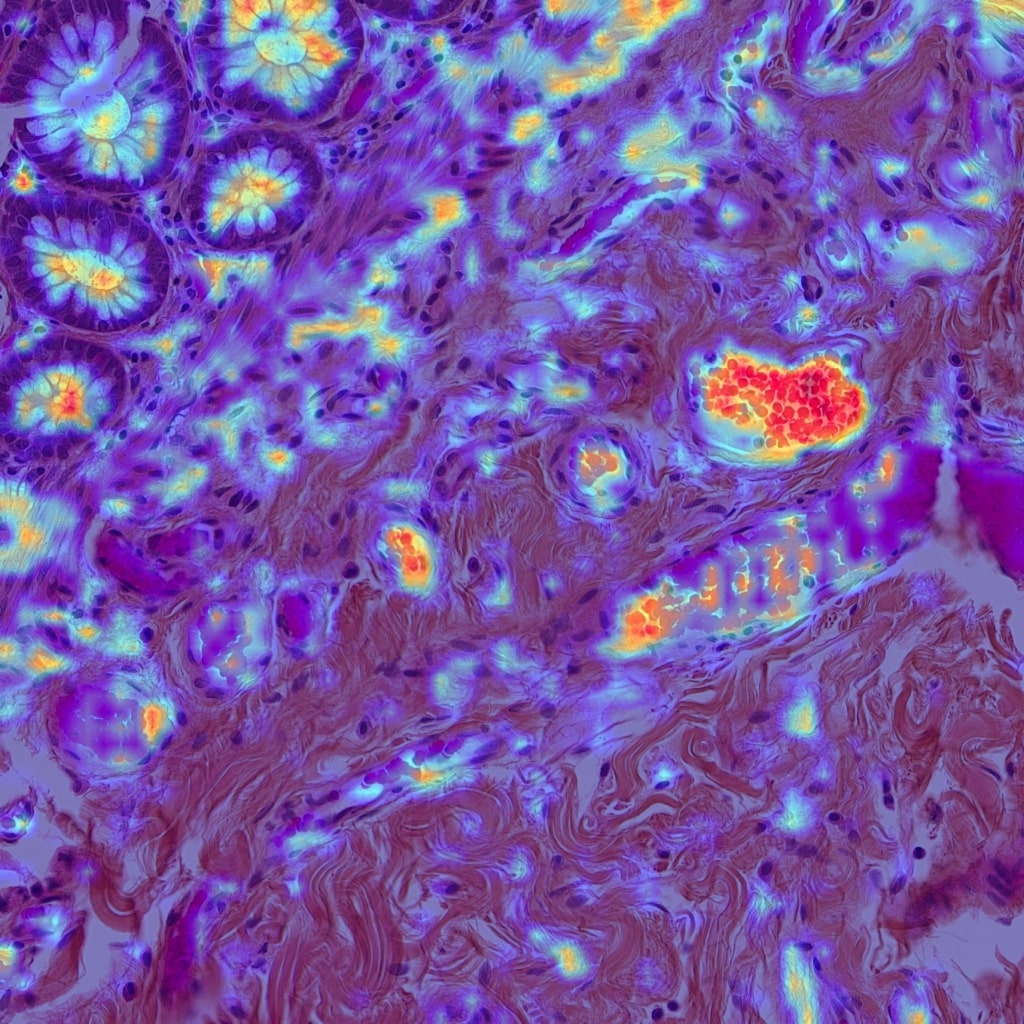}\\
RBC
\end{minipage} &
\begin{minipage}{0.13\linewidth}
\centering
\includegraphics[width=\linewidth]{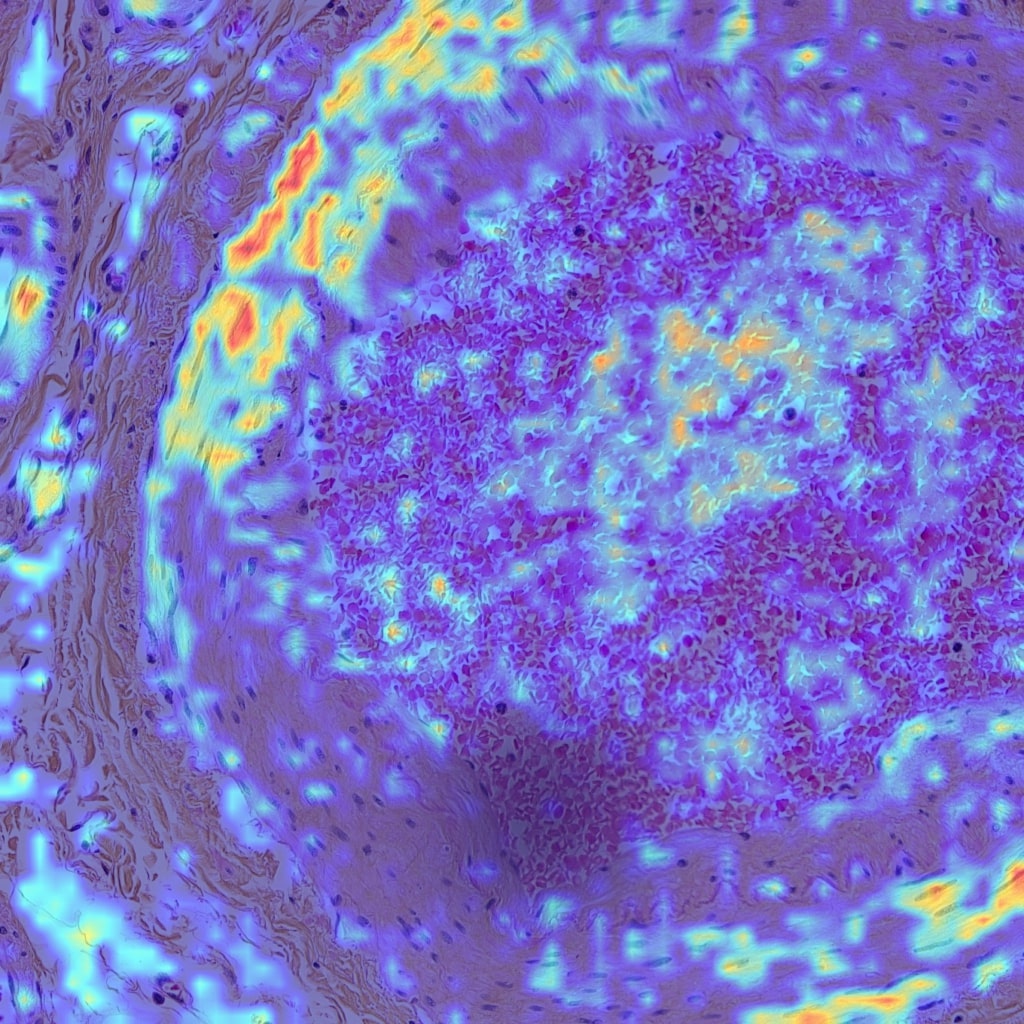}\\
V
\end{minipage} &
\begin{minipage}{0.13\linewidth}
\centering
\includegraphics[width=\linewidth]{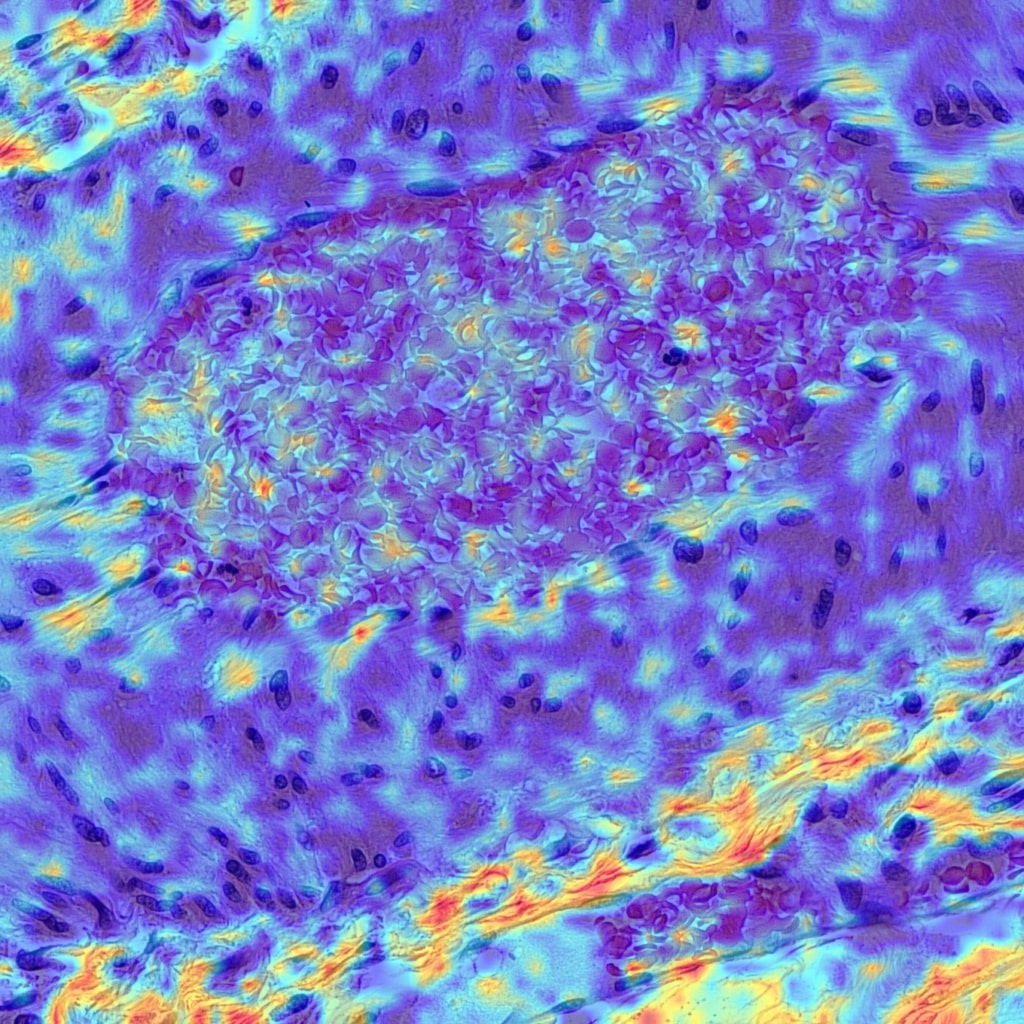}\\
EF
\end{minipage} &
\begin{minipage}{0.13\linewidth}
\centering
\includegraphics[width=\linewidth]{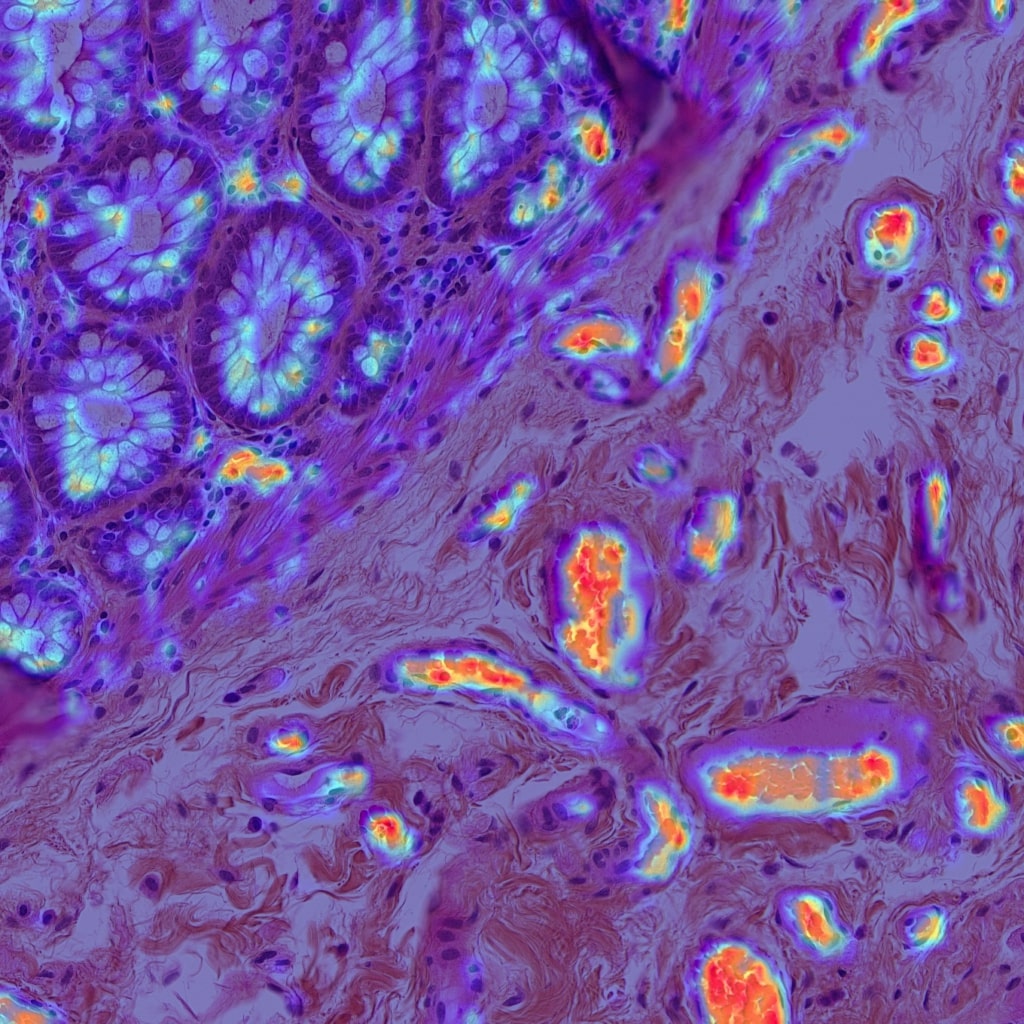}\\
LC
\end{minipage} &
\begin{minipage}{0.13\linewidth}
\centering
\includegraphics[width=\linewidth]{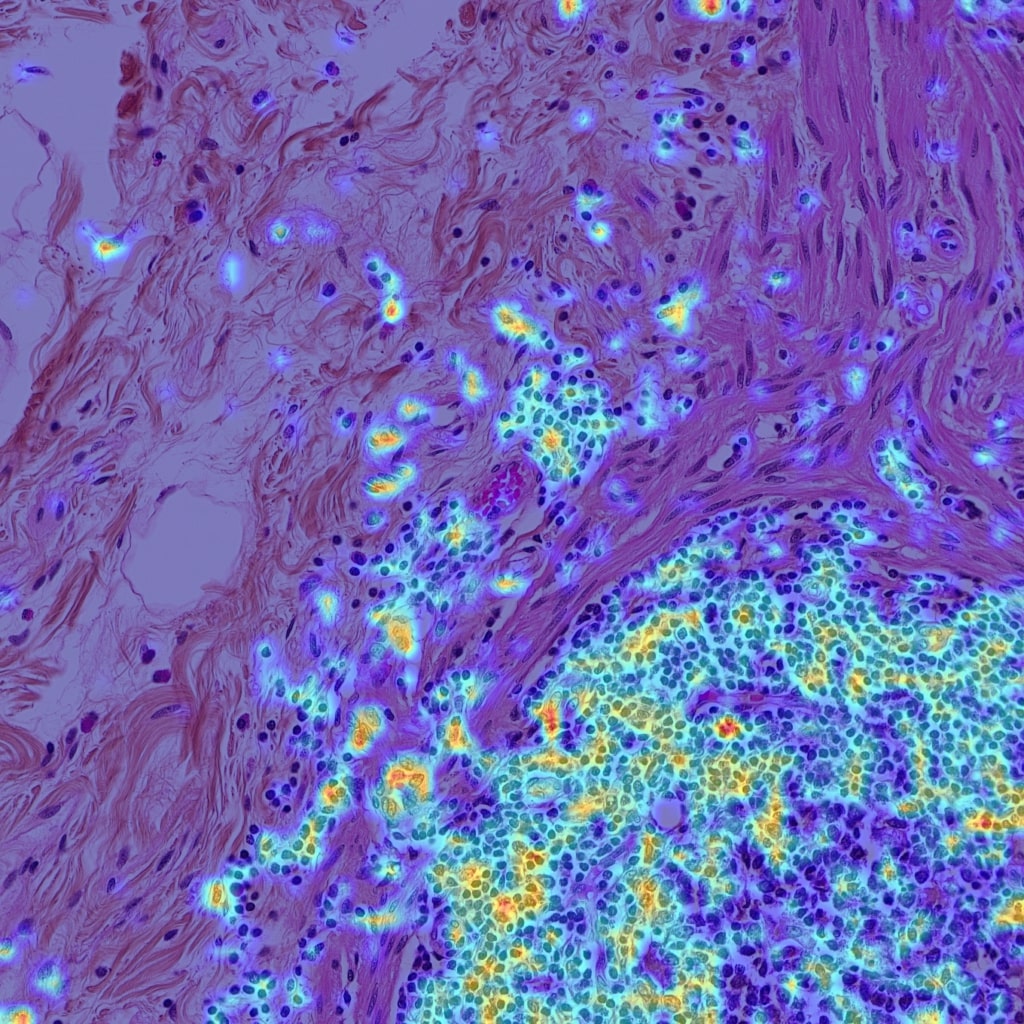}\\
LY
\end{minipage} &
\begin{minipage}{0.13\linewidth}
\centering
\includegraphics[width=\linewidth]{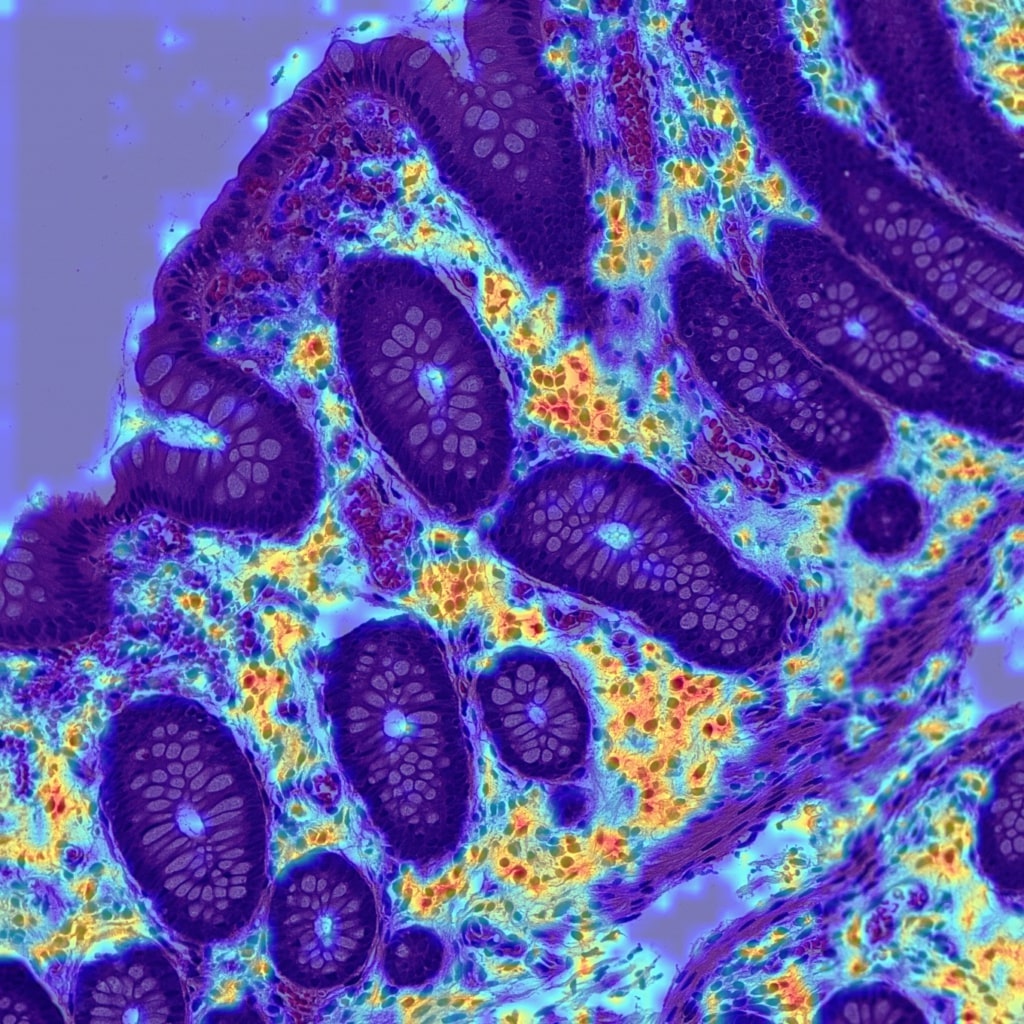}\\
PC
\end{minipage} &
\begin{minipage}{0.13\linewidth}
\centering
\includegraphics[width=\linewidth]{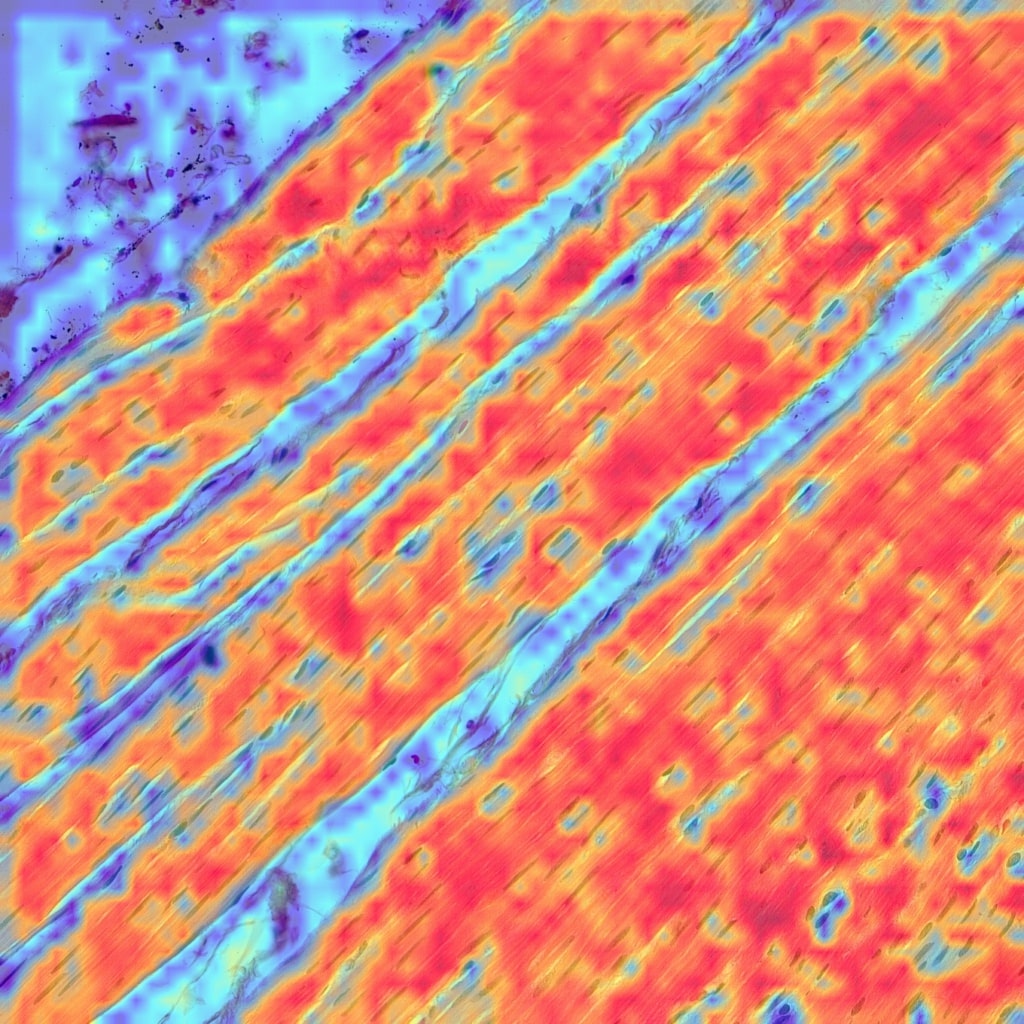}\\
SMC
\end{minipage} \\

\begin{minipage}{0.13\linewidth}
\centering
\includegraphics[width=\linewidth]{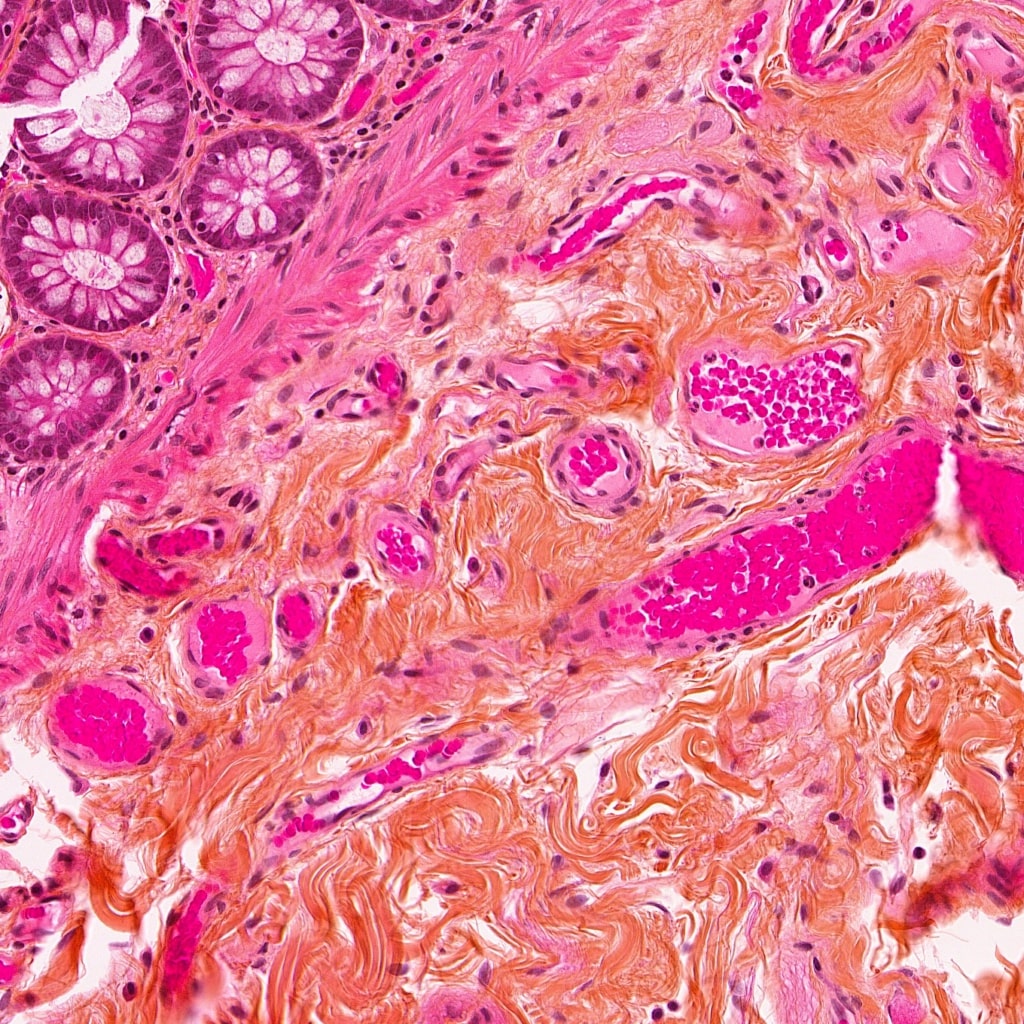}\\
RBC (orig)
\end{minipage} &
\begin{minipage}{0.13\linewidth}
\centering
\includegraphics[width=\linewidth]{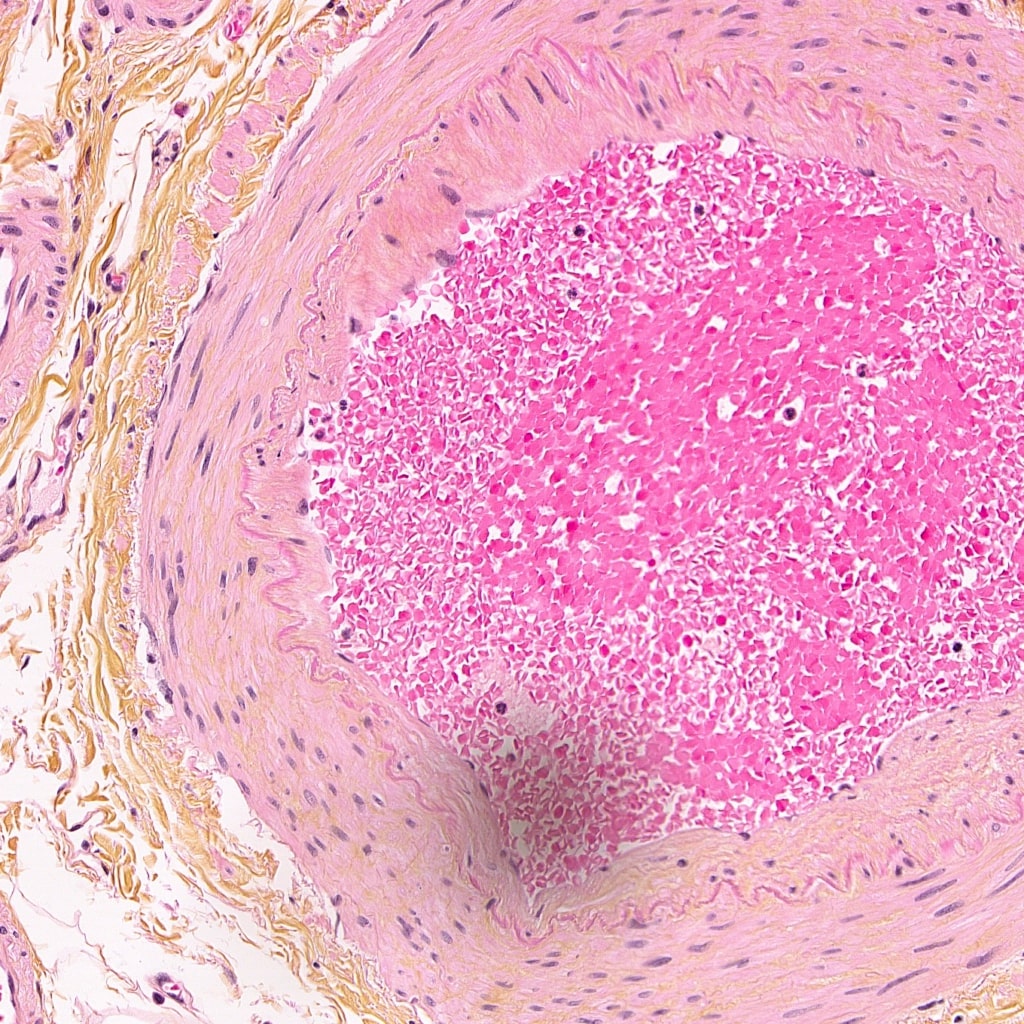}\\
V (orig)
\end{minipage} &
\begin{minipage}{0.13\linewidth}
\centering
\includegraphics[width=\linewidth]{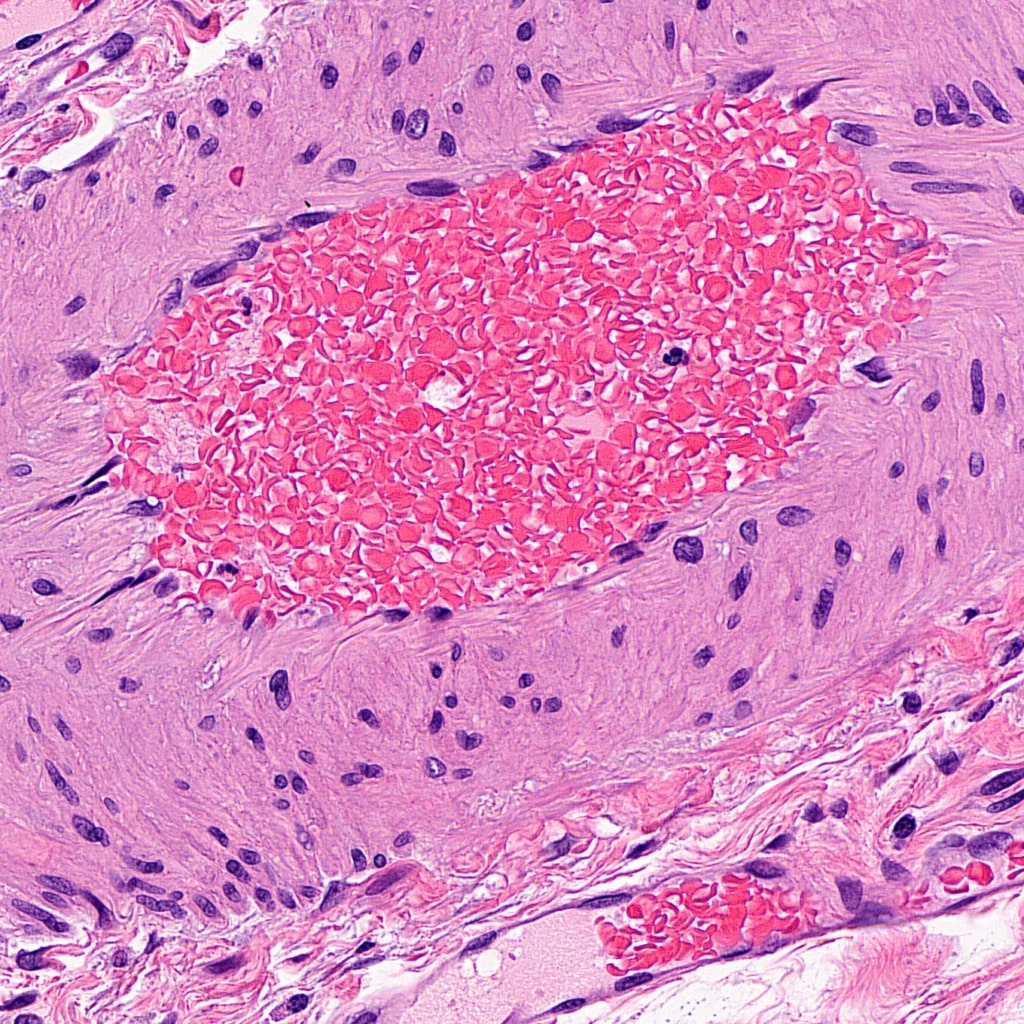}\\
EF (orig)
\end{minipage} &
\begin{minipage}{0.13\linewidth}
\centering
\includegraphics[width=\linewidth]{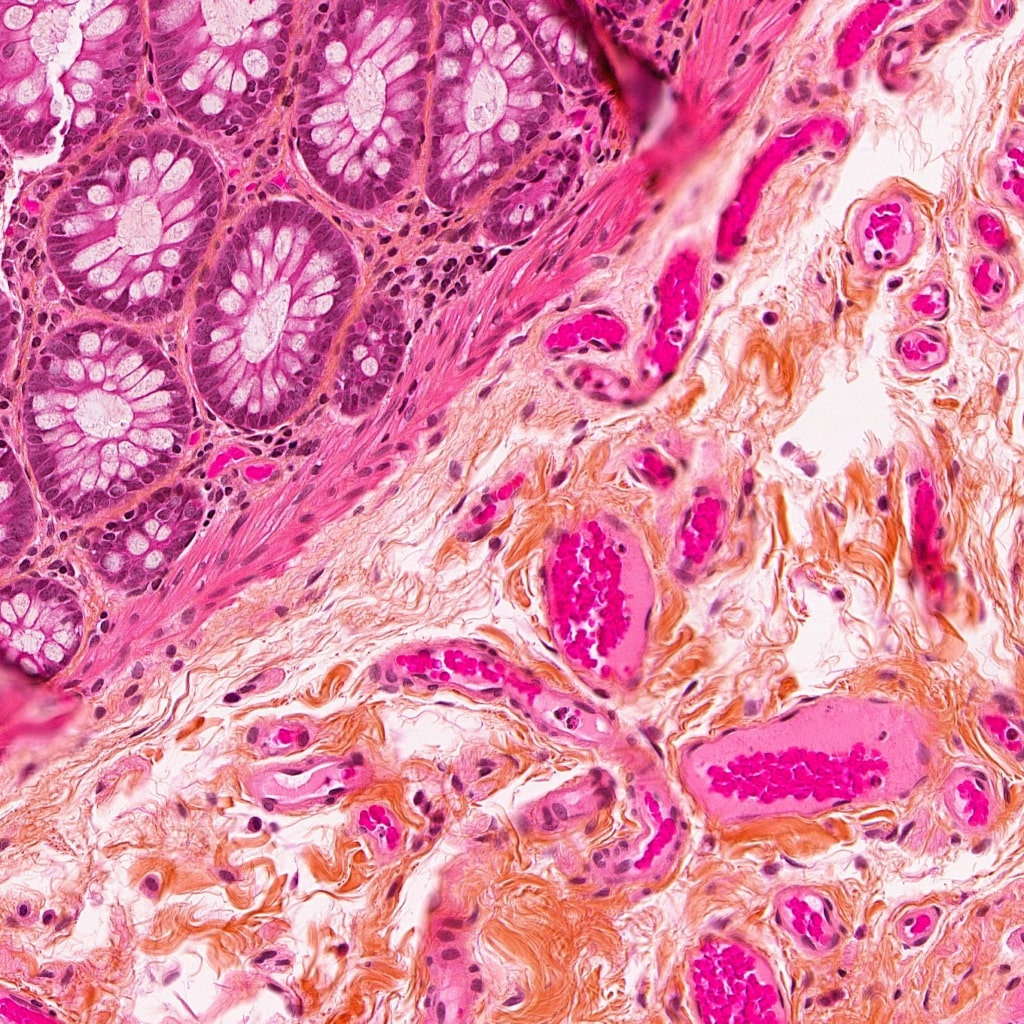}\\
LC (orig)
\end{minipage} &
\begin{minipage}{0.13\linewidth}
\centering
\includegraphics[width=\linewidth]{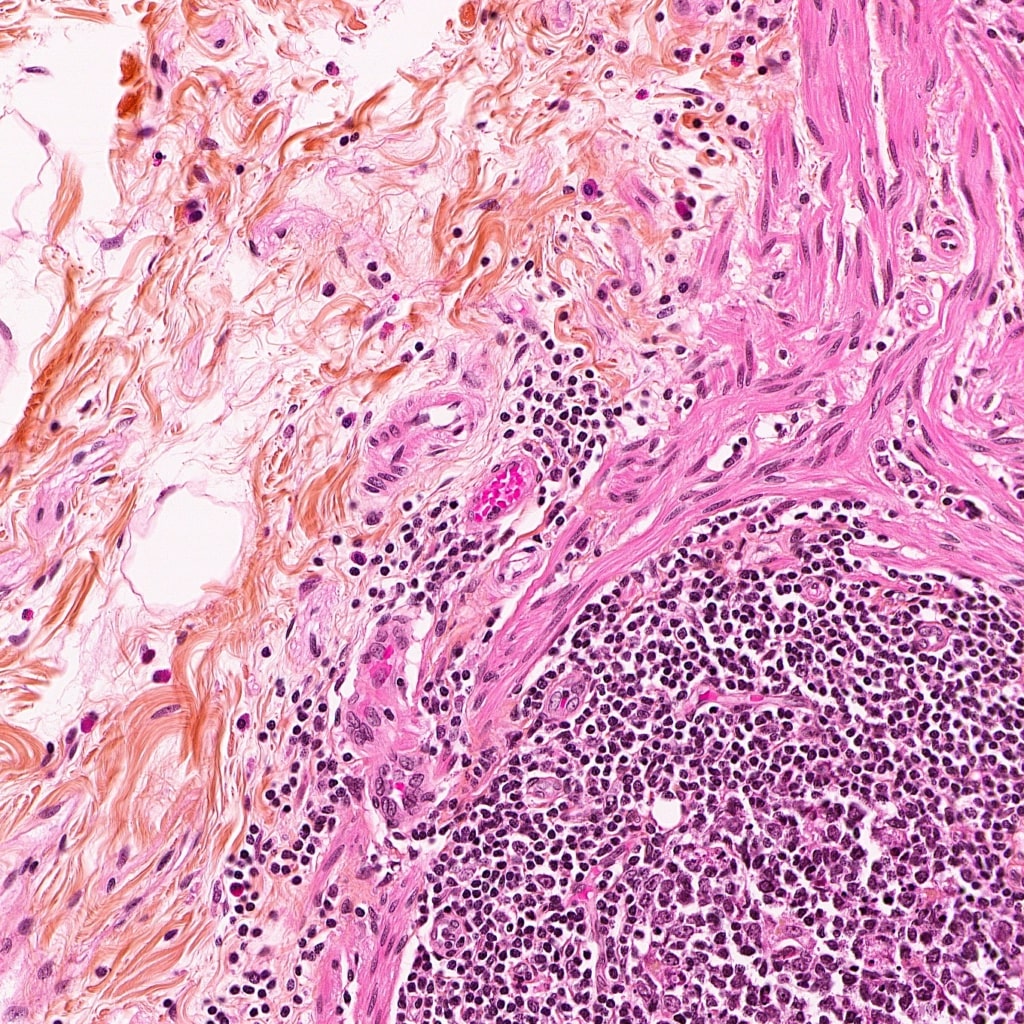}\\
LY (orig)
\end{minipage} &
\begin{minipage}{0.13\linewidth}
\centering
\includegraphics[width=\linewidth]{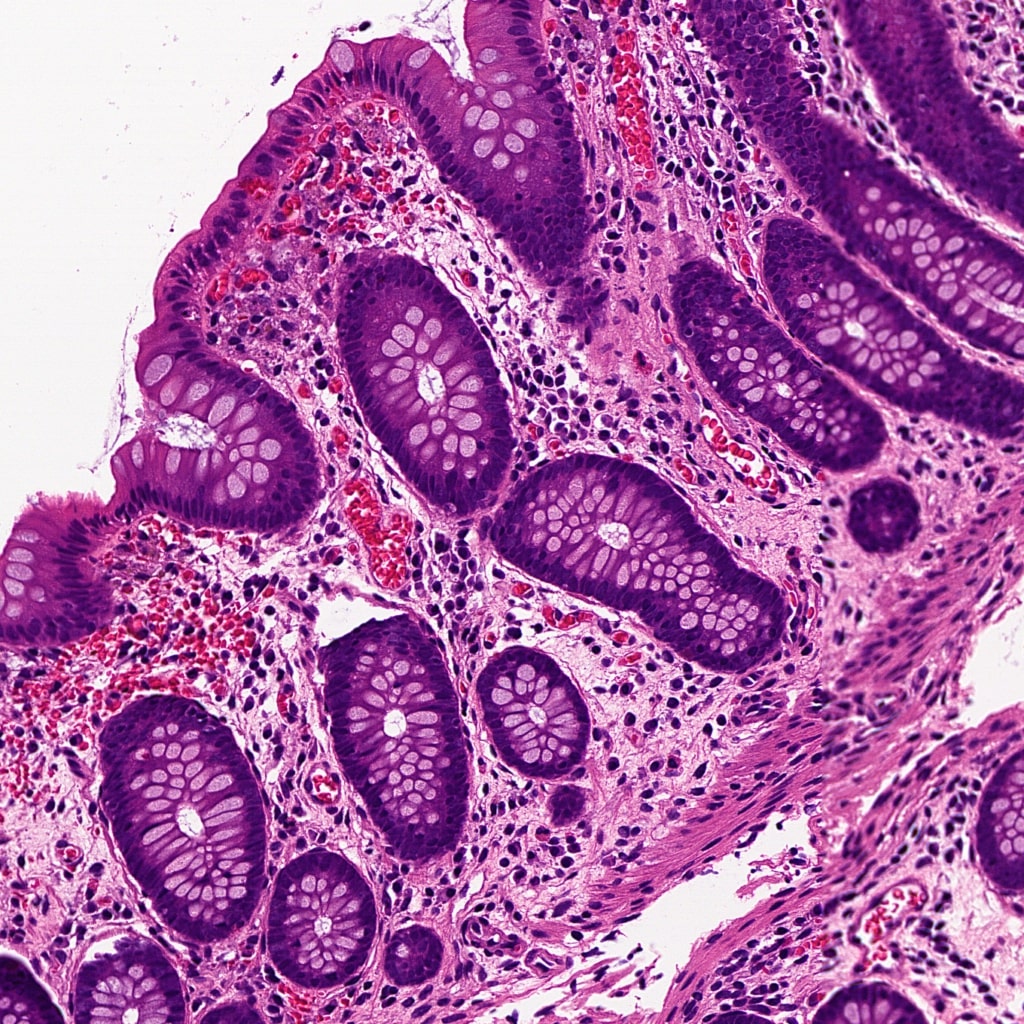}\\
PC (orig)
\end{minipage} &
\begin{minipage}{0.13\linewidth}
\centering
\includegraphics[width=\linewidth]{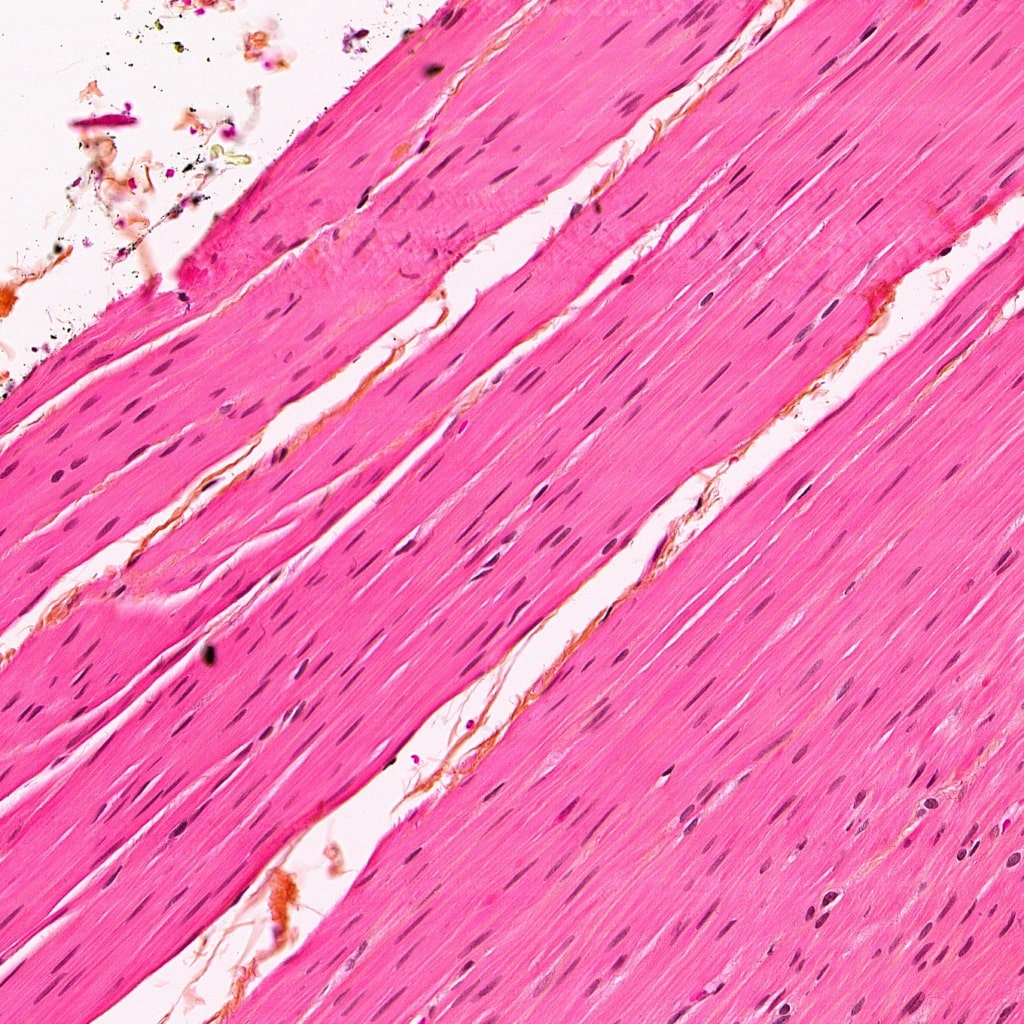}\\
SMC (orig)
\end{minipage} \\

\end{tabular}
} 
\caption{GradCAM vs. Original images across 14 HTTs from ADPv2. Rows 1 and 3 show GradCAM heatmaps; rows 2 and 4 show corresponding originals.}
\label{fig:gradcam_visualizations_pairs}
\end{figure}

We also leverage GradCAM to identify the salient features used by our model during prediction. GradCAM is performed on the patch level to show how granular tissue morphologies contribute to model's prediction of particular HTTs. For demonstration, we visualize the class activation maps of all 14 HTTs used in training on test image patches in our ADPv2 dataset, as in Figure \ref{fig:gradcam_visualizations_pairs}. 

Whereas for the majority of HTTs, activation maps as seen with GradCAM have mostly focused directly on the HTT of interest (e.g., LA, lymphocytes; GD, glands; SMC, smooth muscle cells), in other instances, salient features most used by the model as visualized by the activation maps include surrounding non-HTT tissues (e.g., for AT, adipose tissue) or interfaces/limits between HTT-positive and non-HTT regions (e.g., for V, large vessels).

\subsection{Confidence Score Distribution Analysis}

\begin{figure}[h]
\centering
\begin{subfigure}{0.45\textwidth}
\centering
\includegraphics[width=\textwidth]{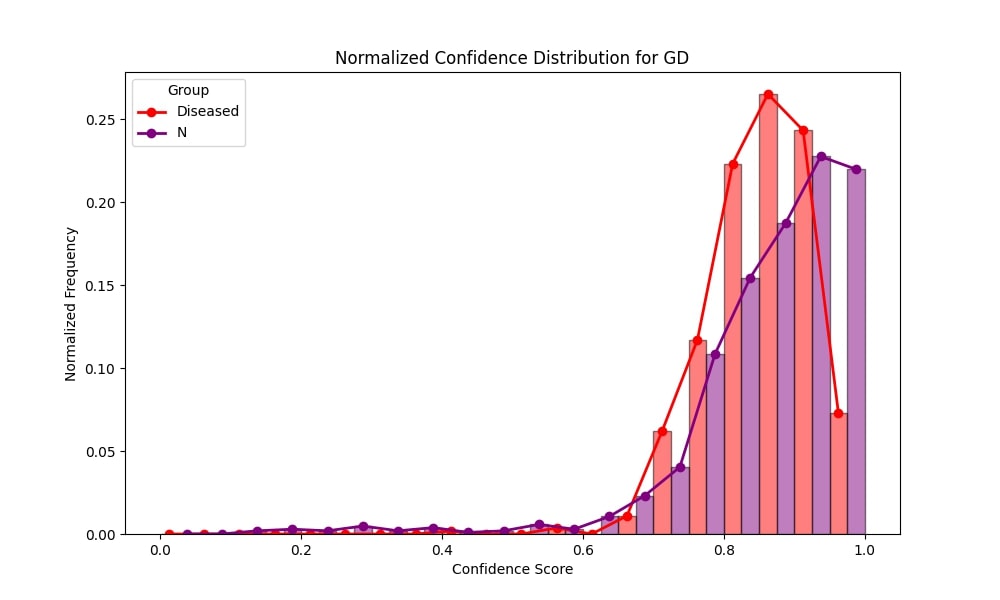}
\caption{Healthy vs Diseased}
\label{fig:dist_plots_h_vs_d}
\end{subfigure}
\begin{subfigure}{0.45\textwidth}
\centering
\includegraphics[width=\textwidth]{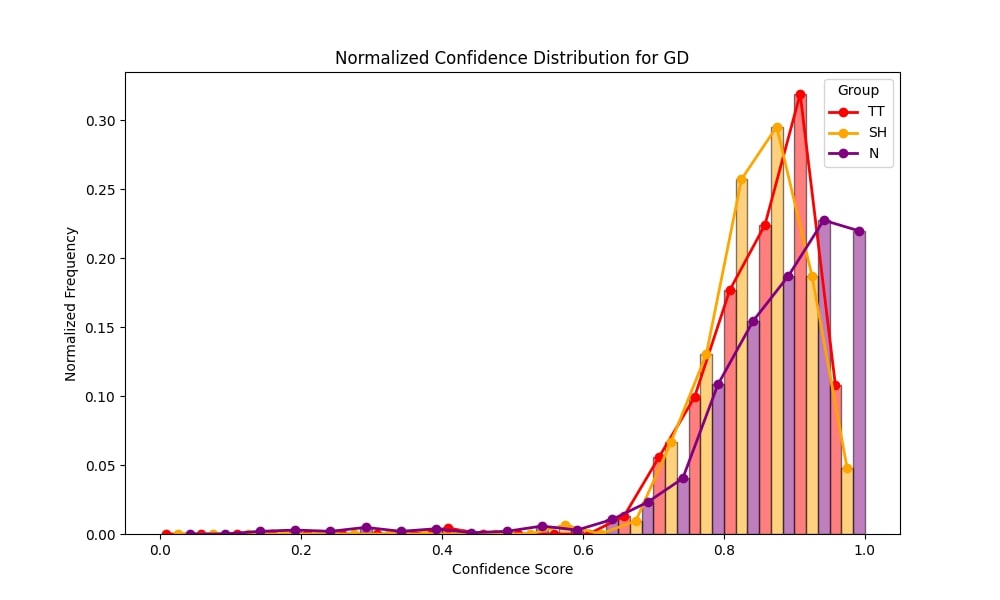}
\caption{Healthy vs SH vs TT}
\label{fig:dist_plots_sn_vs_tt}
\end{subfigure}
\caption{Predicted confidence score histogram for glands. (a) The histogram for healthy patches and diseased patches, where diseased patches are four diseases combined. (b) The histogram for healthy patches and two precursor pathways of colon cancer development, where SH is SSL and HP combined, and TT is TA and TVA combined.
Abbreviations: GD = glands; N = normal; SH = sessile serrated lesions and hyperplastic polyps;  TT = tubular and tubulovillous adenomas.}
\label{fig:dist_plots}
\end{figure}

In Figure \ref{fig:dist_plots}, we present the distribution of the predicted confidence score on patches extracted from gland areas on slides of five categories. From Figure \ref{fig:dist_plots_h_vs_d}, we observe a clear separation between the diseased distribution and the healthy distribution, where the diseased confidence scores are shifted to the left, which could be related to the subtle glandular distortions as interpreted by the model. Further, in Figure \ref{fig:dist_plots_sn_vs_tt}, there is a separation between TT (TA + TVA) and SH (SSL + HP), suggesting that the model's confidence is affected by underlying morphological glandular differences between categories of colorectal precursor lesions as they diverge from normal mucosa (e.g., SSL+HP: luminal epithelial serrations, and, in most cases, crypt elongation, small basally located nuclei; TA+TVA: dysplasia, by definition, with prominent pseudostratification in most cases).\cite{33961885}\cite{35457279}\cite{21585430} Of note, a decrease in interrater agreement regarding colorectal polyp classification can be often similarly seen among both general and expert pathologists, highlighting the complexity of 'diseased' glandular interpretation and potential decrease in confidence scores.\cite{21585430}\cite{23611360}

We quantify the distribution shifts using t-tests. As summarized in Table~\ref{tab:gd_ttest}, sessile serrated lesions+hyperplastic polyps elicit a pronounced reduction in model confidence relative to normal tissue (Welch $t=-6.98$, $p_{\text{adj}}<0.001$), whereas the effect is more modest for traditional tubular+tubulovillous adenomas
($t=-1.99$, $p_{\text{adj}}=0.047$).
TT and SH distributions also differ significantly from each other
($t=3.88$, $p_{\text{adj}}<0.001$), with a shift in classifier confidence scores \emph{within the same tissue compartment}.
These pathway-specific confidence signatures indicate that the network is recognising subtle, biologically meaningful glandular alterations highlighting their potential as \emph{computational biomarkers} for precursor lesion characterization in colorectal cancer screening. 

\begin{table}[ht]
    \centering
    \caption{Welch two-sample $t$-tests on logit-transformed confidence scores for glandular tissue (GD).\newline
             $p$-values are Holm–Bonferroni adjusted to control the family-wise error rate.}
    \label{tab:gd_ttest}
    \begin{tabular}{lcc}
        \toprule
        \textbf{Comparison} & \textbf{$t$-statistic} & \textbf{Adjusted $p$-value} \\
        \midrule
        TT (TA + TVA) vs SH (SSL + HP) &  \phantom{$-$}3.88 & $<0.001$\\
        TT (TA + TVA) vs Normal        &  $-1.99$          & $0.047$\\
        SH (SSL + HP) vs Normal        &  $-6.98$          & $<0.001$\\
        \bottomrule
    \end{tabular}
\end{table} 

\begin{figure}[ht]
    \centering
    \includegraphics[width=0.8\textwidth]{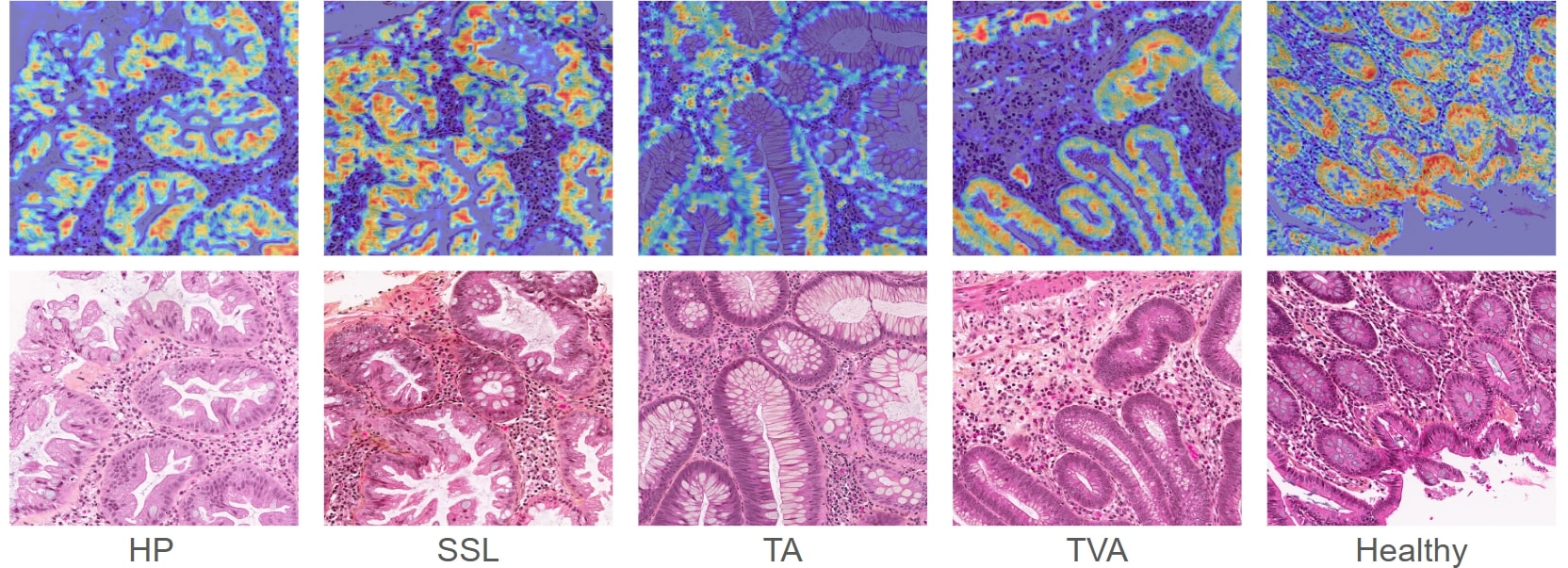}
    \caption{Sampled gland patches from each disease. The model is focused on the shape and structure of the glands when making predictions, which is also the major factor used by pathologists to identify colon cancers.}
    \label{fig:gradcam_dist}
\end{figure}

To further understand distribution shifts in terms of the model's prediction behavior, we visualize the model's attention on diseased patches using GradCAM heatmaps, as shown in Figure \ref{fig:gradcam_dist}. Through the heatmaps, we observe a predominant mucosal epithelial contribution with a minor component from the lamina propria surrounding the 'diseased' glands in cases of HP, SSL, TA and TVA. In healthy tissues, the 'gland' (GD) activation map is relatively uniform and diffuse across the glandular epithelium component, with slight local variations from contributions from epithelial cell nuclei and cytoplasm. In HP and SSL cases, we note strong activations along the glandular epithelium, but most specifically at the sawtooth luminal borders,and branching crypt outlines, with increased contribution from cell cytoplasm, whereas TA and TVA cases elicit model activation hotspots mostly on atypical nuclei (instead of cytoplasm), which can be elongated and pseudostratified, within the dysplastic epithelium, and, to a lesser extent, the lamina propria surrounding the basal aspect of the crypts.

The heatmaps demonstrate that our model classifier focuses diagnostically relevant features, reflecting biologically meaningful deviations from normal glandular architecture. Our model appears to attend to disease‑specific glandular alterations rather than background artifacts to visually anchor its grouping of serrated versus adenomatous lesions. These saliency maps could therefore provide an interpretable bridge between deep‑learning predictions and diagnostic pathology, pending further validation.

\section{Conclusion and Discussion}

In this work, we introduced the ADPv2 dataset, a carefully curated repository of 20,004 annotated image patches from healthy colon biopsies, enriched with a 32-label hierarchical taxonomy. Beyond, we developed and fine-tuned a deep learning pipeline that combines Barlow Twins self-supervised pretraining with the VMamba architecture, demonstrating both the robustness of our data and the efficacy of this novel model design. In addition, we presented an in-depth confidence distribution analysis, revealing characteristic shifts in the model’s predicted confidence scores when encountering various disease groups. These shifts—lower peak sharpness, leftward displacement, and wider spread—could signal underlying structural abnormalities interpreted by the model and, specifically in the case of glandular analysis, might reflect known pathological changes in colorectal precursor lesions. 

Together, these findings significantly advance the field of computational pathology by providing a high-quality dataset, a powerful modeling strategy, and actionable biological insights. Leveraging the ADPv2 dataset and our self-supervised learning-based paradigm can bolster clinical diagnostics, as the observed distribution shifts in confidence scores may be eventually harnessed as image-based markers of disease. By spotlighting subtle deviations from healthy tissue, the framework holds promise for improving not only automated detection and classification but also the potential discovery of novel histopathological biomarkers. Overall, our contributions underscore the growing synergy between data-centric solutions and sophisticated deep learning techniques in supporting pathologists' workflow and enhancing patient care.

\section{Declaration of generative AI and AI-assisted technologies in the writing process}
During the preparation of this work, the authors used ChatGPT for English writing consistency and fluency. After using this tool, the author reviewed and edited the content as needed and take full responsibility for the content of the published article.

\section*{Declaration of competing interests}
The authors have nothing to declare.

\section*{Funding Sources}
This study was funded by the Government of Ontario under the ORF-RE grant (ORF-Re10\_026), and by the NSERC-Discovery Grant Program (RGPIN-2022-05378).

\section*{Acknowledgements}
We thank the members of Huron Digital Pathology for their continual technical support and maintenance of our online ADP platform. 

\bibliographystyle{elsarticle-num}
\biboptions{super,sort&compress}
\bibliography{cas-refs}
\appendix

\section{Annotation Tool Layout Design}

\begin{figure}[!ht]
    \centering
    \includegraphics[width=0.8\textwidth]{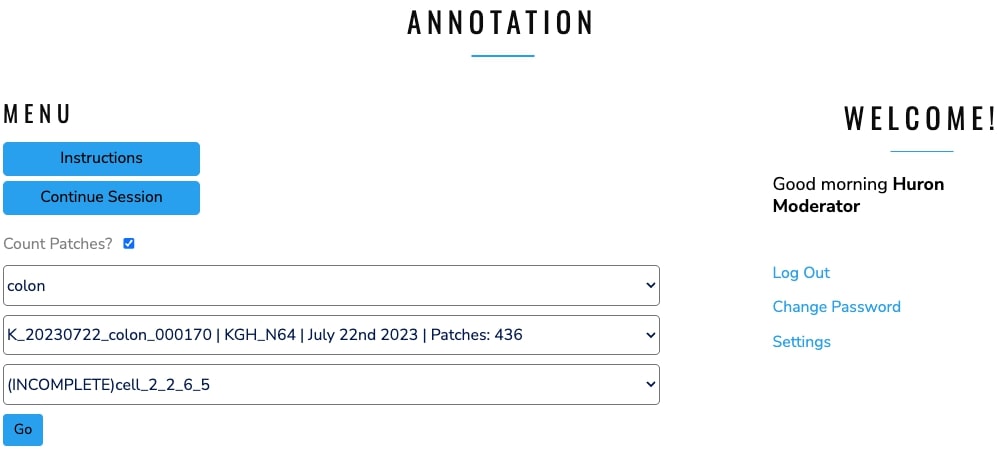}
    \caption{Patch selection page of the ADP Annotation Tool. This selection page allows users to specify the organ, the slide, and the image patch they wish to annotate. }
    \label{fig:adp_drop_down}
\end{figure}

Upon accessing the ADP annotation tool, the annotator should navigate to the patch selection page on the upper menu. An example of this page is shown in figure \ref{fig:adp_drop_down}. Here, annotators select from three drop down menus to begin annotation. The first menu is the organ group, which for ADPv2 is colon. The second menu is the slide selection, of which there are 461 slides to select from. Here, we display the name of the slide, the institute name, the upload date, and optionally, the number of patches that are available for annotation in this slide. Lastly, the annotator selects a patch for labeling. It will be denoted '(INCOMPLETE)' or '(COMPLETE)' depending on whether the annotator has previously finished annotating it already.

\begin{figure}[!ht]
    \centering
    \includegraphics[width=0.8\textwidth]{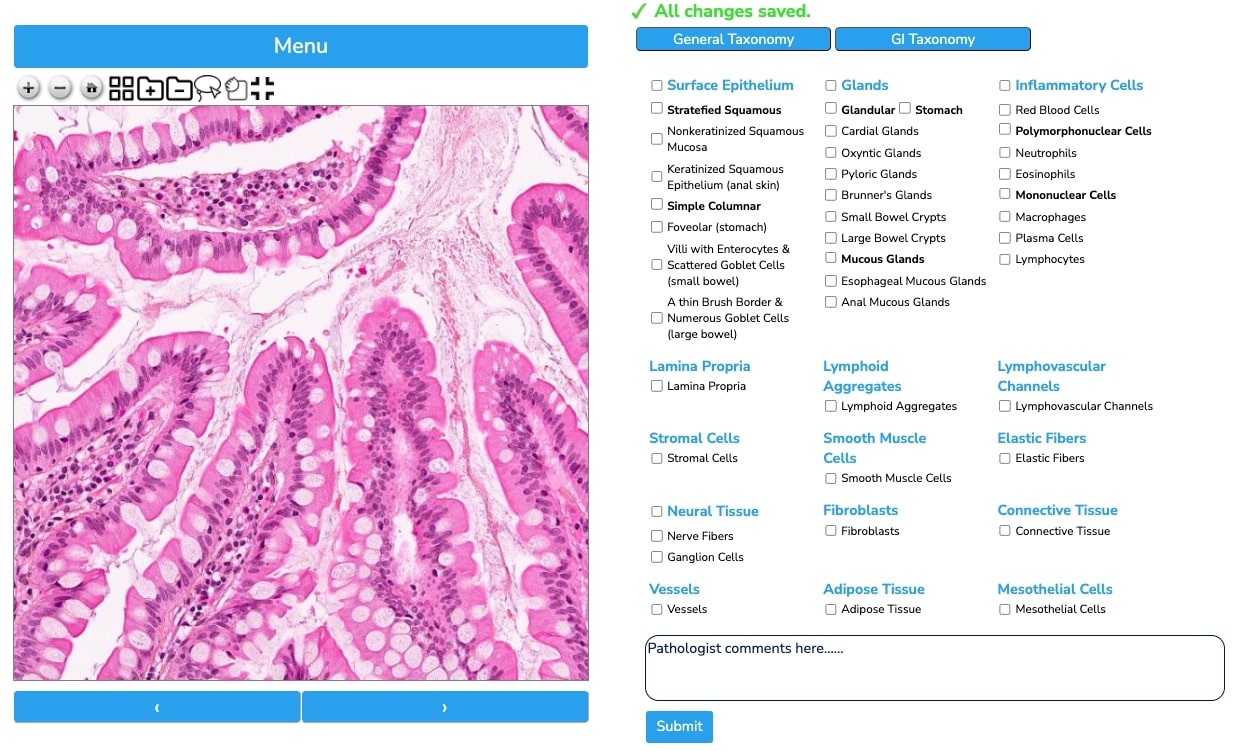}
    \caption{Annotation page of the ADP Annotation Tool. This page enables users to annotate each patch according to the hierarchical tissue taxonomy. The patch viewer on the right enables users to inspect the image patch and can be zoomed in for greater detail.}
    \label{fig:adp_annotation_menu}
\end{figure}

Clicking the 'go' button will direct the user to the annotation page which is shown in figure \ref{fig:adp_annotation_menu}. Here we have an image viewer which allows the annotator to view the full patch, and if required the annotator may zoom in or out on the image patch to see varying resolutions. On the right is the annotation menu that contains the full taxonomy described in table \ref{htt} of the main paper. Certain HTTs that are shown in blue or emboldened denote parent HTTs (granted more than one child label exists) which are available for selection during annotation. When the annotator selects a child HTT, its parents are also automatically selected, though it is an option to select a parent HTT and not a child HTT when it is ambiguous.

\begin{figure}[!ht]
    \centering
    \includegraphics[width=0.8\textwidth]{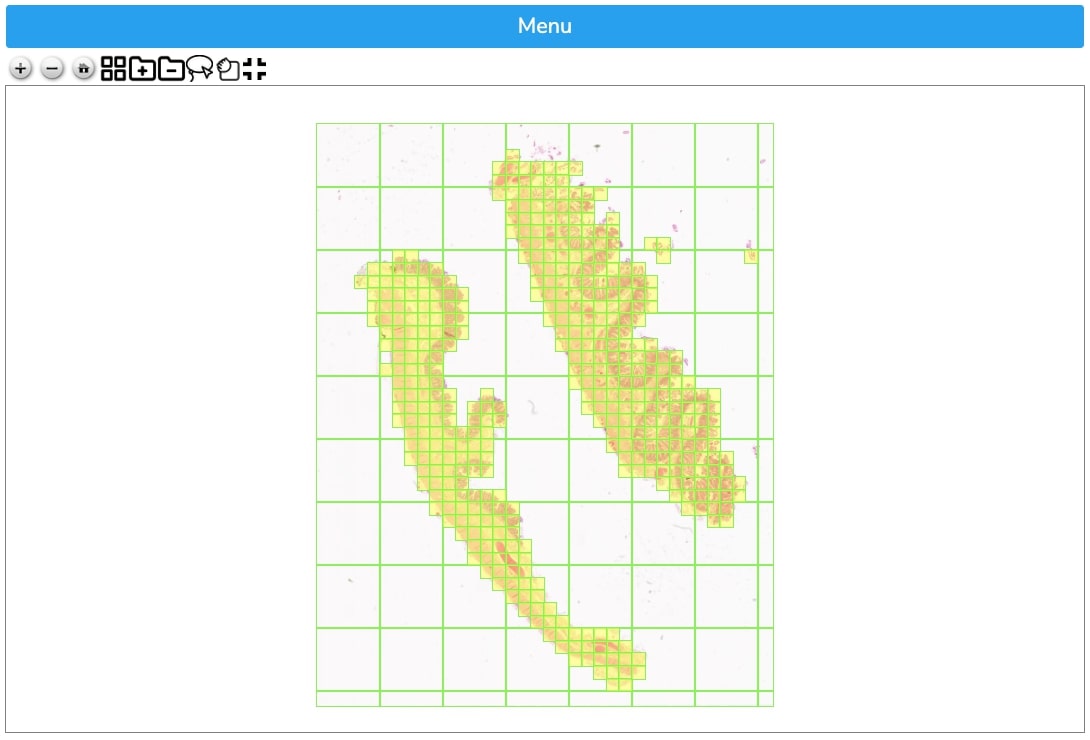}
    \caption{WSI viewer of the ADP Annotation Tool. This viewer allows users to select and deselect patches to add or remove them as potential patches to annotate. In this example, all non-background patches are available for annotation}
    \label{fig:adp_wsi_viewer}
\end{figure}

Clicking on the icon with four squares in the upper menu of the annotation page will direct the user to the zoomed-out WSI where the image patch comes from. From here, users may select another patch within the WSI to annotate. Alternatively, one can add or remove patches as needed, using the 'plus folder' and 'minus folder' icons.

\section{Implementation Details}
In this section, we provide the details of the hyperparameters in multilabel representation learning. Table \ref{tab:hyperparameters_selected} shows the detailed information for Barlow Twins pretraining. Table \ref{tab:finetuning_hyperparameters} shows the information for multilabel fine-tuning. Table \ref{tab:slide_sets} shows the statistics of the patches used in distribution analysis.

\clearpage

\begin{table}[ht]
\label{pretrain_hyperparams}
\centering
\scriptsize
\resizebox{\linewidth}{!}{%
\begin{tabular}{p{4cm} p{2.5cm} p{5.5cm}}
\hline
\textbf{Parameter} & \textbf{Default Value} & \textbf{Description} \\
\hline
\multicolumn{3}{c}{\textbf{Training Hyperparameters}} \\
Epochs & 250 & Total number of training epochs. \\
Batch Size & 448 & Mini-batch size per training step. \\
Learning Rate & 6e-05 & Initial learning rate for the optimizer. \\
Weight Decay & 1e-6 & Weight decay factor. \\
Lambd & 0.0051 & Weight on off-diagonal terms in the Barlow Twins loss. \\
\hline
\multicolumn{3}{c}{\textbf{Scheduler Information (CosineLRScheduler)}} \\
$t_{initial}$ & 250 & Total epochs (matches training epochs). \\
$lr_{min}$ & 1e-6 & Minimum learning rate. \\
$warmup_{t}$ & 10 & Number of warmup epochs. \\
$warmup_{lr_{init}}$ & 1e-6 & Initial learning rate during warmup. \\
$t_{in\_epochs}$ & True & Scheduler time unit is in epochs. \\
\hline
\multicolumn{3}{c}{\textbf{Projector Size}} \\
Projector MLP & 8192-8192-8192 & MLP configuration appended to the end of the VMamba encoder \\
\hline
\multicolumn{3}{c}{\textbf{Augmentation (Main Transform)}} \\
RandomResizedCrop & 1024 & Crop size with bicubic interpolation. \\
RandomHorizontalFlip & 0.5 & Horizontal flip probability. \\
RandomRotation & 180° & Maximum rotation angle. \\
RandomGrayscale & 0.2 & Probability to convert image to grayscale. \\
Solarization & 0.0 & Solarization probability. \\
RandStainNA (std\_hyper) & -0.3 & Stain normalization hyperparameter. \\
RandStainNA (probability) & 0.5 & Probability of applying RandStainNA. \\
Normalize (mean) & {[}0.8710, 0.6560, 0.7565{]} & Mean for image normalization. \\
Normalize (std) & {[}0.1524, 0.2428, 0.1715{]} & Standard deviation for normalization. \\
\hline
\end{tabular}%
}
\caption{Selected hyperparameters for VMamba Barlow Twins pretraining.}
\label{tab:hyperparameters_selected}
\end{table}

\begin{table}[!ht]
\centering
\caption{Dataset configuration (General Statistics) for Barlow Twins Pretraining}
\begin{tabular}{l r}
\hline
\multicolumn{2}{c}{\textbf{General Statistics}} \\
\hline
Training Images     & 115,413 \\
Original Image Size & 1360 $\times$ 1360 \\
Image Origin & Unlabelled patches extracted from ADPv2 \\
GPUs               & 4 $\times$ NVIDIA Tesla V100 (32GB VRAM) \\
\hline
\label{tab:general_stats}
\end{tabular}
\end{table}

\clearpage

\begin{table}
\centering
\scriptsize
\resizebox{\linewidth}{!}{%
\begin{tabular}{p{4cm} p{2.5cm} >{\footnotesize\raggedright\arraybackslash}p{5.5cm}}
\hline
\textbf{Parameter} & \textbf{Default Value} & \textbf{Description} \\
\hline
\multicolumn{3}{c}{\textbf{Training Hyperparameters}} \\
Epochs & 70 & Total number of training epochs. \\
Batch Size & 64 & Mini-batch size per training step. \\
Learning Rate & 0.001 & Initial learning rate for AdamW. \\
Weight Decay & 0.1 & Weight decay factor for AdamW. \\
Image Size & 1024 & Dimensions of resized images. \\
\hline
\multicolumn{3}{c}{\textbf{Scheduler Information (CosineLRScheduler)}} \\
$t_{initial}$ & 70 & Total scheduler duration in epochs \\
$lr_{min}$ & 0.0002 & Minimum learning rate. \\
$warmup_{t}$ & 10 & Number of warmup epochs. \\
$warmup_{lr_{init}}$ & 0.0002 & Initial learning rate during warmup. \\
\hline
\multicolumn{3}{c}{\textbf{Loss Hyperparameters (Asymmetric Loss)}} \\
$\gamma_{n}$ & 2 & Gamma for negative samples \\
$\gamma_{p}$ & 1 & Gamma for positive samples \\
Clip & 0.05 & Clip loss from easy negative samples\\
\hline
\multicolumn{3}{c}{\textbf{Augmentation (Training Transform)}} \\
Resize & 1024 & Resizes images to 1024$\times$1024 pixels. \\
RandomHorizontalFlip & 0.5 & Horizontal flip probability. \\
RandomVerticalFlip & 0.5 & Vertical flip probability. \\
RandomRotation & 180° & Maximum rotation angle in degrees. \\
ColorJitter (RandomApply) & brightness=0.2, contrast=0.2, saturation=0.1, hue=0.05 (p=0.2) & Random color jitter applied with probability 0.2. \\
Normalize (mean) & {[}0.8627, 0.6328, 0.7579{]} & Mean values for image normalization. \\
Normalize (std) & {[}0.1758, 0.2738, 0.1852{]} & Standard deviations for image normalization. \\
RandomErasing & True & Random erasing is applied. \\
\hline
\end{tabular}%
}
\caption{Selected hyperparameters for VMamba finetuning.}
\label{tab:finetuning_hyperparameters}
\end{table}

\clearpage

\begin{table}[ht]
\centering
\scriptsize
\begin{tabular}{lrrrrrr}
\hline
\textbf{Subtype} & \textbf{Images} & \textbf{Annotations} & \textbf{Patches} & \textbf{MPP} & \textbf{Magnification}  & \textbf{Patch Size} \\
\hline
Normal & 200 & 1132 & 2102 & 0.4 & 20x  & 1360 $\times$ 1360 \\
TVA & 217 & 843 & 26,258 & 0.4 & 20x  & 1360 $\times$ 1360 \\
TA  & 207 & 465 & 5875 & 0.4 & 20x & 1360 $\times$ 1360 \\
HP  & 212 & 284 & 2120 & 0.4 & 20x & 1360 $\times$ 1360 \\
SSL & 201 & 548 & 6510 & 0.4 & 20x & 1360 $\times$ 1360 \\
\hline
\end{tabular}
\caption{Summary of Selected Slide Sets For Distribution Analysis}
\label{tab:slide_sets}
\end{table}

\section{HTT Visual Descriptions}

\begin{longtable}{@{}p{0.3\textwidth} p{0.7\textwidth}@{}}
\caption{HTT Visual Descriptions}
\label{visual_htt}\\
\toprule
\textbf{Term} & \textbf{Visual Description} \\
\midrule
\endfirsthead
\multicolumn{2}{@{}l}{\ldots continued from previous page} \\
\toprule
\textbf{Term} & \textbf{Visual Description} \\
\midrule
\endhead

\bottomrule
\multicolumn{2}{@{}r}{\ldots continued on next page} \\
\endfoot

\bottomrule
\endlastfoot

\textbf{A thin brush border \& numerous goblet cells (large bowel) (BBNGC)} &
The crypts are lined by a single layer of cells, primarily goblet cells. The edges of the colonic crypts have many goblet cells with pink cytoplasm and dark eccentric nuclei. \\

\textbf{Large bowel crypts (LBC)} &
The colonic crypts appear as tubular structures with dark purple nuclei and pink cytoplasm for the epithelial cells. In cross-sectional cuts, the crypts appear as round or oval structures arranged in clusters. \\

\textbf{Lamina propria (LP)} &
Layer of loose connective tissue found beneath the epithelial lining of the colon and between the colonic crypts. It contains a mix of immune cells and blood vessels. \\

\textbf{Lymphoid aggregates (LA)} &
Various in size and are often round or oval. Mainly contain dark blue-stained lymphocytes. \\

\textbf{Lymphovascular channels (LC)} &
Small circular channels that may be surrounded by muscle. Thinner than vessels and lacks the elastic fibers found in vessels. \\

\textbf{Vessels (V)} &
Big circular channels surrounded by muscle cells containing elastic fibers and filled with red blood cells. \\

\textbf{Smooth muscle cells (SMC)} &
Elongated and spindle-shaped, with pink cytoplasm and dark purple, elongated nuclei. They are found in large amounts and are also located in the walls of blood vessels. \\

\textbf{Elastic fibers (EF)} &
They appear as fine, thin, and wavy structures with pink-stained connective tissue matrices. \\

\textbf{Nerve fibers (NF)} &
Typically appear as bundles composed of several wavy neuronal axons with admixture of Schwann cells and fibroblasts. \\

\textbf{Ganglion cells (GC)} &
Cells with large prominent, dark purple nuclei, that is centrally located, with a visible nucleolus that creates a "bullseye" appearance. Their pink-red cytoplasm surrounds their nucleus, providing a sharp contrast between the two. \\

\textbf{Connective tissue (CT)} &
Stroma typically appears as background non-epithelial tissue in between other tissues - yellow under HPS staining and pink under H\&E staining. Abundant between other tissues and within lamina propria.  \\

\textbf{Neutrophils (NTR)} &
Have multilobed nucleus (3 to 5 lobes) with pink intracytoplasmic granules. \\

\textbf{Eosinophils (ES)} &
Have bright pink cytoplasm with large eosinophilic granules and a dark blue, bilobed nucleus. They are medium-sized and often found in the context of allergic reactions or parasitic infections. \\

\textbf{Macrophages (MA)} &
Medium to large cells with a pink cytoplasm and a dark blue, irregularly shaped nucleus. They may contain phagocytosed material that appears as pale pink areas within the cytoplasm. \\

\textbf{Plasma cells (PC)} &
Have a deeply basophilic cytoplasm with a distinctive perinuclear halo and an eccentrically placed nucleus. They are characterized by a large, round nucleus with a dense chromatin pattern. \\

\textbf{Lymphocytes (LY)} &
Small, round cells with a dense, dark blue nucleus and scant cytoplasm. They are found throughout lymphoid tissues. \\

\textbf{Red blood cells (RBC)} &
Small pink/red cells, typically anucleated, which are abundant in vessels and lymphovascular channels. \\

\textbf{Adipose tissue (AT)} &
White areas are due to intracellular fat content. The cytoplasm of adipocytes appears white and their nuclei are noncentrally located. \\
\end{longtable}

\section{Training Curves}

\begin{figure}[ht]
\centering

\includegraphics[width=1.0\textwidth]{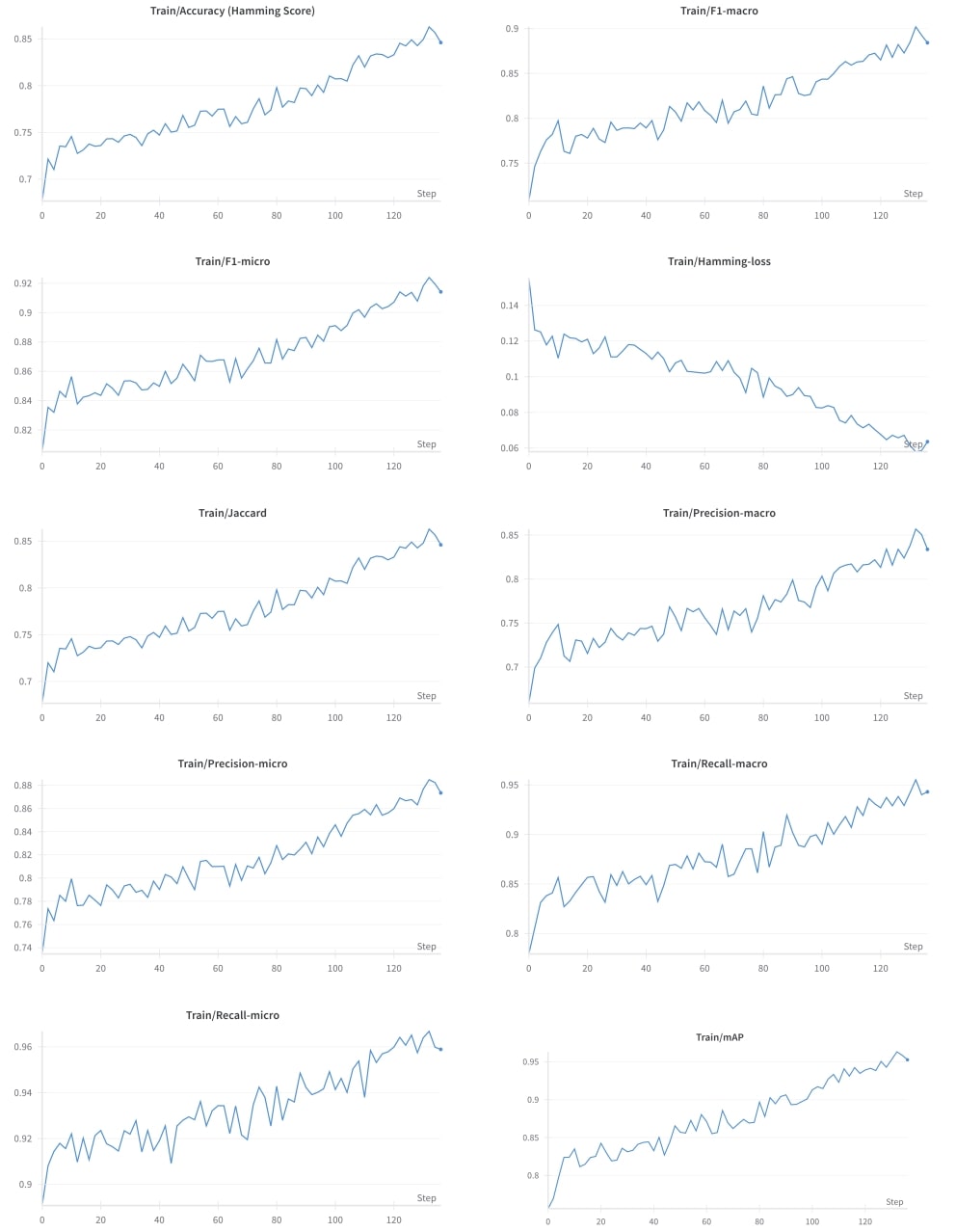}
\caption{VMamba Finetuning Training Curves}
\label{fig:vmamba_training_curves}
\end{figure}

\begin{figure}[ht]
\centering
\includegraphics[width=1.0\textwidth]{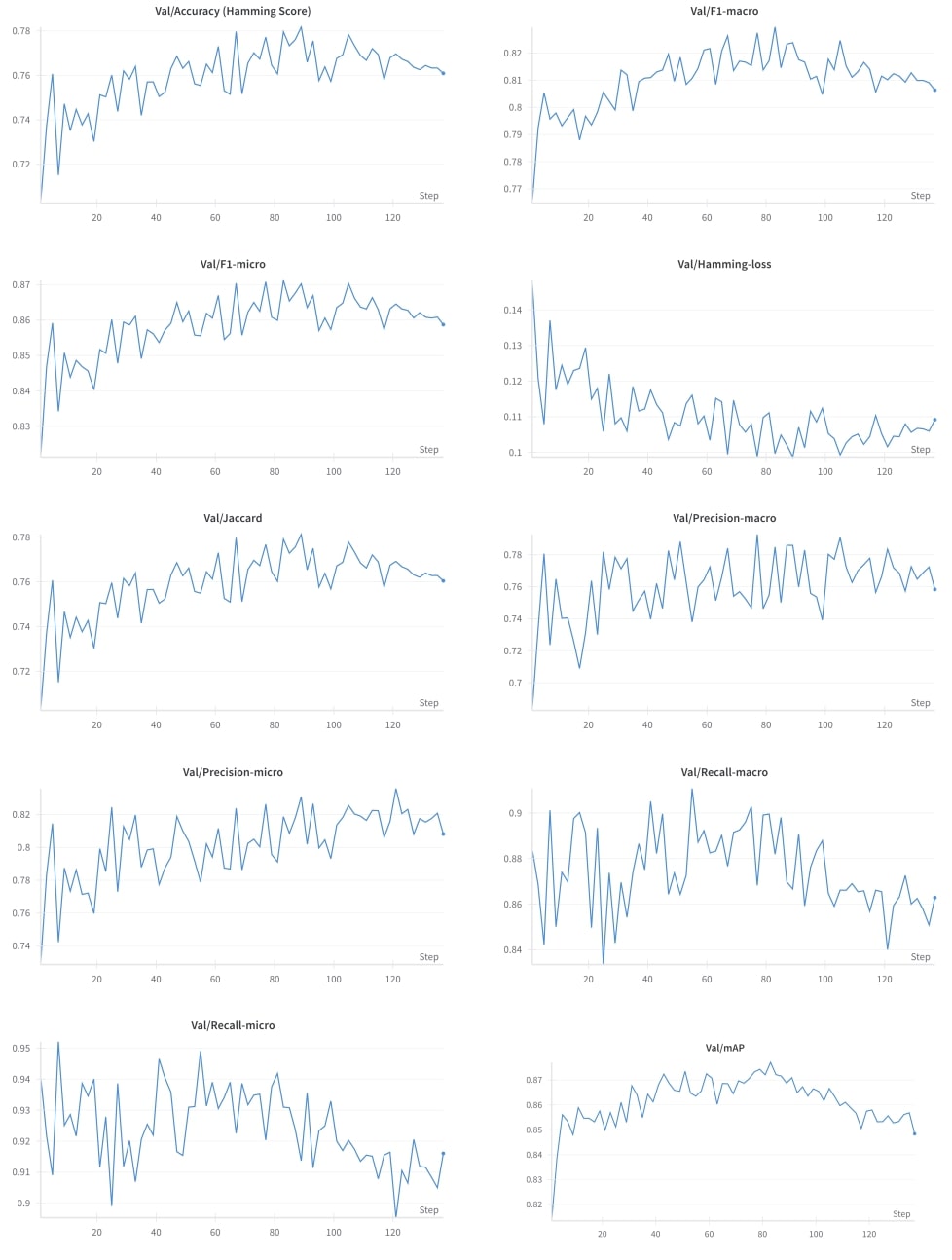}
\caption{VMamba Finetuning Validation Curves}
\label{fig:vmamba_validation_curves}
\end{figure}

\begin{figure}[ht]
    \centering
    \includegraphics[width=1.0\textwidth]{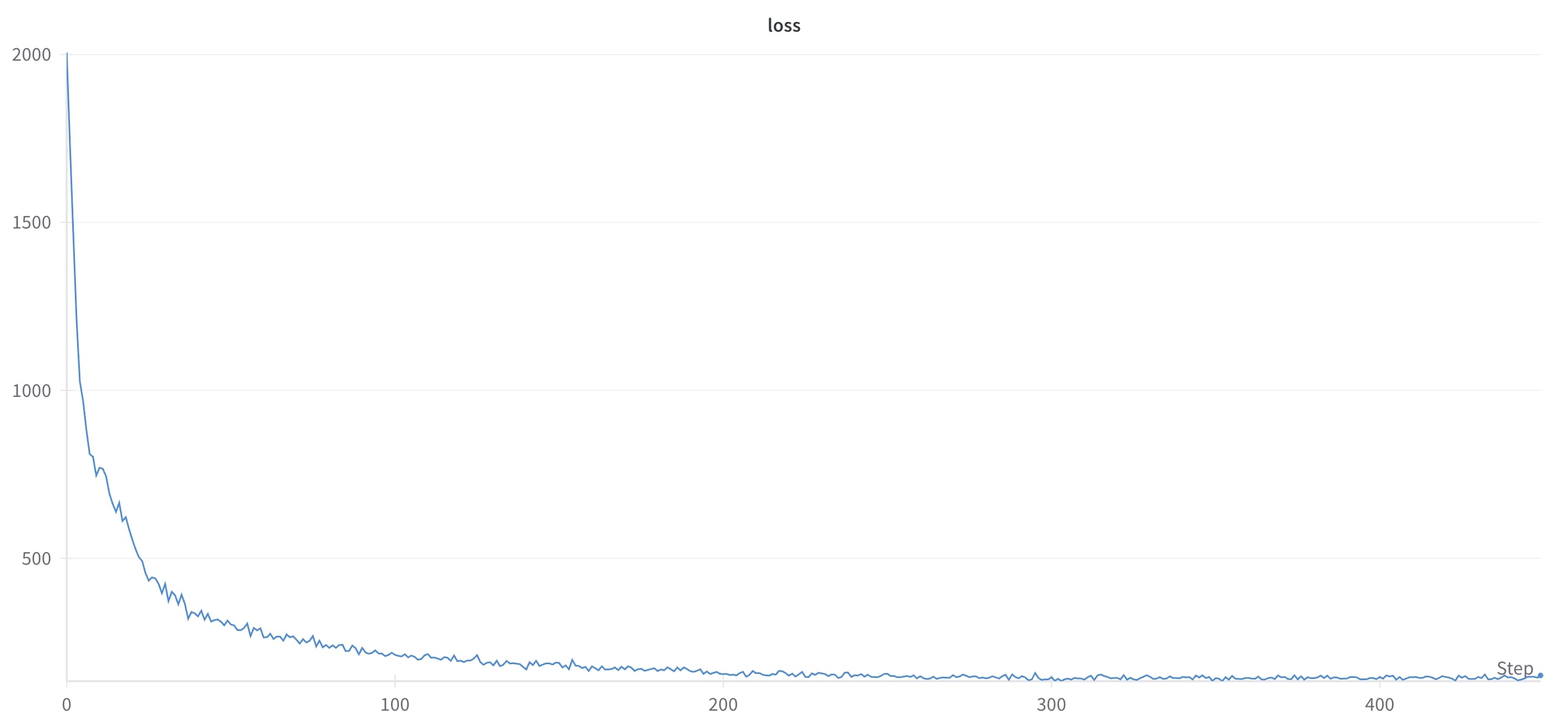}
    \caption{Pretraining Curve Vmamba Encoder via Barlow Twins}
    \label{fig:Pretrain}
\end{figure}

\begin{figure}[ht]
    \centering
    \includegraphics[width=1.0\textwidth]{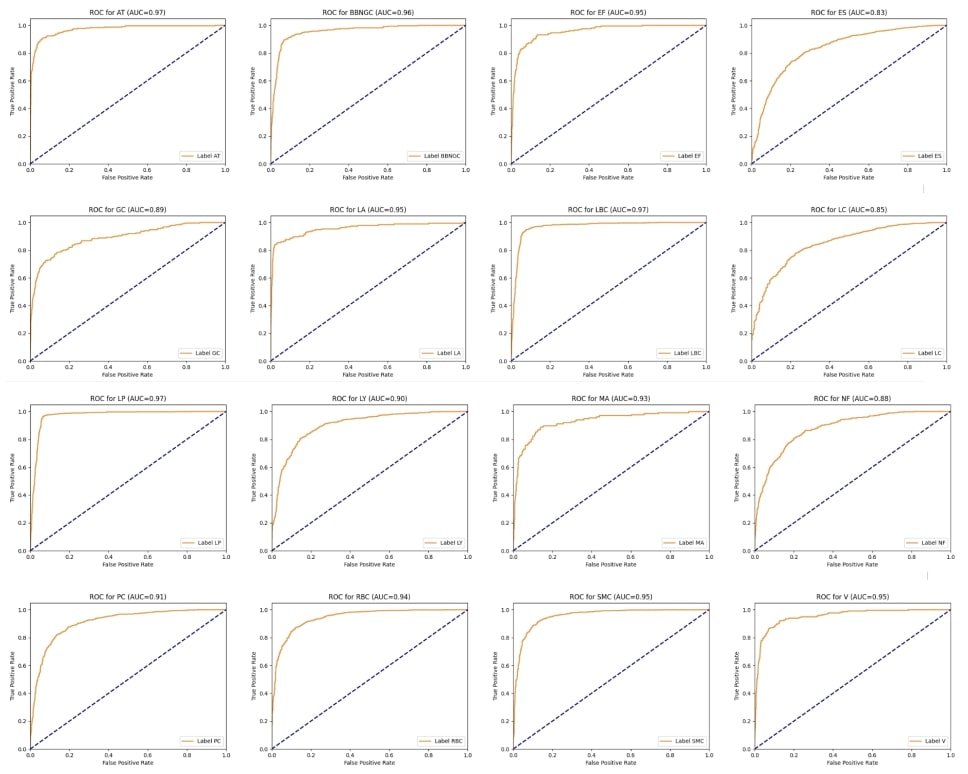}
    \caption{Receiver operating characteristic (ROC) curves of each HTT, based on test set performance}
    \label{fig:ROC}
\end{figure}

\end{document}